\documentclass[twocolumn,reprint,amsmath,superscriptaddress,amssymb,aps,nofootinbib,a4paper]{revtex4-1}\usepackage[utf8]{inputenc}
\usepackage{graphicx}
\usepackage{dcolumn}
\usepackage{comment}

\usepackage{braket}
\usepackage{bbold}
\usepackage{xcolor}
\usepackage{color,soul}
\usepackage{graphicx}
\usepackage{amssymb}
\usepackage{amsmath}
\usepackage{physics}
\usepackage{xfrac}
\makeatletter 
\renewcommand\onecolumngrid{% <<<<<<
\do@columngrid{one}{\@ne}%
\def\set@footnotewidth{\onecolumngrid}% <<<<<<<<<<<<<<<<
\def\footnoterule{\kern-6pt\hrule width 1.5in\kern6pt}%
}
\makeatother
\usepackage{geometry}
\begin{document}
\title{Multiplexed Entanglement of Multi-emitter Quantum Network Nodes}

\author{A.~Ruskuc}
\thanks{These authors contributed equally to this work.}
\affiliation{Thomas J. Watson, Sr, Laboratory of Applied Physics, California Institute of Technology, Pasadena, CA, USA}
\affiliation{Kavli Nanoscience Institute, California Institute of Technology, Pasadena, CA, USA}
\affiliation{Institute for Quantum Information and Matter, California Institute of Technology, Pasadena, CA, USA}
\affiliation{Present address: Department of Physics, Harvard University, Cambridge, MA, USA.}
\author{C.-J.~Wu}
\thanks{These authors contributed equally to this work.}
\affiliation{Thomas J. Watson, Sr, Laboratory of Applied Physics, California Institute of Technology, Pasadena, CA, USA}
\affiliation{Kavli Nanoscience Institute, California Institute of Technology, Pasadena, CA, USA}
\affiliation{Institute for Quantum Information and Matter, California Institute of Technology, Pasadena, CA, USA}
\affiliation{Division of Physics, Mathematics and Astronomy, California Institute of Technology, Pasadena, CA, USA}
\author{E.~Green}
\affiliation{Thomas J. Watson, Sr, Laboratory of Applied Physics, California Institute of Technology, Pasadena, CA, USA}
\affiliation{Kavli Nanoscience Institute, California Institute of Technology, Pasadena, CA, USA}
\affiliation{Institute for Quantum Information and Matter, California Institute of Technology, Pasadena, CA, USA}
\author{S.~L.~N.~Hermans}
\affiliation{Thomas J. Watson, Sr, Laboratory of Applied Physics, California Institute of Technology, Pasadena, CA, USA}
\affiliation{Kavli Nanoscience Institute, California Institute of Technology, Pasadena, CA, USA}
\affiliation{Institute for Quantum Information and Matter, California Institute of Technology, Pasadena, CA, USA}
\author{W.~Pajak}
\affiliation{Thomas J. Watson, Sr, Laboratory of Applied Physics, California Institute of Technology, Pasadena, CA, USA}
\affiliation{Kavli Nanoscience Institute, California Institute of Technology, Pasadena, CA, USA}
\affiliation{Institute for Quantum Information and Matter, California Institute of Technology, Pasadena, CA, USA}
\author{J.~Choi}
\affiliation{Department of Electrical Engineering, Stanford University, Stanford, CA, USA}
\author{A.~Faraon}
\email{faraon@caltech.edu}
\affiliation{Thomas J. Watson, Sr, Laboratory of Applied Physics, California Institute of Technology, Pasadena, CA, USA}
\affiliation{Kavli Nanoscience Institute, California Institute of Technology, Pasadena, CA, USA}
\affiliation{Institute for Quantum Information and Matter, California Institute of Technology, Pasadena, CA, USA}
\maketitle
\renewcommand{\figurename}{Fig.}

\textbf{
Quantum networks that distribute entanglement among remote nodes will unlock transformational technologies in quantum computing, communication, and sensing \cite{Gottesman2012,Komar2014,Ekert2014,Jiang2007}. However, state-of-the-art networks \cite{Stephenson2020,krutyanskiy2023,Daiss2021,VanLeent2022,Jing2019,Lago-Rivera2021,Liu2021,pompili2021,Delteil2016,knaut2024} utilize only a single optically-addressed qubit per node; this constrains both the quantum communication bandwidth and memory resources, greatly impeding scalability. Solid-state platforms, \cite{Zaporski_2023,Christle2015,Lukin2020a,Higginbottom2022,komza2022,Bayliss2020a,Rose2018,Martinez2022,Katharina2024,Rosenthal2023}, provide a valuable resource for multiplexed quantum networking where multiple spectrally-distinguishable qubits can be hosted in nano-scale volumes. Here we harness this resource by implementing a two-node network consisting of several rare-earth ions coupled to nanophotonic cavities \cite{Utikal2014,Kindem2020a,Xia:22,Deshmukh:23,Gritsch:23,Yang2023,Uysal2024b}. This is accomplished with a protocol that entangles distinguishable $^{171}$Yb ions through frequency-erasing photon detection combined with real-time quantum feedforward. This method is robust to slow optical frequency fluctuations occurring on timescales longer than a single entanglement attempt: a universal challenge amongst solid-state emitters. We demonstrate the enhanced functionality of these multi-emitter nodes in two ways. First, we mitigate bottlenecks to the entanglement distribution rate through multiplexed entanglement of two remote ion pairs \cite{Simon2007,Dam2017}. Secondly, we prepare multipartite W-states comprising three distinguishable ions as a resource for advanced quantum networking protocols \cite{Lipinska2018,Hondt2005}. These results lay the groundwork for scalable quantum networking based on rare-earth ions.
}

A key challenge in long-range quantum networking is overcoming the entanglement rate bottleneck caused by information propagation between distant nodes. With one qubit per node, waiting for photon transmission and classical signalling limits entanglement rates to $c/L$, where $c$ is the speed of light and $L$ is the node separation.  Multi-emitter nodes resolve this issue through multiplexing \cite{Simon2007,Dam2017}, whereby multiple photons travelling through an optical fibre are entangled with separate qubits \cite{Lago-Rivera2023,Krutyanskiy2024,Hartung2024}. This boosts entanglement rates to $Nc/L$, where $N$ is the number of emitters per node.

Solid-state systems host dense ensembles of optically-addressed qubits, providing a natural implementation of multi-emitter nodes. This is enabled by crystalline disorder leading to optical frequency shifts, facilitating independent, spectrally-resolved qubit control \cite{chen2020} (Fig.~1a,b). The key challenge in utilizing multi-emitter nodes for quantum networking is to efficiently and robustly entangle spectrally-distinguishable qubits whose frequencies also fluctuate. This task is particularly challenging due to the random nature of spontaneous photon emission used for heralding, leading to stochastic phases and maximally mixed states \cite{Vittorini2014}.

Here we show how a feedforward operation conditioned on photon emission times corrects these random phases, thereby retrieving pure, entangled quantum states (Fig.~1c). Remarkably, due to the narrow optical frequency distribution in our $^{171}$Yb:YVO$_4$ platform, we can entangle any qubits, regardless of their frequency difference. As we will show, this provides an efficient paradigm for multiplexed quantum networking, substantially increasing entanglement rates and the quantum resources available per node (Fig.~1d).

\begin{center}
\begin{figure*}[htb]
\includegraphics[width=0.8\textwidth]{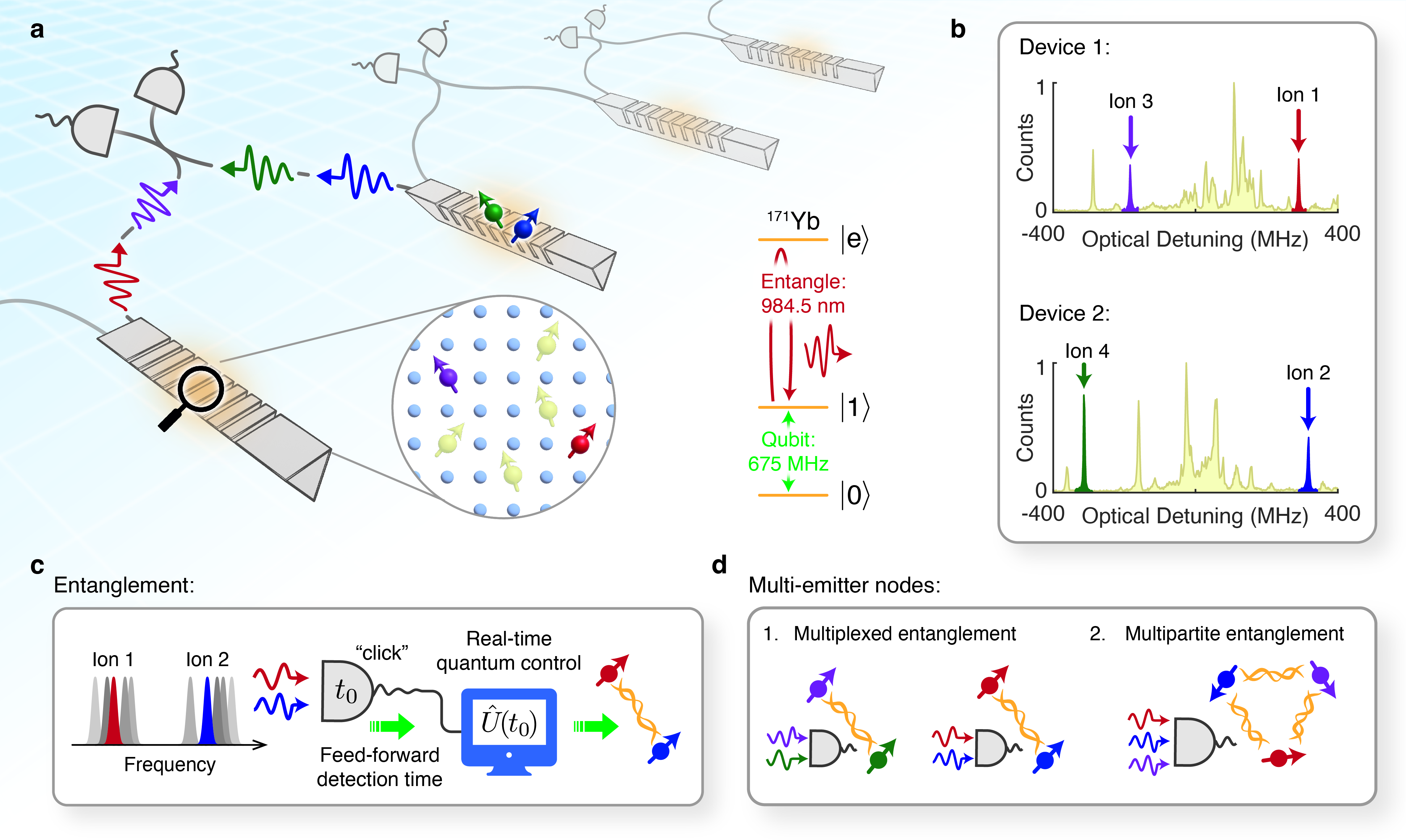}
\caption{{\bf Schematic of a quantum network link based on multiple $^{171}$Yb qubits in nanophotonic cavities}. {\bf a,}
Individual $^{171}$Yb ions in YVO$_4$ have an optical transition ($\ket{1}\leftrightarrow \ket{\text{e}}$) at 984.5~nm for readout and entanglement heralding, and a hyperfine spin transition ($\ket{0}\leftrightarrow\ket{\text{1}}$) at $2\pi\times675~$MHz for long-term memory, referred to as the qubit. Photonic emission entangled with $^{171}$Yb qubits in two remote nanophotonic cavities is interfered and measured at a central beamsplitter, forming a quantum network link. {\bf b,} Each node contains an ensemble of $^{171}$Yb ions with $\approx2\pi\times200$~MHz inhomogeneous optical frequency distribution, as shown in the normalized photoluminescence excitation spectra. In this work, four spectrally-resolved ions indicated with arrows are utilized. {\bf c,} We achieve scalable remote entanglement distribution with heralding protocols that compensate for static shifts and dynamic fluctuations in ions’ optical transition frequencies. For each heralding event, this is achieved by precisely measuring the photon emission time, $t_0$, and applying a measurement-conditioned feedforward operation, thus retrieving pure entangled states. {\bf d,} We extend these protocols to include multiple qubits per node. First, we utilize temporal multiplexing across two remote ion pairs to increase the entanglement distribution rate. Then, we utilize the expanded local Hilbert space of each node to prepare W states comprising three frequency-resolved qubits.
}
\end{figure*}
\end{center}

\subsection*{$^{171}$Yb:YVO$_4$ Quantum Network Nodes}

Our quantum networking platform consists of two nanophotonic cavities (Devices~1 and 2), fabricated from YVO$_4$ crystals which host $^{171}$Yb$^{3+}$ ions (Fig.~1a,b). These ions fulfil key requirements for quantum networking, including a ground state spin qubit (hyperfine states $\ket{0}$ and $\ket{1}$ separated by $2\pi\times675~$MHz) \cite{Kindem2018,Kindem2020a} and a coherent, cycling optical transition ($\ket{1}\leftrightarrow\ket{\text{e}}$ at 984.5~nm) verified via Hong-Ou-Mandel interferometry (Supplementary Information Section~C, Extended Data Fig.~1).

There are $\approx20$ $^{171}$Yb$^{3+}$ ions in each cavity (Fig.~1b) with a static optical inhomogeneous distribution of $\approx 2\pi\times200~$MHz. In Device~1, we use two of the seven spectrally-resolved ions (Ions~1 and 3), with $\approx2~\mu$s Purcell-enhanced lifetimes, while in Device~2, we use two of the eight spectrally-resolved ions (Ions~2 and 4) with $\approx1~\mu$s lifetimes. Each ion has a long-term optical linewidth of $\approx2\pi\times1~$MHz (defined as the standard deviation of the frequency distribution) which is $\approx10$ times broader than the lifetime limit. Using optical echo measurements, we obtain near lifetime-limited decays, confirming that linewidths are dominated by slow spectral wandering. We use a delayed echo measurement to extract an optical spectral diffusion correlation timescale of $\tau_c=1.42\pm0.04~$ms (Extended Data Fig.~2, Supplementary Information Section~D). We note that lifetime-limited emission has been observed in several rare-earth platforms \cite{Kindem2020a,Merkel2020,Ulanowski2024}.

The devices are mounted inside a $^3$He cryostat, with photons collected into separate optical fibres, combined on a beamsplitter and sent to photon detectors (Methods Section~A, Extended Data Fig.~3). Photonic emission from Ions 1--4 is detected with efficiencies of $1.100\pm0.002\%$, $0.460\pm0.005\%$, $1.396\pm0.002\%$ and $0.521\pm0.001\%$, respectively.

\subsection*{Remote Entanglement Distribution}
First, we develop a single-photon heralding protocol \cite{cabrillo1999} to efficiently and robustly entangle distinguishable ions (Fig.~2a, Extended Data Fig.~4). We demonstrate this protocol by preparing Bell states between Ions 1 and 2 in different devices, separated by an optical frequency difference of $\approx2\pi\times31~$MHz.

We initialize both ions in $\ket{0}$. Then, the optical phase drift between the device paths is measured and compensated by a laser phase rotation (Methods Section~B, Extended Data Fig.~5). Microwave pulses applied to each qubit prepare
\begin{equation*}
\ket{\psi_\text{init}}=\\(\sqrt{1-\alpha_1}\ket{0}+\sqrt{\alpha_1}\ket{1})\otimes(\sqrt{1-\alpha_2}\ket{0}+\sqrt{\alpha_2}\ket{1}),
\end{equation*}
consisting of a weak superposition with small probability, $\alpha_i$, in the $\ket{1}$ state of Ion $i$. Subsequently, each ion is optically excited with a resonant $\pi$ pulse from $\ket{1}$ to $\ket{\text{e}}$. An entangled state is heralded when a photon is detected from spontaneous emission at random time $t_0$. This eliminates the optically dark contribution, $\ket{00}$; furthermore, using small $\alpha_i$ mitigates the likelihood of double photon emission associated with $\ket{11}$. $\alpha_1$ and $\alpha_2$ are chosen to maximize the entanglement fidelity with values of 0.062 and 0.078, respectively (Methods Section~C).

\begin{center}
\begin{figure*}[htb]
\includegraphics[width=0.96\textwidth]{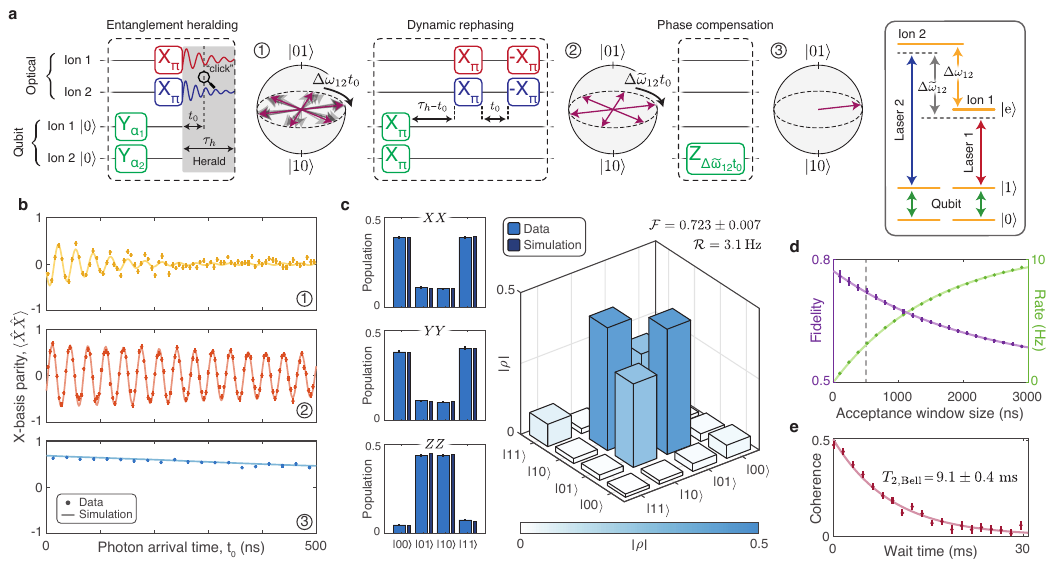}
\caption{{\bf Remote entanglement generation between two $^{171}$Yb qubits}. {\bf a,} During entanglement heralding, a photon detection at random time $t_0$ heralds the two-qubit state $\ket{\psi(t_0)}=1/\sqrt{2}(\ket{10}+e^{-i\phi(t_0)}\ket{01})$ with stochastic phase $\phi(t_0)=\Delta\omega_{12}t_0$ depending on $t_0$ and the fluctuating optical frequency difference, $\Delta\omega_{12}$ (Bloch sphere \textcircled{1}). Next, a dynamic rephasing sequence mitigates the contribution from frequency fluctuations, yielding a residual phase $\Delta\widetilde{\omega}_{12}t_0$, where $\Delta\widetilde{\omega}_{12}$ is the frequency difference between lasers used to drive the ions (Bloch sphere \textcircled{2}). We compensate this phase, leading to a deterministic Bell state, $1/\sqrt{2}(\ket{01}+\ket{10})$, independent of $t_0$ (Bloch sphere \textcircled{3}). Operations depend on the photon arrival time and are dynamically adjusted via real-time feedforward.
{\bf b,} Entangled state coherence between Ions 1 and 2 is correlated with photon measurement time, $t_0$, at three different points (\textcircled{1}, \textcircled{2} and \textcircled{3}, see Fig.~2a). Panel \textcircled{1}: without phase correction the coherence exhibits parity oscillation at the optical frequency difference, $\approx2\pi\times31~$MHz, and decay from optical frequency fluctuation. Panel \textcircled{2}: after dynamic rephasing, coherence is extended and oscillates at the static frequency difference between the lasers used to drive Ions 1 and 2, $\Delta\widetilde{\omega}_{12}$. Panel \textcircled{3}: residual phase variation, $\phi(t_0)=\Delta\widetilde{\omega}_{12}t_0$, is compensated yielding a deterministic Bell state.
{\bf c,} Quantum state tomography of the phase-corrected Bell state with 500~ns photon acceptance window yields a fidelity and rate of $\mathcal{F}=0.723\pm0.007$ and 3.1~Hz respectively. Populations in 9 two-qubit Pauli bases are measured for the tomography; $XX$, $YY$ and $ZZ$ results are plotted in addition to the absolute values of the resulting density matrix, $|\hat{\rho}|$.
{\bf d,} As the acceptance window is varied from 100~ns to 2900~ns, the entanglement fidelity and rate range from $\mathcal{F}=0.758\pm0.016$ and $\mathcal{R}=0.73~$Hz, to $\mathcal{F}=0.588\pm0.004$ and $\mathcal{R}=9.4$~Hz, respectively.
{\bf e,} The entangled state is stored using XY-8 dynamical decoupling with $1/e$ memory time $T_{2,\text{Bell}}=9.1\pm0.4~$ms. In {\bf b} solid lines are from simulations, in {\bf d} and {\bf e} they are fits to exponential decays.}
\end{figure*}
\end{center}

The heralded state has the form
\begin{equation}
\label{ramsey_state}
\ket{\psi(t_0)} = a(t_0) \ket{10} + b(t_0) \ket{01} e^{-i\phi(t_0)},
\end{equation}
with the random, $t_0$-dependent phase $\phi(t_0) = \Delta \omega_{12}\times t_0$. $\Delta\omega_{12}=\omega_2-\omega_1$ is the optical frequency difference between the ions \cite{Hermans2023}. Here, $a(t_0)$ and $b(t_0)$ are real coefficients (Methods Section~C); for simplicity, we consider $a(t_0) = b(t_0) = 1/\sqrt{2}$ in the subsequent discussion.

We characterize entangled state coherence by measuring a parity oscillation of $\ket{\psi(t_0)}$ correlated with the stochastic photon detection time, $t_0$ (Fig.~2b, Panel \textcircled{1}). Oscillations between $\ket{\psi^+}=1/\sqrt{2}(\ket{01}+\ket{10})$ and $\ket{\psi^-}=1/\sqrt{2}(\ket{01}-\ket{10})$ at $\Delta\omega_{12} \approx 2\pi \times 31$MHz are observed via measurement of the $X$-basis parity expectation $\bra{\psi(t_0)}\hat{X}\hat{X}\ket{\psi(t_0)}$ as a function of $t_0$. Over time-integrated measurement, $\Delta\omega_{12}$ fluctuates around its mean value, $\langle\Delta\omega_{12}\rangle$, causing a Gaussian decay with $1/e$ timescale $185\pm15$~ns. Due to the long optical frequency correlation time ($\tau_c=1.42\pm0.04~$ms, Extended Data Fig.~2), $\Delta\omega_{12}$ is quasi-static during a single heralding attempt and treated as a Gaussian-distributed shot-to-shot random variable.

We combat the oscillation and decay of coherence with the subsequent stages of this protocol. First, we develop a dynamic rephasing sequence to cancel the effect of optical and qubit frequency fluctuations. Within the heralding window of duration $\tau_h=2.9~\mu$s, the ions spent a duration $t_0$ in a superposition of $\ket{0}$ and $\ket{\text{e}}$ undergoing optical dephasing. After spontaneous emission, they spent the remaining $\tau_h-t_0$ in a superposition of $\ket{0}$ and $\ket{1}$ undergoing qubit dephasing (Supplementary Information Section~F). The amount of dephasing changes from shot-to-shot due to variable optical and qubit transition frequencies {\it and} random photon emission times, $t_0$. In the dynamic rephasing protocol, the durations of subsequent evolution periods are adjusted in real-time based on the previously measured photon emission time. Specifically, after the heralding window, we apply $\pi$ pulses to each qubit and wait for $\tau_h-t_0$, thereby rephasing qubit coherence. Then, we transfer population from $\ket{1}$ to $\ket{\text{e}}$ with optical $\pi$ pulses applied to both ions, wait for $t_0$ to rephase the optical coherence, and coherently transfer population back to $\ket{1}$ with a second pair of $\pi$ pulses. This requires optical transition frequency stability between entanglement heralding and dynamic rephasing, which is ensured by the long correlation timescale ($\tau_c=1.42\pm0.04~$ms) and holds true even for long-range entanglement links (Supplementary Information Section~F).

Consequently, parity oscillations persist to longer photon detection times, verifying that the effect of optical frequency fluctuations has been mitigated (Fig.~2b, Panel \textcircled{2}). Unlike previously where the oscillation frequency was determined by the fluctuating difference between ions' optical frequencies, here, it is stably set by the driving laser frequency difference, $\Delta\widetilde{\omega}_{12}$. At later photon detection times, the contrast decays exponentially due to spontaneous emission during optical rephasing (Extended Data Fig.~6): ions must remain in $\ket{e}$ for a time $t_0$, and any undetected emission destroys the entangled state coherence. Hence, the decay timescale is determined by the ions' optical lifetimes (Supplementary Information Section~F).

While $\Delta\widetilde{\omega}_{12}$ is static, the residual phase, $\phi(t_0)=\Delta\widetilde{\omega}_{12} t_0$, is random due to the stochastic nature of $t_0$, requiring real-time phase compensation \cite{Vittorini2014}. The second stage of our protocol counteracts this by applying a $Z$-rotation to Ion 2's qubit (between $\ket{0}$ and $\ket{1}$) by an angle $\phi(t_0)$. Panel \textcircled{3} in Fig.~2b shows that the phase is successfully compensated, and a {\it deterministic} Bell state, $\ket{\psi^+}$, is heralded regardless of $t_0$.

The entangled state is verified by measuring populations in nine cardinal two-qubit Pauli bases along the $X$-, $Y$- and $Z$-axes and performing maximum likelihood tomography to reconstruct the density matrix, $\hat{\rho}$ \cite{James2001}. The $XX$, $YY$ and $ZZ$ populations, and magnitude of $\hat{\rho}$, all obtained with a 500~ns acceptance window (i.e. accepting photons if $0<t_0<500~$ns), are plotted in Fig.~2c. All results presented in this work are corrected for readout infidelity (Supplementary Information Section~P). The resulting entanglement fidelity is $\mathcal{F}=0.723\pm0.007$ with a heralding rate of $\mathcal{R}=3.1$~Hz. Simulation results with only one free parameter (a slight correction to Ion 1's photon collection efficiency) predict a fidelity of $\mathcal{F}_\text{sim}=0.729\pm0.004$, showing agreement within standard error (Supplementary Information Section~H). There is an anticorrelation between fidelity and heralding rate  with photon acceptance window size; values range from $\mathcal{F}=0.758\pm0.016$ and $\mathcal{R}=0.73~$Hz to $\mathcal{F}=0.588\pm0.004$ and $\mathcal{R}=9.4~$Hz, respectively, as the window size is increased from 100~ns to 2900~ns (Fig.~2d). Finally, we probe the entangled state storage time, by applying XY-8 dynamical decoupling, yielding a 1/e decoherence time of $T_{2,\text{Bell}}=9.1\pm0.4~$ms (Fig.~2e) limited by the spin coherence of individual qubits.

Our simulation identifies three dominant errors limiting entanglement fidelities: photon emission from weakly coupled ions (noise counts) causing heralding events uncorrelated with the qubit states; spontaneous emission during optical $\pi$ pulses; and lifetime-limited decay of entangled state coherence leading to reduced fidelities for later photon detection events. These errors contribute roughly equally (Extended Data Fig.~7a, Supplementary Information Section~H); we propose mitigation strategies in the Outlook.

The entanglement rate is determined by the 12.3~kHz experiment repetition rate and $(2.83\pm0.02)\times10^{-4}$ success probability (Extended Data Fig.~7b, Supplementary Information Section~I). The repetition rate is limited by ion initialization which takes 70$\%$ of each attempt's duration; the next section shows how entanglement multiplexing addresses this.

Note that the qubit readout scheme, which is designed to mitigate against photon loss, also post-selects outcomes where ions occupy the qubit manifold. Hence, erroneous occupation of auxiliary ground state levels would lead to a slight over-estimate of the entanglement fidelity (by a factor of $\lesssim1.06$: Supplementary Information Section~J). To resolve this, we implement a two-photon entanglement protocol \cite{Barrett2005}, where auxiliary state occupation is carved out at the heralding stage. This leads to an unconditional (non-post-selected) fidelity of $\mathcal{F}=0.81\pm0.02$, albeit with a reduced rate of $\mathcal{R}=0.049~$Hz (Supplementary Information Section~K, Extended Data Fig.~8). This protocol also eliminates infidelity from $\ket{11}$ and, due to higher bright-state populations, reduces the relative impact of noise counts.

In Supplementary Information Section~L, we extend this two-photon scheme to probabilistically teleport quantum states between the two ions, achieving an average state transfer fidelity of $0.834\pm0.011$, which exceeds the classical bound of $2/3$.

\subsection*{Multiplexed Rate Enhancement}

Next, we demonstrate how multiplexing boosts quantum communication rates by parallelizing the limiting step of an entanglement protocol across multiple remote ion pairs. For long-distance networking, this removes bottlenecks associated with propagation time between nodes \cite{Simon2007,Dam2017}; however, for our short-range network it mitigates limitations imposed by the ion initialization time.

We use four optically distinguishable qubits in a pairwise configuration: Pair 1 consists of Ions 1 and 2; Pair 2 consists of Ions 3 and 4, separated by  $\approx 2\pi\times104~$MHz (Fig.~3a). We initialize all ions in parallel followed by consecutive entanglement attempts on Pair 2 and then Pair 1 using the protocol in Fig.~2. If a heralding photon is detected during either attempt, the experiment proceeds exclusively with the corresponding ion pair. Specifically, we perform dynamic rephasing and phase compensation to prepare a deterministic entangled state, followed by two-qubit parity measurements in the $X$, $Y$ or $Z$ bases to characterize the fidelity. Microwave pulses are globally applied to qubits within the same device; thus, driving frequencies are selected to be intermediate between co-located qubits.

Figure~3b compares the multiplexed entanglement rates and fidelities with non-multiplexed measurements that exclusively entangle Pair 1 {\it or} Pair 2 with resonant qubit pulses. The multiplexed rate and fidelity range from $\mathcal{F}_\text{mult}=0.714\pm0.008$ and $\mathcal{R}_\text{mult}=1.6~$Hz to $\mathcal{F}_\text{mult}=0.571\pm0.002$ and $\mathcal{R}_\text{mult}=23$~Hz, as the photon acceptance window increases from 100~ns to 2900~ns. Across this range, multiplexing boosts the entanglement rate by a factor of $\approx1.9$. This exceeds the theoretical prediction of 1.7, which is explained by the off-resonant nature of dynamical decoupling pulses applied to Pair 1 qubits prior to entanglement heralding, leading to larger probabilities of $\ket{1}$ state occupation, $\alpha_1$ and $\alpha_2$. This increases the rate, albeit at the expense of reduced fidelity (Supplementary Information Section~M).

With a $500~$ns photon acceptance window, we measure a multiplexed fidelity of $\mathcal{F}_\text{mult}=0.682\pm0.004$, composed of $\mathcal{F}_\text{mult}^{(1)}=0.700\pm0.005$ and $\mathcal{F}_\text{mult}^{(2)}=0.668\pm0.006$ for Pair~1 and Pair~2 Bell states, respectively (Extended Data Fig.~9). These compare to independently measured fidelities of $\mathcal{F}_1=0.745\pm0.006$ for Pair~1 and $\mathcal{F}_2=0.656\pm0.007$ for Pair~2. Supplementary Information Section~M shows that the multiplexed entanglement fidelity reduction mostly arises from the off-resonant qubit pulses. We also confirm negligible cross-talk between ions in the same device.

\begin{figure}[h!]
\includegraphics[width=0.48\textwidth]{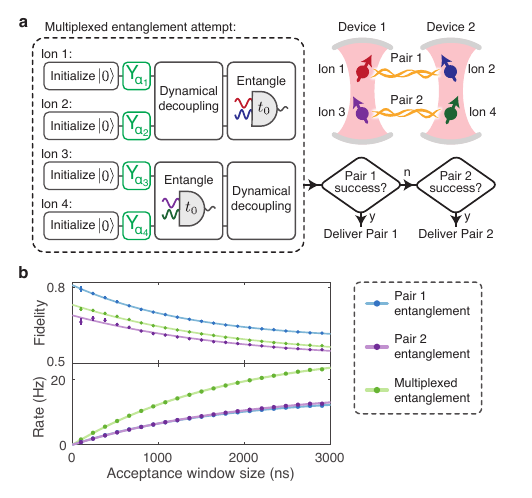}
\caption{{\bf Multiplexed entanglement rate enhancement using two pairs of $^{171}$Yb spin qubits}. {\bf a,} After initializing all four ions, two consecutive entanglement attempts are performed, first on Pair 2 (consisting of Ions 3 and 4), then on Pair 1 (consisting of Ions 1 and 2). If either entanglement attempt is successful, the sequence proceeds exclusively with the corresponding ion pair. We apply the dynamic rephasing and phase compensation protocols, before reading out the two-qubit state parity in the the $X$, $Y$ or $Z$ bases. Since the ion initialization time bottlenecks our entanglement rate, parallelizing this step across all four ions leads to delivery of entangled states at an enhanced rate. A detailed experiment sequence can be found in Extended Data Fig.~9. {\bf b,} As the photon acceptance window is varied from 100~ns to 2900~ns, the entanglement rate and Bell state fidelity for the multiplexed protocol range from $\mathcal{R}=1.6~$Hz and $\mathcal{F}=0.714\pm0.008$ to $\mathcal{R}=23$~Hz and $\mathcal{F}=0.571\pm0.002$, respectively (purple markers). Compared to optimized (non-multiplexed) entanglement experiments for Pair 1 or Pair 2 (blue and green markers, respectively), we observe similar fidelities for the multiplexed protocol which increases the rate by a factor of $\approx1.9$. Solid lines are fits to exponential decays.
}
\end{figure}

\subsection*{Tripartite W State Generation}
We demonstrate the versatility and scalability of our protocol and multi-qubit nodes by heralding multipartite W states on three optically distinguishable qubits (Ions~1, 2 and 3). These are an important class of entangled states for quantum networking \cite{Lipinska2018,Hondt2005} that can be distributed over long distances using repeater protocols \cite{Ramiro2023}.

To generate these states, each qubit starts in a weak superposition, suppressing contributions with multiple spin excitations (e.g. $\ket{110}$, $\ket{111}$). All three ions are resonantly optically excited, and a photon detection carves out the optically dark state, $\ket{000}$. Neglecting contributions from multiple spin excitations (Methods Section~E) leads to a heralded W-state:
\begin{equation*}
\ket{\psi(t_0)}=\frac{1}{\sqrt{3}}(\ket{100}+\ket{010}e^{-i\Delta\omega_{12}t_0}+\ket{001} e^{-i\Delta\omega_{13}t_0}),
\end{equation*}
(Fig.~4a and 4b), where $t_0$ is the photon detection time and $\Delta\omega_{ij}$ is the fluctuating optical frequency difference between ions $i$ and $j$.

We perform dynamic rephasing to counteract these fluctuations, leading to
\begin{equation*}
    \ket{\psi(t_0)}=1/\sqrt{3}(\ket{100} + \ket{010}e^{-i\Delta\widetilde{\omega}_{12}t_0} + \ket{001}e^{-i\Delta\widetilde{\omega}_{13}t_0}).
\end{equation*}
Two residual stochastic phases need to be compensated: $\Delta\widetilde{\omega}_{12}t_0$ and $\Delta\widetilde{\omega}_{13}t_0$, where $\Delta\widetilde{\omega}_{ij}$ is the static frequency difference between lasers used to drive Ions $i$ and $j$. As before, phase differences between ions in different devices are compensated by local $Z$-rotations. However, for qubits within the same device, global microwave driving precludes this approach. Instead, we utilize a differential AC-Stark shift, enabled by optical inhomogeneity (Methods Sections~D,~E, ref.~\cite{chen2020}).

The phase-stabilized, three-qubit W-state, $\ket{\text{W}}=1/\sqrt{3}\left(\ket{100}+\ket{010}+\ket{001}\right)$, is verified by measuring populations in 27 cardinal three-qubit Pauli bases (along $X$, $Y$ and $Z$) and performing maximum likelihood tomography to reconstruct the density matrix (Fig.~4c). With a photon acceptance window size of $300~$ns, the W-state generation rate and fidelity are $\mathcal{R}=2.0~$Hz and $\mathcal{F}=0.592\pm0.007$, respectively.

\subsection*{Outlook}
We have shown that multi-emitter nodes are a powerful tool for quantum networks. They enable scaling of remote entanglement distribution rates---through multiplexing---and generation of multipartite entangled states. This was achieved with a protocol that counteracts fluctuating optical frequency differences, facilitating efficient entanglement of any combination of distinguishable emitters. With technological advancements in detector timing resolution \cite{Korzh2020}, this protocol will be applicable to solid-state platforms with larger optical frequency separations.

By mitigating the three dominant sources of error, we predict an increase in entangled state fidelity to $\mathcal{F}=0.883\pm0.004$ (Supplementary Information Section~H). Specifically, photon emission from weakly coupled ions can be counteracted using YVO$_4$ samples with lower concentrations of $^{171}$Yb or host materials with fewer rare-earth impurities \cite{Gritsch:23,Uysal2024b,Xia:22,Yang2023}. Control errors originating from spontaneous emission during optical pulses can be mitigated using larger optical Rabi frequencies. Finally, undetected secondary spontaneous emission events during optical rephasing that destroy the entangled state coherence can be counteracted by temporarily extending the optical lifetime using a dynamically tunable cavity \cite{Casabone2021,Xia:22,Yu2023,Yang2023}. By eliminating these undesirable photon emissions, we could also increase the heralding window size without reducing the entanglement fidelity, thereby boosting efficiency.

We will further increase entanglement rates by fabricating critically-coupled cavities with optimized fibre coupling, targeting single-pair rates of $\sim230$~Hz (Supplementary Information Section~I). Using all spectrally-resolved ions we can extend the multiplexing protocol to 7 pairs (Supplementary Information Figs.~S1 and S2) and boost rates by a factor of 4. Further expansion can be achieved by tailoring the optical inhomogeneous distribution and increasing $^{171}$Yb concentration \cite{Ulanowski2024}. Ultimately, the multiplexing capacity will be determined by the detector timing resolution (restricting separations to $\lesssim2\pi\times2~$GHz) and the ion linewidths ($\approx2\pi\times1~$MHz).

Entanglement multiplexing using these multi-emitter nodes will be crucial in overcoming the entanglement rate bottleneck associated with photon propagation times over long distances \cite{Simon2007,Dam2017}. To do this, we will demonstrate single-photon conversion to telecom wavelengths, \cite{knaut2024,stolk2024}, enabling entanglement distribution over distances $>10$~km. Due to the high spectral stability of our emitters (with correlation timescale, $\tau_c=1.42\pm0.04~$ms), our protocol is robust to extremely long photon propagation distances with $<1.5\%$ fidelity reduction (Supplementary Information Section~F).

We will also combine these multiplexed links with nuclear spin quantum memories \cite{Ruskuc2022} and intra-node entanglement between $^{171}$Yb qubits (Extended Data Fig.~10) to implement a multiplexed quantum repeater, outlined in Supplementary Information Section~Q. With these advancements, multi-emitter nodes will greatly expand scaling opportunities, enabling a new generation of advanced, long-range quantum networks.

While completing this manuscript, we became aware of related theoretical work proposing a similar protocol \cite{Uysal2024}.

\begin{figure}[h!]
\includegraphics[width=0.48\textwidth]{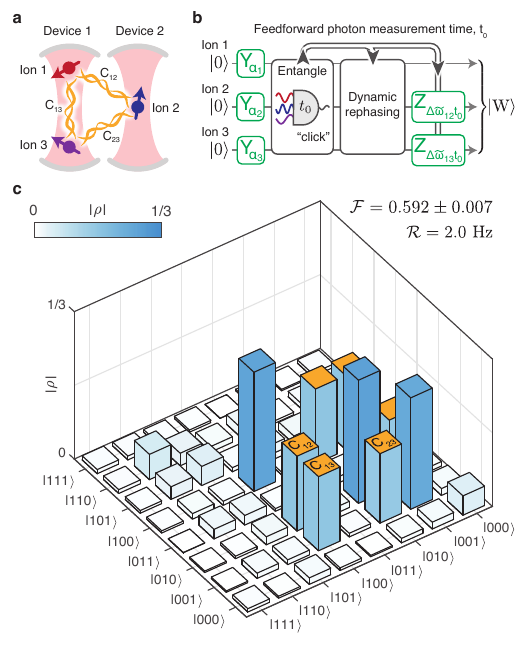}
\caption{{\bf Tripartite W-state generation between three $^{171}$Yb spin qubits}. {\bf a,} W states are prepared on three distinguishable ions, whereby Ions 1 and 3 are co-located in Device 1 and Ion 2 is in Device 2. $C_{ij}$ is the two-body qubit coherence between Ions $i$ and $j$. {\bf b,} These tripartite W states are generated using a similar single-photon protocol as Fig.~2. Each ion is prepared in a weak superposition state, subsequently, all three ions are optically excited. Entanglement is heralded by the detection of a single photon at time $t_0$, thereby preparing the tripartite W state $\ket{\psi(t_0)}=1/\sqrt{3}(\ket{100} + \ket{010}e^{-i\Delta\omega_{12}t_0} + \ket{001}e^{-i\Delta\omega_{13}t_0})$ where $\Delta\omega_{ij}$ is the fluctuating optical frequency difference between Ions $i$ and $j$. We dynamically rephase optical and spin decoherence on all three ions leading to $\ket{\psi(t_0)}=1/\sqrt{3}(\ket{100} + \ket{010}e^{-i\Delta\widetilde{\omega}_{12}t_0} + \ket{001}e^{-i\Delta\widetilde{\omega}_{13}t_0})$ with $\Delta\widetilde{\omega}_{ij}$ the frequency difference between lasers used to drive Ions $i$ and $j$. The two $t_0$-dependent stochastic phases are compensated in real-time using single qubit $Z$-rotations based on a combination of a drive phase shift and an AC stark shift of the qubit transitions (Methods Section~E, ref.~\cite{chen2020}). Ultimately, this generates a canonical W state with the form $\ket{\text{W}}=1/\sqrt{3}(\ket{100} + \ket{010} + \ket{001})$. {\bf c,} After performing the entire protocol, we measure the quantum state in all 27 three-qubit Pauli bases along $X$, $Y$ and $Z$. With a $300~$ns photon acceptance window we measure an entanglement rate of $\mathcal{R}=2.0~$Hz, we use maximum likelihood quantum state tomography to extract the density matrix with a measured fidelity of $\mathcal{F}=0.592\pm0.007$.}
\end{figure}

\clearpage

\bibliographystyle{naturemag}
\bibliography{ybybpaper.bib}

\clearpage
\newpage
\onecolumngrid
\setcounter{figure}{0}    
\renewcommand{\figurename}{Extended Data Fig.}

\section*{Extended Data}
\begin{figure}[h!]
\includegraphics[width=140mm]{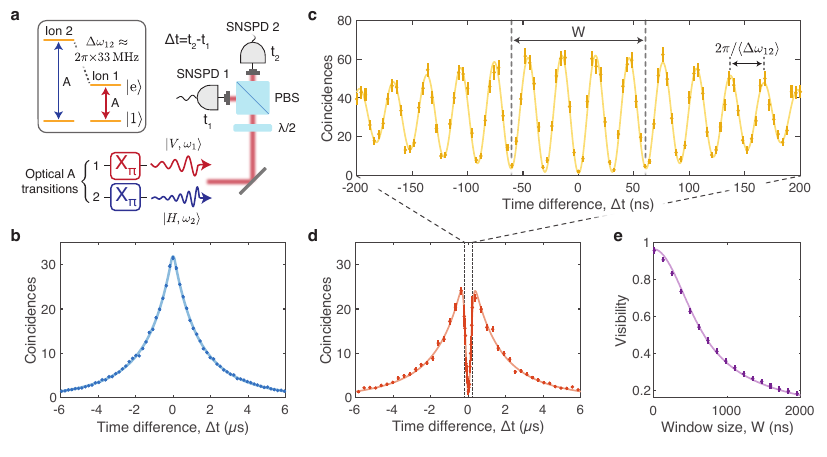}
\caption{{\bf Hong-Ou-Mandel indistinguishability measurement on photons emitted by two remote $^{171}$Yb ions}. {\bf a,} Experimental setup for Hong-Ou-Mandel (HOM) measurements. Orthogonally polarized photons emitted from the $A$ transitions of two ions with frequency difference $\omega_2-\omega_1=\Delta\omega_{12}\approx2\pi\times33~$MHz (denoted $\ket{V,\omega_1}$ and $\ket{H,\omega_2}$) are incident on a polarizing beamsplitter (PBS). Photon detections on two superconducting nanowire single photon detectors (SNSPD 1 and SNSPD 2 with detection times $t_1$ and $t_2$, respectively) are correlated. We maximize the HOM contrast by rotating the half waveplate ($\lambda$/2). The ions' optical frequencies drift on week-long timescales, hence $\Delta\omega_{12}$ is slightly different here compared to Fig.~2. {\bf b,} After optical excitation of both ions with resonant $\pi$ pulses, photon coincidences are histogrammed with respect to the detection time difference ($\Delta t=t_2-t_1$) using a 160~ns bin size. We don't see any signature of HOM interference due to the $\approx2\pi\times33~$MHz frequency difference, rendering the photons distinguishable. {\bf c,} Frequency information is erased by reducing the bin size below $1/\Delta\omega_{12}$. We use a bin size of $4~$ns and observe a quantum beat in coincidences, i.e. an oscillation between photon bunching and anti-bunching, at the optical frequency difference. Note that the number of coincidences has been re-normalized by the bin width to provide a direct comparison with {\bf b}. {\bf d,} We plot the number of coincidences using a $4~$ns bin size but only at the trough of each oscillation (i.e. at the points of maximum anti-bunching), thereby recovering a conventional Hong-Ou-Mandel dip. Dip width is limited by optical Ramsey decoherence of the two ions between consecutive emission events. {\bf e,} We extract a HOM visibility from the oscillation contrast, this is averaged over a two-sided window with width $W$ i.e. $-W/2<\Delta t<W/2$. For the smallest window size of 6~ns, a $96\pm2\%$ visibility is achieved. Solid lines are fits to a model detailed in Supplementary Information Section~C.
}
\end{figure}

\begin{figure}[h!]
\includegraphics[width=180mm]{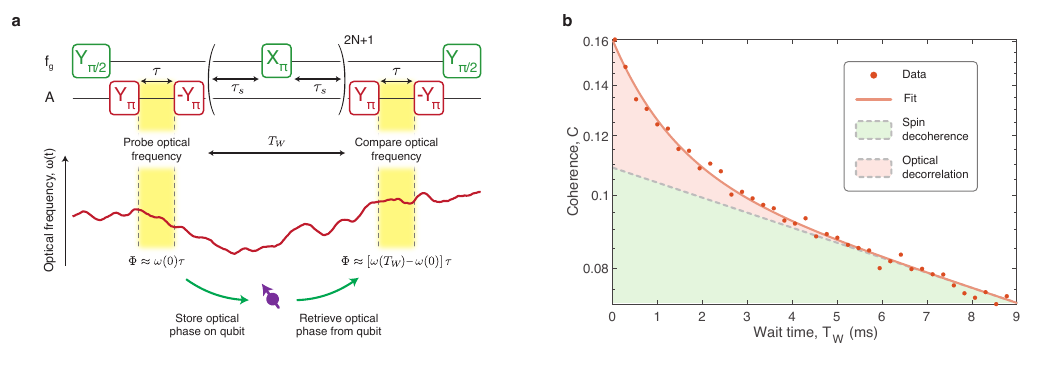}
\caption{{\bf Measuring the optical spectral diffusion correlation timescale of a $^{171}$Yb ion}. {\bf a,} Optical spectral diffusion on timescales longer than the optical lifetime is measured using a delayed echo pulse sequence. First, a superposition between the $\ket{0}$ and $\ket{e}$ states, $\ket{\psi}=1/\sqrt{2}(\ket{0}+\ket{e})$, is used to probe the initial optical frequency, $\omega(0)$. A free evolution time, $\tau$, yields the state $\ket{\psi}\approx\frac{1}{\sqrt{2}}(\ket{0}+\ket{e}e^{-i\omega(0)\tau})$. Next, the state is mapped to the qubit manifold, $\ket{\psi}\approx\frac{1}{\sqrt{2}}(\ket{0}+\ket{1}e^{-i\omega(0)\tau})$, and stored for a duration $T_W$ using an XY-8 pulse sequence with an odd number of pulses ($2N+1$) that have a separation of $2\tau_s=5.8~\mu$s. During this wait time the optical frequency undergoes spectral diffusion with correlation timescale $\tau_c$ (red line). Finally, the optical frequency is probed for a second time by mapping the qubit state back on to the optical transition. After a free evolution time, $\tau$, the quantum state is $\ket{\psi}\approx\frac{1}{\sqrt{2}}(\ket{0}+\ket{1}e^{-i\left[\omega(T_W)-\omega(0)\right]\tau})$. When the wait time is much less than the optical correlation timescale ($T_W\ll\tau_c$), $\omega(T_W)-\omega(0)\approx0$ and coherence is preserved. However, when $T_W\gg\tau_c$, $\omega(T_W)$ and $\omega(0)$ are uncorrelated, optical coherence isn't rephased and the measurement contrast reduces. {\bf b,} Experimental results plotting the final state coherence, $C$, against wait time, $T_W$, with logarithmic $y$ axis and fixed $\tau=1.5~\mu$s. We fit a decay profile with form $\log(C)=a-b(1-e^{-T_W/\tau_c})-\frac{T_W}{T_{2,s}}$ (solid line) where $a$, $b$ and $\tau_c$ are free parameters, and $T_{2,s}$ is the spin XY-8 coherence time, obtained from independent measurements. Note that the initial coherence contrast, $a$, is limited by optical decay during the two free evolution periods of duration $\tau$. The red and green regions correspond to decays attributed to optical decorrelation and spin decoherence, respectively. We extract an optical frequency correlation timescale of $\tau_c=1.42\pm0.04~$ms.}
\end{figure}

\begin{figure}[h!]
\includegraphics[width=147mm]{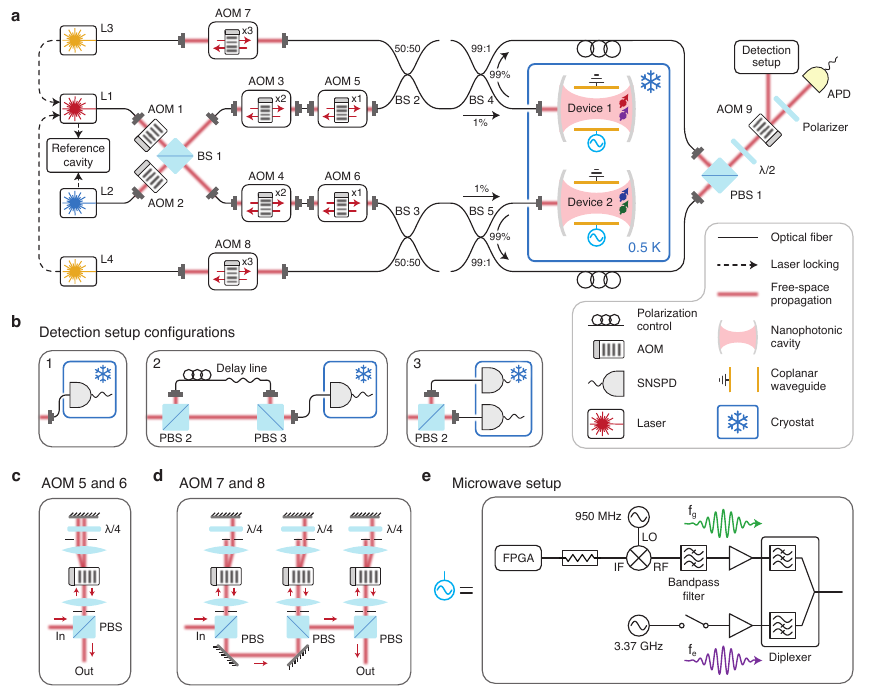}
\caption{{\bf Experimental setup for remote entanglement generation}. {\bf a,} Nanophotonic cavities (Devices 1 and 2) are cooled by a $^3$He cryostat to 0.5 K. Optical control of ions' $A$ and $F$ transitions is achieved using lasers L1, L3 and L4 which are  modulated by a series of acousto-optic modulator (AOM) setups for pulse generation and frequency tuning (AOM~3--8). All lasers are frequency-locked to a reference cavity. Photons exiting the devices are combined on a polarizing beamsplitter (PBS~1), AOM~9 routes photons towards the Detection setup (depicted in {\bf b}). Heterodyne phase measurement of the optical path difference between Devices 1 and 2 is achieved using light pulses from laser L2 which are routed by AOM~9 to an avalanche photodiode (APD) for measurement. {\bf b,} There are three possible detection setups used for entanglement heralding and readout. Setup 1 consists of a single superconducting nanowire single photon detector (SNSPD) and heralds entanglement of two ions within the same device. Setup 2 uses a single SNSPD combined with a time-delayed interferometer to entangle two ions with different optical frequencies located in separate devices. Setup 3 uses two SNSPDs for entanglement experiments involving more than two ions. {\bf c,} Setups AOM~5 and 6 each consist of a single AOM in double pass configuration and enable simultaneous, phase-stable driving of ions in the same device. {\bf d,} Setups AOM~7 and 8 consist of three acousto-optic modulators, each in double pass configuration, enabling pulse generation with a $2\pi\times600$MHz optical frequency tuning range. {\bf e,} Microwave setup for driving the ground and excited state spin transitions at $\approx2\pi\times675$~MHz and $\approx2\pi\times3.37$~GHz, respectively. The ground state (qubit) pulses are generated by heterodyne modulation combined with filters for image rejection before being amplified, combined with the excited state pulses on a diplexer and sent to the device.}
\end{figure}

\begin{figure}[h!]
\includegraphics[width=180mm]{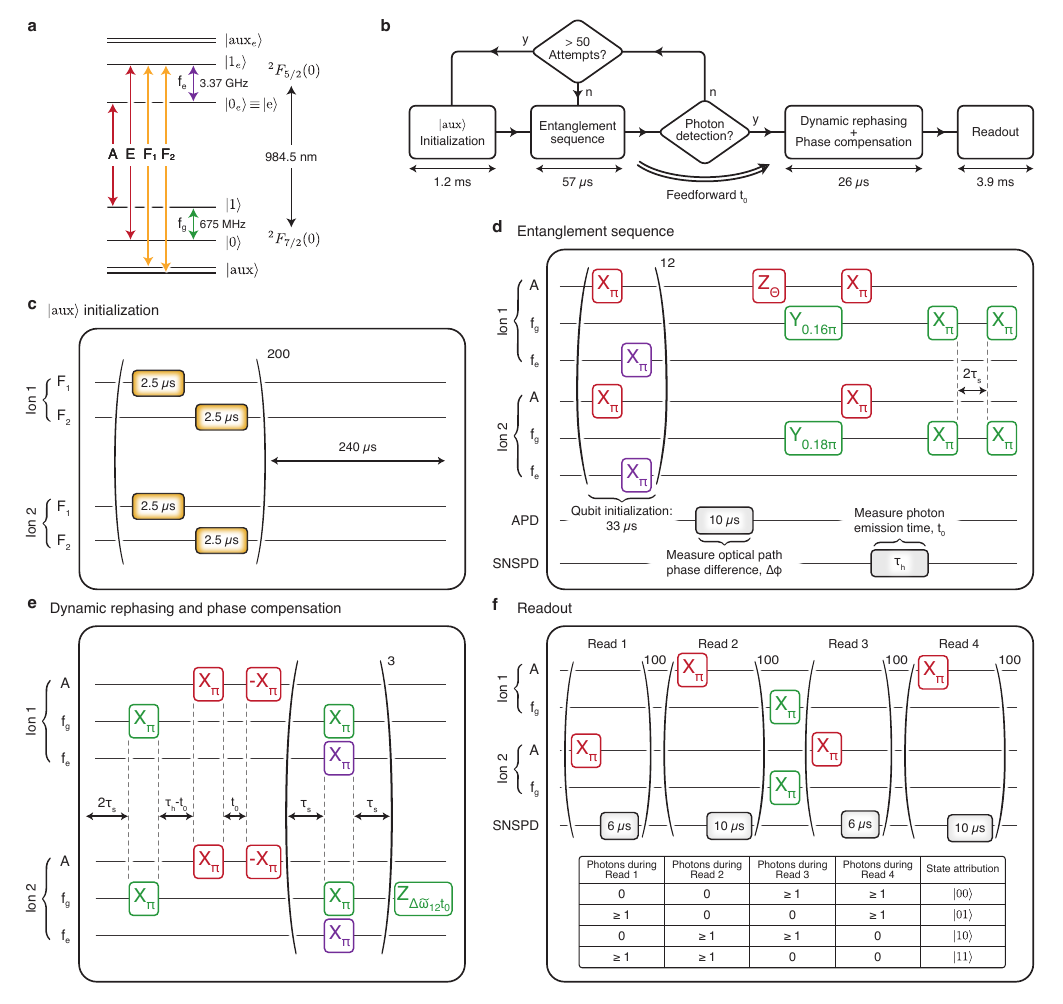}
\caption{{\bf Pulse sequence for remote entanglement generation between two $^{171}$Yb spin qubits}. {\bf a,} Energy levels of $^{171}$Yb ions at $\text{B}=0$ with optical and spin transitions indicated. $A$ and $E$ optical transitions have polarization aligned with the cavity and are Purcell enhanced. The $\ket{\text{e}}$ state referred to in the main text was labelled $\ket{0_e}$ in previous work. {\bf b,} Flow chart showing steps for remote entanglement generation between Ions 1 and 2.  {\bf c,} $\ket{\text{aux}}$ initialization involves 400 pulses, each $2.5~\mu$s long, applied alternately to the $F_1$ and $F_2$ transitions of the two ions, emptying $\ket{\text{aux}}$ into the qubit manifold via decay on $E$. {\bf d,} Initialization into $\ket{0}$ is achieved with pairs of consecutively applied $\pi$ pulses on $A$ and $f_e$ followed by decay on $E$. The optical path phase difference, $\Delta\Phi$, is measured and compensated by a $Z$ rotation, $\Theta$, on Ion 1's $A$ transition (defined in Methods Section~B). A weak superposition is prepared on each qubit, the ions are optically excited and entanglement is heralded via detection of a single photon at stochastic time $t_0$ in a window of duration $\tau_h$. {\bf e,} After several periods of dynamical decoupling (inter-pulse separation $2\tau_s=5.8~\mu$s) the spin coherence is rephased for a duration of $\tau_h-t_0$. Optical coherence is rephased for a duration $t_0$ between two $A$ transition $\pi$ pulses. A $Z$ rotation is applied to Ion 2's qubit by angle $\Delta\widetilde{\omega}_{12}t_0$ where $\Delta\widetilde{\omega}_{12}$ is the $A$ transition driving frequency difference between the two ions. {\bf f,} Each Ion's state is read out via repeated excitation on the $A$ transition followed by photodetection. We use four readout periods: the first two determine whether the ions are in $\ket{1}$. A qubit $\pi$ pulse is applied and the final two periods measure $\ket{0}$. States are ascribed according to conditions in the table, all other photodetection outcomes are discarded.}
\end{figure}

\begin{figure}[h!]
\includegraphics[width=161.25mm]{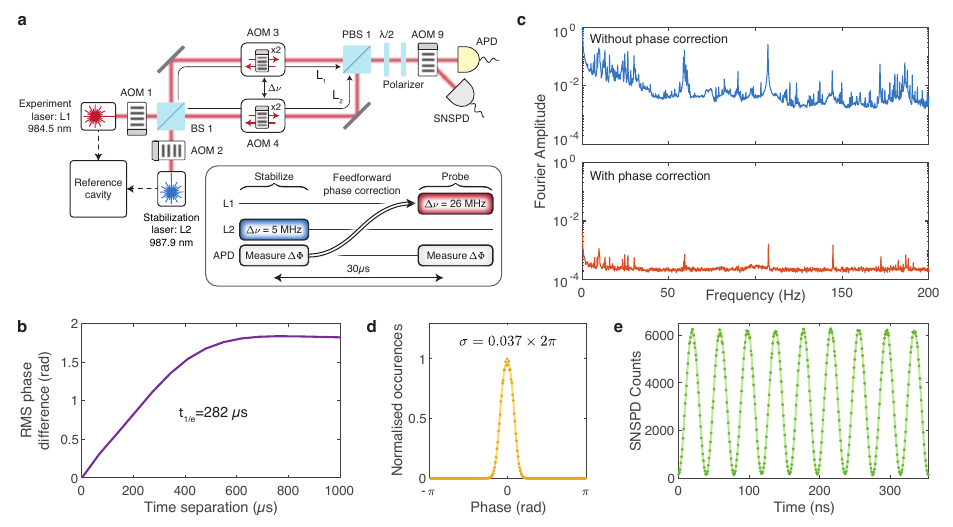}
\caption{{\bf Stabilizing the relative path length between two remote quantum network nodes}. {\bf a,} Simplified optical setup and pulse sequence. The stabilization laser, L2, at 987.9 nm measures the relative optical path phase. This is achieved using AOM~2, 3 and 4 to generate $10~\mu$s long pulses with a $2\pi\times5~$MHz frequency difference that travel along the two separate device paths with distances $L_1$ and $L_2$. AOM~9 routes the pulses to an avalanche photodetector (APD) which detects the resulting beat-note phase, thereby measuring the optical path phase difference, $\Delta\Phi$. After a delay of $30~\mu$s, the optical phase is probed using L1 at 984.5 nm using pulses generated by AOM~1, 3 and 4 with a $2\pi\times26$~MHz frequency difference. However, this time, the driving phase of AOM~3 is adjusted to correct the optical phase difference by an angle $\Theta$ (defined in Methods Section~B), thereby counteracting phase drift and rendering the optical phase constant between consecutive probe periods. The optical phase during each probe period is measured using the APD. {\bf b,} With the phase correction turned off ($\Theta=0$), we measure the optical phase stability by plotting the RMS difference between values of $\Delta\Phi$ in probe periods of varying separation, yielding a 1/e phase correlation timescale of $\approx282~\mu$s. {\bf c,} The resulting Fourier transform of the optical phase noise with the correction, $\Theta$, turned off and on are shown in the top and bottom panels, respectively. {\bf d,} With the phase stabilization turned on, a normalized histogram of the optical phases during the probe periods yields a Gaussian distribution with standard deviation $0.037\times2\pi$~rad, corresponding to an entangled state fidelity limitation of $\mathcal{F}<0.987$. {\bf e,} The probe pulses are attenuated to the single-photon level and detected using a superconducting nanowire single photon detector (SNSPD), the resulting contrast of 0.944 verifies that SNSPD timing jitter does not contribute significantly to these measurements.}
\end{figure}

\begin{figure}[h!]
\includegraphics[width=90mm]{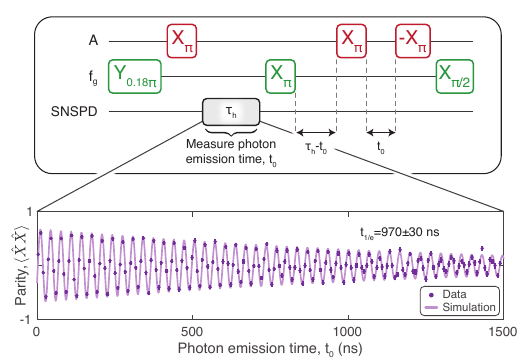}
\caption{{\bf Lifetime-limited entanglement using the dynamic rephasing protocol}. After heralding an entangled state between Ions 1 and 2 and applying the dynamic rephasing protocol described in the main text, the entangled state coherence is probed via a measurement of the $X$-basis parity expectation value, $\langle\hat{X}\hat{X}\rangle$, correlated with the photon emission time, $t_0$. The coherence oscillates at the relative difference in driving laser frequencies,$\Delta\widetilde{\omega}_{12}$, and undergoes an exponential decay with a $1/e$ timescale of $970\pm30~$ns, limited by the Purcell-enhanced optical lifetimes of the two ions. Markers and solid lines correspond to experimental data and simulations, respectively, as detailed in Supplementary Information Section~G.}
\end{figure}

\begin{figure}[h!]
\includegraphics[width=140mm]{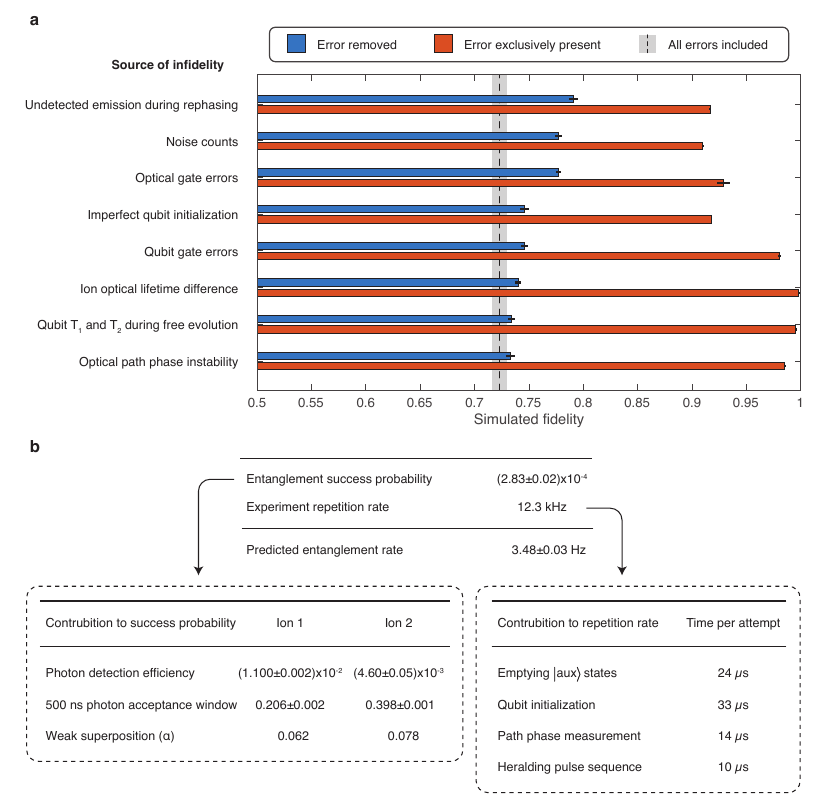}
\caption{{\bf Analysis of fidelities and rates when entangling two remote $^{171}$Yb spin qubits}. {\bf a,} Entanglement fidelity analysis. We characterize the entanglement fidelity by performing simulations based on independently measured experimental parameters. This leads to a simulated fidelity of $\mathcal{F}=0.729\pm0.004$ (dashed vertical line with grey region for error bar), consistent with experimental results. We characterize the individual fidelity contributions in two ways: first by simulating the fidelity when a given error source is removed (blue bars), and secondly, by simulating the fidelity when the error source is exclusively present (orange bars). We find three dominant sources of error: undetected spontaneous emission during the dynamic rephasing protocol; noise counts i.e. photons that don't originate from our two $^{171}$Yb ions; and errors in the optical gates applied to the ions. Error bars are obtained by performing 50 simulation repetitions, each with input parameters sampled from Gaussian
distributions with mean and standard deviation given by the experimentally determined values and errors presented in Supplementary Information Section~E. More detail on the fidelity estimation can be found in Supplementary Information Sections~G and H. {\bf b,} Entanglement rate analysis. The predicted entanglement rate of $\mathcal{R}=3.48\pm0.03$~Hz is obtained from the experiment repetition rate, 12.3~kHz, multiplied by the entanglement success probability, $(2.83\pm0.02)\times10^{-4}$. A slight deviation from the measured result is due to slow drift of experimental parameters on the timescale of several days between measurement and calibration. The dominant limitation to the experiment repetition rate arises from the qubit initialization time, requiring $33~\mu$s per attempt. We also list contributions to the success rate consisting of: the photon detection efficiency, the fraction of photons within the 500~ns acceptance window, and the weak superposition states used in the single photon heralding protocol. More detail on this estimation can be found in Supplementary Information Section~I.}
\end{figure}

\begin{figure}[h!]
\includegraphics[width=134mm]{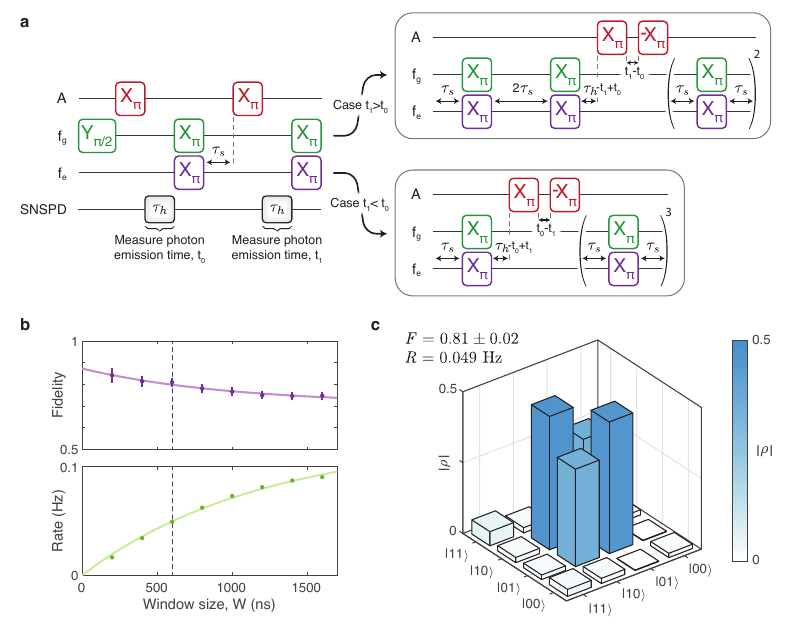}
\caption{{\bf Entanglement heralding between two remote $^{171}$Yb ions using a two photon protocol}. {\bf a,} Pulse sequence for remote entanglement generation. Each qubit is prepared in a superposition state, $1/\sqrt{2}(\ket{0}+\ket{1})$, and optically excited in two rounds, separated by a qubit $\pi$ pulse. Photon detections in both rounds at stochastic times $t_0$ and $t_1$ (measured relative to the start of their respective heralding windows) carve out $\ket{00}$ and $\ket{11}$, respectively, thereby heralding entanglement. Subsequently, a dynamical decoupling sequence is applied with inter-pulse spacing $2\tau_s=5.8~\mu$s. In cases where $t_1<t_0$ optical and spin coherences are rephased for durations $\tau_h-t_0+t_1$ and $t_0-t_1$, respectively, after an even number of dynamical decoupling periods. In cases where $t_1>t_0$ optical and spin coherences are rephased for durations $\tau_h-t_1+t_0$ and $t_1-t_0$, respectively, after an odd number of dynamical decoupling periods. Note that we also apply a $Z$ rotation to Ion 2's qubit by an angle $\Delta\widetilde{\omega}_{12}(t_0-t_1)$ (not labelled). {\bf b,} Entanglement rate and fidelity plotted against window size, W, where heralding events are accepted under the condition $|t_1-t_0|<W/2$. The smallest window size, $W=200~$ns, leads to a fidelity and rate of $\mathcal{F}=0.84\pm0.03$ and $\mathcal{R}=17$~mHz, respectively. The largest window size of $W=1.6~\mu$s corresponds to a fidelity and rate $\mathcal{F}=0.75\pm0.02$ and $\mathcal{R}=91$~mHz, respectively. {\bf c,} Density matrix obtained via maximum likelihood tomography with a window size $W=600~$ns (vertical dashed line in \textbf{b}) leading to a fidelity and rate of $\mathcal{F}=0.81\pm0.02$ and $\mathcal{R}=49$~mHz, respectively.}
\end{figure}

\begin{figure}[h!]
\includegraphics[width=134mm]{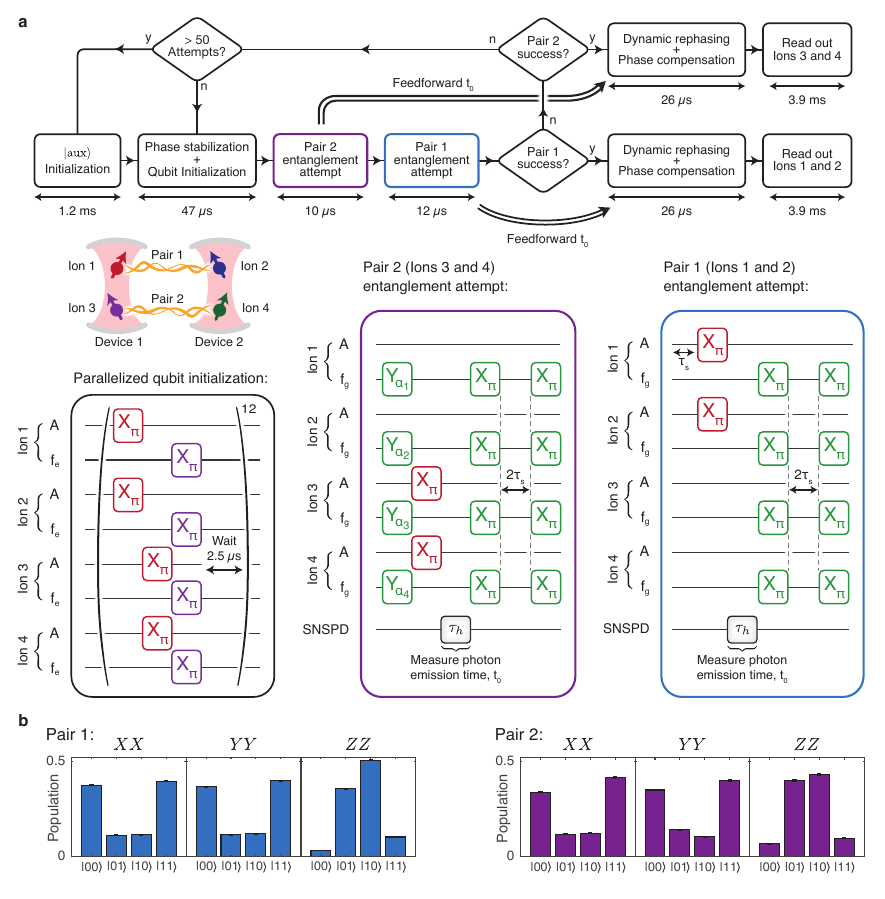}
\caption{{\bf Multiplexed entanglement: sequence detail and parity measurements}. {\bf a,} After parallelized initialization of all four ions, two consecutive entanglement attempts are performed, first on Pair 2 (consisting of Ions 3 and 4), then on Pair 1 (consisting
of Ions 1 and 2), as depicted in the pulse sequences. We note that all four ions have frequency-resolved optical transitions enabling independent optical control. However, due to the narrow distribution of qubit transition frequencies, all microwave control pulses are globally applied to ions in the same device. If either entanglement attempt is successful, the sequence proceeds to apply dynamic rephasing and phase compensation protocols exclusively to the corresponding ion pair. Details of these pulse sequences can be found in Extended Data Fig.~4. Finally, we measure two-qubit populations of the entangled pair in the $XX$, $YY$ or $ZZ$ bases. {\bf b,} We segregate our multiplexed experimental results into two separate data sets corresponding to successful heralding attempts for Pair 1 and Pair 2. For each data set we plot population histograms in the $XX$, $YY$ and $ZZ$ bases for a photon acceptance window size of 500~ns with corresponding fidelities of $\mathcal{F}=0.700\pm0.005$ and $\mathcal{F}=0.668\pm0.006$ for Pair 1 and Pair 2, respectively.}
\end{figure}

\begin{figure}[h!]
\includegraphics[width=170mm]{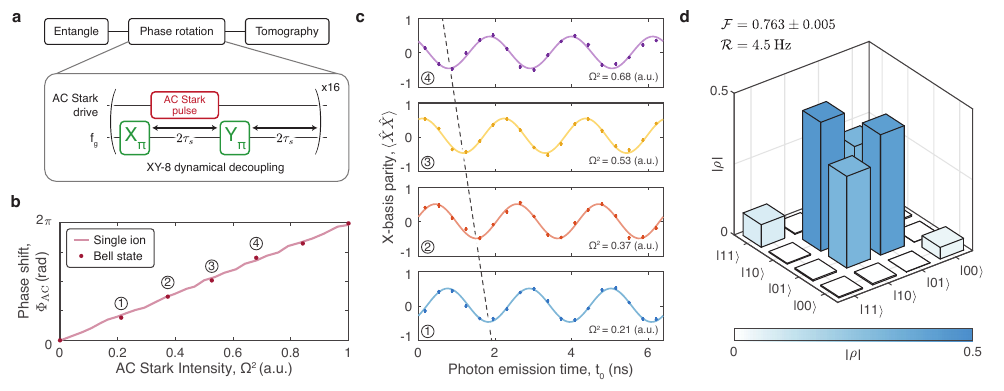}
\caption{{\bf Heralded entanglement between two $^{171}$Yb ions in the same nanophotonic cavity}. {\bf a,} The entanglement protocol requires single-qubit $Z$ rotations which cannot be achieved using global microwave driving. Instead, we utilize a differential AC Stark shift of the qubits generated by an optical pulse with Rabi frequency $\Omega$, oppositely detuned from the ions' optical transition frequencies. We embed these pulses in alternate periods of an XY-8 dynamical decoupling sequence, preserving qubit coherence whilst simultaneously accumulating a differential phase \cite{chen2020}. The sequence consists of $32~\pi$ pulses separated by $2\tau_s=5.8~\mu$s; we vary the differential phase by adjusting the AC Stark intensity which scales with $\Omega^2$. {\bf b,} We verify the AC Stark control in two ways. First, we independently initialize Ions 1 and 3 in superposition states, apply the AC Stark sequence, and measure the spin populations $\langle\hat{X}\rangle$ and $\langle\hat{Y}\rangle$. We calculate the difference between the qubits' phase rotations, $\Phi_\text{AC}$, and plot this for different AC Stark intensities, $\Omega^2$ (solid line). We also probe the differential phase using heralded Bell states, detailed in {\bf c} (markers), these independently derived measurements exhibit close correspondence. {\bf c,} After heralding an entangled state on Ions 1 and 3, we apply an AC Stark sequence prior to measurement. This adds a Bell state phase and corresponding shift to the parity oscillation of coherence with photon emission time, $t_0$. We measure this shift for different AC Stark intensity values, $\Omega^2$; the four parity oscillations correspond to the labelled markers in  {\bf b}. {\bf d,} We dynamically adjust the AC Stark intensity in each experiment to counteract the stochastic heralding phase $\Delta\widetilde{\omega}_{13}t_0$, rendering the entangled state coherence independent of $t_0$. We accept photons in a $500~$ns window and perform maximum likelihood tomography on the quantum state. We plot the resulting density matrix which has a  fidelity of $\mathcal{F}=0.763\pm0.005$, the heralding rate is $\mathcal{R}=4.5$~Hz.}
\end{figure}

\clearpage
\twocolumngrid
\section*{Methods}
\subsection{Setup}
Extended Data Fig.~3 provides a detailed schematic of the experimental setup.

Each device consists of a YVO$_4$ chip with a nanophotonic cavity fabricated via focused ion beam milling, and microwave coplanar waveguides  fabricated by electron-beam lithography \cite{Zhong:16}. The devices are mounted in separate setups on the still plate of a $^3$He cryostat (Bluefors LD-250He) with base temperature of $\approx0.5$~K. These setups enable optical coupling to fibre, microwave coupling to coax lines, nitrogen-condensation tuning of the optical resonances, and magnetic field cancellation via superconducting coils \cite{Kindem2020a}.

Optical control is achieved with the setup in Extended Data Fig.~3a. The Ions' $A$ transitions are addressed by a continuous-wave titanium sapphire laser (L1, M2 Solstis), divided into two paths at a beamsplitter (BS~1). Each path is modulated by a pair of shutter setups (AOM~3 and 5 for Device 1; AOM~4 and 6 for Device 2) for spectrally multiplexed control of ions within the same device. These setups enable two different modes of operation: 1) they can generate optical pulses at any frequency in the optical inhomogneous distribiution for addressing individual ions, 2) they can generate two-tone pulses for simultaneous optical excitation when entangling ions in the same device (as depicted in Extended Data Fig.~3c). A laser at 987.9~nm (L2, Moglabs Cateye) is used for phase stabilization; it is distributed into the two optical paths via the second input port of BS 1. AOM~1 and AOM~2 select between L1 and L2, depending on the required wavelength. Additional optical transitions ($F_1$ and $F_2$, Extended Data Fig.~4a) are driven via lasers L3 and L4 (Toptica DL Pro) modulated by AOM~7 and AOM~8 for Device 1 and Device 2, respectively (Extended Data Fig.~3d). Lasers L1 and L2 are stabilized to a reference cavity (Stable Laser Systems) via Pound-Drever-Hall locking \cite{Drever1983}. L3 and L4 are stabilized via frequency-offset locking to L1.

 The photodetection (output) paths associated with each device are separated from the input via 99:1 beamsplitters, they pass through electronic polarization controllers (OZ Optics EPC) and are combined into the same spatial mode by a polarizing beamsplitter (PBS~1), albeit in orthogonal polarization states. AOM~9 routes light to a detection setup for entanglement heralding and readout or an avalanche photodetector (APD, Laser Components A-CUBE-S500-240) to measure the path phase difference.

There are three different detection setup configurations used in this work (Extended Data Fig.~3b). Setup 1 consists of a single superconducting nanowire single photon detector (SNSPD, Photonspot) and is used to herald entangled states between two ions within the same device (Supplementary Information Section~N). Setup 2 is used for bipartite remote entanglement distribution; it uses a time-delayed interferometer combined with a single SNSPD and can herald entangled states with the same efficiency as more conventional two detector setups \cite{Moehring2007}. Specifically, the orthogonally polarized photonic modes from the two devices are mixed via a 45 degree rotation and second polarizing beamsplitter (PBS~2), leading to a spin-photon entangled state:
\begin{equation}
\begin{split}
    \ket{\psi(t)}&=1/2(\ket{10}+\ket{01}e^{-i\Delta\omega_{12}t})\ket{H}\\
    &+1/2(\ket{10}-\ket{01}e^{-i\Delta\omega_{12}t})\ket{V},
    \end{split}
\end{equation}
where $\ket{V}$ is the photonic mode associated with the upper (fibre coupled) arm of the interferometer, and $\ket{H}$ corresponds to the lower (free space) arm. At this stage, a common approach is to place two SNSPDs at the output ports of PBS~2: depending on which detector clicks, opposite parity Bell states would be heralded. However, due to the optical frequency difference between Ions 1 and 2, a propagation delay of duration $\Delta t=\pi/\Delta\omega_{12}$ prior to photodetection can be used to invert the entangled state parity. We introduce a calibrated fibre delay-line into the upper arm of the interferometer before combining the two paths on a third polarizing beamsplitter (PBS~3). At this stage the spin-photon entangled state is given by:
\begin{equation}
\begin{split}
    \ket{\psi(t)}&=1/2(\ket{10}+\ket{01}e^{-i\Delta\omega_{12}t})(\ket{H}+\ket{V})
    \end{split}
\end{equation}
where $\ket{H}$ and $\ket{V}$ correspond to orthogonally polarized photonic modes in the same spatial output port of PBS~3. We can now deterministically herald Bell states:
\begin{equation}
\ket{\psi(t)}=1/\sqrt{2}(\ket{10}+\ket{01}e^{-i\Delta\omega_{12}t})
\end{equation}
using a single detector. Setup 3 is used for experiments involving more than two ions, it consists of two SNSPDs after PBS~2, depending on which detector clicks, we apply a conditional $Z$ rotation to Ion 2 or Ion 4's qubit, thereby eliminating  detector-dependent parity phase shifts.

The microwave control setup is depicted in Extended Data Fig.~3e. Microwave pulses at $\approx2\pi\times3.37$ GHz, used to drive the excited state spin transition, are generated with RF signal generators (SRS SG380), microwave switches, and amplifiers (Minicircuits ZHL-16W-43-S+). Control pulses at the $\approx2\pi\times675$~MHz ground state qubit transition frequencies are generated via heterodyne mixing of a target waveform at $\approx2\pi\times275$~MHz with a local oscillator at $2\pi\times950$~MHz (Holzworth HS 9002A), bandpass-filtering to remove the image frequencies and amplification (Amplifier Research 10U1000 for Device 1 and Minicircuits ZHL-20W-13SW+ for Device 2).

The experiment is operated by a Quantum Machines OPX control platform which generates all pulses for optical and spin driving. Both SNSPDs and the APD are connected to the two OPX input ports (one SNSPD and the APD share a single port via a diplexer). This enables measurement-based feed-forward using single photon detection times, necessary for the dynamic rephasing and phase compensation protocols described in the main text.

\subsection{Phase Stabilization}
The phase difference between the combined laser excitation and photodetection paths associated with the two devices is imprinted on heralded Bell states and therefore requires active stabilization \cite{cabrillo1999},\cite{Minar2008}. A simplified version of the experimental setup relevant to this section (with the devices excluded) is presented in Extended Data Fig.~5a whereby the two optical paths, with lengths $L_1$ and $L_2$, form an interferometer. The relevant Bell state phase is $\Delta \Phi_{\lambda_1}=2\pi(L_1-L_2)/\lambda_1$ where $\lambda_1=984.5~$nm is the experiment wavelength.

We measure the relative path phase using detuned optical pulses at $\lambda_2=987.9~$nm , yielding $\Delta\Phi_{\lambda_2}=2\pi(L_1-L_2)/\lambda_2$. This off-resonant light reflects from the devices without exciting the cavity mode, thereby minimizing heating. Specifically,  AOM~2, AOM~3 and AOM~4 are used to generate $10~\mu s$ light pulses from laser L2, with a $5~$MHz frequency difference travelling in two separate arms of the interferometer. When recombined on PBS~1 and measured on the APD, the phase of the resulting heterodyne beat note yields $\Delta\Phi_{\lambda_2}$.

We can use $\Delta\Phi_{\lambda_2}$ as an approximate measure for $\Delta\Phi_{\lambda_1}$, under the condition that their difference is static, i.e. $\Delta \Phi_{\lambda_1}-\Delta \Phi_{\lambda_2}=2\pi(L_1-L_2)\left(\frac{1}{\lambda_1}-\frac{1}{\lambda_2}\right)=\text{constant}$. Therefore, with $L_2-L_1\approx3~$m, fluctuation of the optical frequency difference between these two lasers must by much less than $\approx70~$MHz, which is achieved by stabilizing both lasers to a reference cavity via Pound-Drever-Hall (PDH) locking \cite{Drever1983}. Hereafter, we will ignore the static difference between these values and drop the subscripts, i.e. $\Delta\Phi\equiv\Delta\Phi_{\lambda_1,\lambda_2}$.

The time dependence of $\Delta\Phi$ is investigated in Extended Data Fig.~5b where we plot the RMS difference between phase measurements of varying separation, yielding a 1/e correlation timescale of $\approx282~\mu$s. Hence, we need to first measure and then correct $\Delta\Phi$ for each entanglement attempt, which is achieved by adjusting the microwave driving phase of AOM~3 during optical control pulses at $\lambda_1$.

To demonstrate and verify our ability to stabilize the optical phase difference we use the pulse sequence depicted in Extended Data Fig.~5a (inset). Specifically, during experiment repetition $i$, we first measure the optical phase difference, $\Delta\Phi_i$, with a pulse at $\lambda_2$, as described previously. Then, we wait for a duration of $30~\mu$s before probing the phase using an optical pulse at $\lambda_1=984.5~$nm, however this time we apply an optical phase correction via AOM~3 of $\Theta=-\Delta\Phi_i-\beta(\Delta\Phi_i-\Delta\Phi_{(i-1)})$. Note that $\Theta$ includes both a static component ($\Delta\Phi_i$) and a linear extrapolation of the phase trajectory ($\beta(\Delta\Phi_i-\Delta\Phi_{(i-1)})$), where $\beta$ accounts for the difference in delay time between different pulses.

 We use the APD to measure the heterodyne phase during each probe pulse at wavelength $\lambda_1$, the Fourier transform of the resulting time-dependent optical phase provides the phase noise frequency spectrum. We plot this spectrum with and without the phase correction ($\Theta$) applied (Extended Data Fig.~5c), and note the reduction in integrated phase noise between these cases. In Extended Data Fig.~5d, we quantify the effectiveness of our phase stabilization by histogramming the optical phases gathered over 20~minutes of integration time with a resulting standard deviation of $\sigma=0.037\times 2\pi~$rad. This residual phase variation reduces the coherence of our entangled states, thereby limiting the fidelity to $\mathcal{F}<0.987$. Finally, we attenuate the probe pulses to the single-photon level and measure the resulting beat note using an SNSPD (note that we use an optical frequency difference of $2\pi\times26~$MHz between the two interferometer arms during the $\lambda_1$ probe pulse to mimic the ion frequency difference). We histogram the single photon arrival times in Extended Data Fig.~5e with an integration time of 1~minute, the resulting contrast is 0.944. This verifies that SNSPD jitter is not a significant limitation in these experiments.

\subsection{Detailed Entanglement Sequence}
In this section we provide more detail regarding the bipartite remote entanglement sequence. Extended Data Fig.~4a provides a detailed energy level structure with all optical and spin transitions used in this work labelled (see Supplementary Information Section~A for more detail). The experiment flow is depicted in Extended Data Fig.~4b. We start by depleting the $\ket{\text{aux}}$ states of both ions into the qubit manifold via interleaved optical pulses on $F_1$ and $F_2$ followed by cavity-enhanced decay on $E$ (Extended Data Fig.~4c). We estimate an initialization fidelity lower-bounded by 0.98 (see Supplementary Information Section~J). Within each heralding attempt, the probability of transferring population into the $\ket{\text{aux}}$ states is relatively small ($\approx1.2\times 10^{-3}$), we therefore only repeat this step if 50 consecutive heralding attempts have failed.

Next, we initialize the qubits into $\ket{0}$ via consecutive $\pi$ pulses on $A$ and $f_e$ followed by cavity-enhanced decay on $E$. This is repeated twelve times yielding qubit initialization fidelities of $0.9976\pm0.0003$ and $0.9954\pm0.0008$ for Ions 1 and 2, respectively (Extended Data Fig.~4d).

The phase difference between the two optical paths is measured and stabilized, as described in the previous section.

Next, we herald an entangled state using a single photon protocol. We start by preparing an initial state, $\ket{\psi_\text{init}}$, where each qubit is in a weak superposition with probability $\alpha_i$ in the $\ket{1}$ state:
\begin{equation*}
\ket{\psi_\text{init}}=(\sqrt{1-\alpha_1}\ket{0}+\sqrt{\alpha_1}\ket{1})\otimes(\sqrt{1-\alpha_2}\ket{0}+\sqrt{\alpha_2}\ket{1}).
\end{equation*}
Both ions are optically excited and an entangled state is heralded by the detection of a single photon which is measured at stochastic times $t_0$. The resulting conditional density matrix associated with a photon detection in a window of size $\delta t$ centred at $t_0$ is given by \cite{Hermans2023}:
\begin{multline*}
\hat{\rho}(t_0)=\\R_\text{DC}\delta t\!\!\!\!\!\!\sum_{i,j\in\{0,1\}}\!\!\!\!\!\!\left(|\bra{ij}\ket{\psi_\text{init}}|^2\ket{ij}\bra{ij}\right)+\delta t\ket{\psi(t_0)}\bra{\psi(t_0)}\\
+\ket{11}\bra{11}\left(\frac{p_\text{det}^{(1)}}{T_{1,o}^{(1)}}e^{-t_0/T_{1,o}^{(1)}}+\frac{p_\text{det}^{(2)}}{T_{1,o}^{(2)}}e^{-t_0/T_{1,o}^{(2)}}\right)\alpha_1\alpha_2\delta t,
\end{multline*}
where we have only considered terms up to and including $\mathcal{O}(\delta t)$ and where $R_\text{DC}$ is the noise count rate, $T_{1,o}^{(i)}$ is the optical lifetime of Ion $i$ and $p_\text{det}^{(i)}$ is the probability of detecting a photon emitted by Ion $i$, which is assumed to be much less than 1. $\ket{\psi(t_0)}$ is an unnormalized quantum state given by:
\begin{equation}
\ket{\psi(t_0)} = a(t_0)\ket{10}+b(t_0)\ket{01} e^{-i\phi(t_0)},
\end{equation}
with
\begin{align}
a(t_0)&=\sqrt{\frac{p_\text{det}^{(1)}\alpha_1(1-\alpha_2)}{T_{1,o}^{(1)}}}e^{-t_0/2T_{1,o}^{(1)}},\\
b(t_0)&=\sqrt{\frac{p_\text{det}^{(2)}\alpha_2(1-\alpha_1)}{T_{1,o}^{(2)}}}e^{-t_0/2T_{1,o}^{(2)}},\\
\phi(t_0)&=\Delta\omega_{12}\times t_0,
\label{phi_def}
\end{align}
 where $\Delta\omega_{12}$ is the fluctuating optical frequency difference between Ions 1 and 2. Since $\Delta\omega_{12}$ can be regarded as quasi-static on the timescale of a single heralding attempt, we don't explicitly label the time dependence and regard it as a shot-to-shot random variable, following a Gaussian distribution. Note that the conditional density matrix, $\hat{\rho}(t_0)$ is un-normalized, with trace equal to the probability of a photon detection occurring in the window of size $\delta t$. The unconditional density matrix within a photon acceptance window of size $\tau_A$ is given by $\hat{\rho}=\langle\int_0^{\tau_A}dt_0\left(\hat{\rho}(t_0)/\delta t\right)\rangle_{\Delta\omega_{12}}$, and is ensemble-averaged over optical frequency differences. We note that the analytic form of this density matrix incorporates only a limited set of error sources for illustrative purposes. The simulations presented in the main text are derived using a comprehensive numerical model which is presented in Supplementary Information Section~G.

Optimal values of $\alpha_1$ and $\alpha_2$ are identified in two stages. First, we choose a ratio $\xi=\alpha_1/\alpha_2=0.79$ which balances the populations associated with $\ket{10}\bra{10}$ and $\ket{01}\bra{01}$ across our acceptance window, i.e. that satisfies $\int_0^{\tau_A}|a(t_0)|^2dt_0=\int_0^{\tau_A}|b(t_0)|^2dt_0$. Then, we sweep $\alpha_2$ from 0 to 1 (with $\alpha_1$ also varying according to $\alpha_1=\xi\alpha_2$) and identify a maximum entangled state fidelity. For small values of $\alpha_2$, ion emission is suppressed and the entangled state fidelity reduces due to heralding based on noise counts, i.e. $\hat{\rho}(t_0)\rightarrow R_\text{DC}\delta t\sum_{i,j\in\{0,1\}}|\bra{ij}\ket{\psi_\text{init}}|^2\ket{ij}\bra{ij}$, representing an incoherent mixture of the four classical two-qubit states collapsed from the initial state. For large values of $\alpha_2$, occupation of $\ket{11}\bra{11}$ increases which also leads to a reduction in fidelity. Hence, there is an optimal intermediate value of $\alpha_2=0.078$ which balances these two competing sources of infidelity.

After heralding an entangled state, we rephase the coherence by counteracting $\phi(t_0)$ (equation~\eqref{phi_def}), using a pulse sequence with parameters that are dynamically varied based on the photon arrival time, $t_0$ (feedforward). The pulse sequence is described in the main text, however, there are two additional aspects worth noting here. First, we apply two qubit dynamical decoupling periods after the entanglement sequence, prior to rephasing; this provides sufficient time for the experiment control system to determine if any photon detection events occurred and whether to proceed. Secondly, after dynamic rephasing, we mitigate qubit population-basis errors caused by imperfections in the preceding optical $\pi$ pulses. Specifically, we wait for optical decay to deplete the $\ket{e}$ state whilst applying three qubit dynamical decoupling periods to preserve entangled state coherence. The qubit $\pi$ pulses are synchronised with excited state microwave $\pi$ pulses applied to the $f_e$ transition, thereby ensuring that population in the excited state manifold always decays to the correct ground state. Note that this procedure will not improve the coherence of heralded Bell states, only the contrast of diagonal density matrix elements (populations). 

Finally, we read out the two qubit state: we apply 100 consecutive optical $\pi$ pulses to Ion 2, each followed by a $6~\mu$s photon detection window. The process is repeated for Ion 1 with a $10~\mu$s window. Photon detections during these two read periods correspond to population in $\ket{1}$. Then, $\pi$ pulses are applied to both qubits and the process is repeated, this time any photon detections correspond to population that was originally in $\ket{0}$. The total number of photons in each read period are counted, and the two-qubit state is ascribed according to the table in Extended Data Fig.~4f. In cases where one of these four conditions isn't satisfied (e.g. no photons detected), the experiment result is discarded. This leads to a readout efficiency of approximately 0.18 and an average readout fidelity of 0.93.

The duration of each step in the experiment sequence is listed in Extended Data Fig.~4b, the total time per entanglement attempt is $81~\mu$s, this includes the effect of $\ket{\text{aux}}$ initialization which, on average, contributes $24~\mu$s per attempt.

\subsection{Entangling Ions in the same Device}
When generating entangled states that include both Ions 1 and 3 (which are within the same device), both Ions' optical transitions are driven simultaneously. This is achieved using setup AOM~5 which consists of a single AOM in double-pass configuration, driven at half the ion frequency difference ($\approx2\pi\times237.5~$MHz) and aligned to accept both $0^\text{th}$ and $-1^\text{st}$ orders (Extended Data Fig.~3c). This generates two optical tones separated by $\approx2\pi\times475~$MHz for driving Ions 1 and 3 in a manner that's passively phase-stable. We also note that photons emitted by these ions travel through the same optical fibre before reaching the detection setup, hence no active phase stabilization of the optical path difference is required.

Global microwave driving precludes single-qubit $Z$ gates which are necessary for correcting stochastic phases, $\Delta\widetilde{\omega}_{13}t_0$, introduced during the heralding protocol (here $\Delta\widetilde{\omega}_{13}$ is the optical drive frequency difference between Ions 1 and 3, $t_0$ is the photon emission time). Instead, we leverage the optical inhomogeneity to implement a differential AC-stark shift on the two qubits using the protocol detailed in \cite{chen2020}. Specifically, we apply an XY-8 dynamical decoupling sequence consisting of 32 qubit $\pi$ pulses with inter-pulse separation $2\tau_s=5.8~\mu$s (total duration $186~\mu$s). During every alternate free evolution period we apply an optical pulse with Rabi frequency $\Omega$ which is detuned by $\Delta\omega_\text{AC}^{(1)}\approx-2\pi\times50$~MHz from Ion 1 and $\Delta\omega_\text{AC}^{(3)}\approx+2\pi\times425$~MHz from Ion 3 (Extended Data Fig.~10a). At the end of the decoupling sequence, this leads to a relative phase shift between the two qubits of $\Phi_\text{AC}\approx8\Omega^2\tau_s(1/\Delta\omega_\text{AC}^{(3)}-1/\Delta\omega_\text{AC}^{(1)})$. We probe $\Phi_\text{AC}$ by independently measuring the $X$- and $Y$-basis populations for each ion (Extended Data Fig.~10b, solid line). In each heralding experiment, the AC shark shift, which scales with $\Omega^2$, is dynamically adjusted such that the resulting $\Phi_\text{AC}$ counteracts the stochastic phase $\Delta\widetilde{\omega}_{13}t_0$ (Supplementary Information Section~N).

Global microwave driving also complicates tomography, where independent rotations of each qubit are required (for instance when measuring in the $ZX$ basis). These rotations are generated via global $X$ and $Y$ control combined with local $Z$-rotations. For instance, applying the following sequence of pulses:
\begin{equation*}
Y_{\pi/4}~Z^{(1)}_{+\pi/2}~Z^{(3)}_{-\pi/2}~X_{\pi/4},
\end{equation*}
followed by population basis readout will lead to a measurement in the $Z$ basis for Ion 1 and $X$ basis for Ion 3. Here, $\alpha^{(i)}_{\theta}$ corresponds to a $\theta$ pulse applied about the $\alpha$ axis to qubit $i$, if no superscript is given, the pulse is global. Since the qubit frequency difference for Ions 1 and 3 is relatively large ($\approx2\pi\times1.13$~MHz) we implement the local $\pm\pi/2$ $Z$ rotations via a 440~ns wait. The fidelity of this approach is limited by the spin Ramsey coherence time, for smaller frequency differences the AC stark shift approach introduced previously can be used.

We note that independent control of qubits in the same device using AC Stark shifts is extendable beyond $N=2$ ions. For $N>2$, this can be achieved using $N-1$ independently controlled AC Stark tones as detailed in ref.~\cite{chen2020}.

\subsection{Tripartite Entanglement Protocol}
This section provides additional detail related to the three-ion entanglement protocol. After initializing all three qubits in $\ket{0}$, each ion is prepared in a weak superposition state:
\begin{equation*}
\begin{split}
\ket{\psi}=&(\sqrt{1-\alpha_1}\ket{0}+\sqrt{\alpha_1}\ket{1})\otimes\\
&(\sqrt{1-\alpha_2}\ket{0}+\sqrt{\alpha_2}\ket{1})\otimes(\sqrt{1-\alpha_3}\ket{0}+\sqrt{\alpha_3}\ket{1}),
\end{split}
\end{equation*}
with $\alpha_1=0.079$, $\alpha_2=0.078$, and $\alpha_3=0.079$.
All three ions are simultaneously, resonantly optically excited and entanglement is heralded on detection of a single photon. This carves out the optically dark component, $\ket{000}$. Components with more than one qubit spin excitation (for instance $\ket{110}$) are suppressed by choosing $\alpha_1,~\alpha_2,~\alpha_3\ll1$. Note that compared to the bipartite entanglement case, the resulting states will have lower fidelity, i.e., assuming that the same value of $\alpha$ is used for all ions, the two ion Bell state fidelity will be upper-bounded by $1-\alpha$, whereas the three ion W state fidelity will be upper-bounded by $(1-\alpha)^2$. Ignoring this source of infidelity and assuming that the values of $\alpha_i$ are chosen to counteract differences in lifetime and detection efficiency between the three ions, we can approximate the heralded state as:
\begin{multline*}
\ket{\psi(t_0)}=\\\frac{1}{\sqrt{3}}\left(\ket{100}e^{-i\omega_1t_0}+\ket{010}e^{-i\omega_2t_0}+\ket{001}e^{-i\omega_3t_0}\right),
\end{multline*}
where $t_0$ is the photon detection time and $\omega_i$ is the optical frequency of Ion $i$.

To prepare deterministic entangled states, after the heralding window, we apply the dynamic rephasing protocol consisting of a qubit $\pi$ pulse, a delay time $\tau_h-t_0$ for spin rephasing (where $\tau_h$ is the heralding window size), and a pair of optical $\pi$ pulses separated by $t_0$. All of these pulses are applied to the three ions simultaneously. After an additional qubit $\pi$ pulse the quantum state is given by:
\begin{equation*}
\ket{\psi(t_0)}=\frac{1}{\sqrt{3}}(\ket{100} + \ket{010}e^{-i\Delta\widetilde{\omega}_{12}t_0} + \ket{001}e^{-i\Delta\widetilde{\omega}_{13}t_0}),
\end{equation*}
where $\Delta\widetilde{\omega}_{ij}$ is the frequency difference between lasers used to drive Ions $i$ and $j$. We can see that there are two residual stochastic quantum phases that need to be compensated. First we apply an AC Stark sequence, described in the previous section and in ref.~\cite{chen2020}, to correct the stochastic phase $\Delta\widetilde{\omega}_{13}t_0$ between Ions 1 and 3, leading to:
\begin{equation*}
\ket{\psi(t_0)}=\frac{1}{\sqrt{3}}\left(\ket{100} + \ket{010}e^{-i\phi(t_0)} + \ket{001}\right),
\end{equation*}
where
\vspace{-2mm}
\begin{multline*}
\phi(t_0)=t_0\left(\Delta\widetilde{\omega}_{12}+\Delta\widetilde{\omega}_{13}\frac{\Delta\omega_\text{AC}^{(3)}}{\Delta\omega_\text{AC}^{(1)}-\Delta\omega_\text{AC}^{(3)}}\right),\\
\end{multline*}
and $\Delta\omega_\text{AC}^{(i)}$ is the detuning of the optical AC Stark tone from Ion $i$.
Then, we apply a phase shift to Ion 2's qubit drive by an angle $\phi(t_0)$ to cancel the residual stochastic phase, leading to the canonical, tripartite W state:
\begin{equation*}
\ket{\text{W}}=\frac{1}{\sqrt{3}}\left(\ket{100} + \ket{010} + \ket{001}\right).
\end{equation*}
Independent basis rotations are applied to the three qubits prior to readout, the process for performing these rotations on qubits within the same device is also described in the previous section. Additional discussion and experimental results can be found in Supplementary Information Section~O.
\newpage

\setcounter{subsection}{0}
\clearpage
\onecolumngrid
\renewcommand{\theequation}{S\arabic{equation}}
\setcounter{equation}{0}
\section*{Supplementary Information}
\subsection{Energy Levels}
Extended Data Fig.~4a depicts the energy level structure and transitions relevant to this work at zero magnetic field. We consider 4f--4f transitions between the $^2F_{7/2}$ ground state multiplet and the $^2F_{5/2}$ excited state multiplet. Specifically, between the lowest lying crystal field levels ($^2F_{7/2}(0)\leftrightarrow^2F_{5/2}(0)$) at 984.5~nm. These levels are each Kramers doublets consisting of an effective electronic spin $1/2$. The hyperfine interaction with the $^{171}$Yb nuclear spin with $I=1/2$ lifts the Kramers degeneracy. Of particular relevance are the $F=0,~m_F=0$ singlet and $F=1,~m_F=0$ triplet energy levels in the ground state: $\ket{0}=\frac{1}{\sqrt{2}}(\ket{\uparrow\Downarrow}-\ket{\downarrow\Uparrow})$ and $\ket{1}=\frac{1}{\sqrt{2}}(\ket{\uparrow\Downarrow}+\ket{\downarrow\Uparrow})$, respectively, where $\{\ket{\uparrow},\ket{\downarrow}\}$ is the electronic spin and $\{\ket{\Uparrow},\ket{\Downarrow}\}$ is the nuclear spin. These levels are split by $\approx2\pi\times675~$MHz and form our qubit, which can be driven by an oscillating magnetic field polarized along the crystalline $c$ axis, the $g$ factor for this transition is -6.08. The remaining ground state energy levels with $F=1,~m_F=\pm1$, labelled $\ket{\text{aux}}$ should be degenerate \cite{Kindem2018}, however, we observe a small splitting (between $2\pi\times20$~MHz and $2\pi\times60~$MHz) which we attribute to reduced symmetry from crystalline strain. We note that spin transitions between the $\ket{\text{aux}}$ states and qubit states are polarized perpendicularly to the $c$ axis and therefore cannot be driven with our on-chip coplanar waveguide.

The excited state $^2F_{5/2}(0)$ doublet is also split by the hyperfine interaction, with $\ket{\text{e}}$ and $\ket{1_e}$ corresponding to the $m_F=0$ singlet and triplet states, respectively. The $\ket{\text{e}}\leftrightarrow\ket{1_e}$ transition can be driven with a $c$-directed oscillating magnetic field at $\approx2\pi\times3.37$~GHz. In previous work \cite{Kindem2020a,Ruskuc2022} the $\ket{\text{e}}$ state was labelled $\ket{0_e}$.

We note that energy levels $\ket{0}$, $\ket{1}$, $\ket{\text{e}}$ and $\ket{1_e}$ don't have a magnetic dipole moment, therefore, any transitions connecting them (both optical and microwave) are insensitive to magnetic field noise, to first order. Furthermore, due to the nonpolar site symmetry, optical transitions are also first order insensitive to electric field noise.

Optical transitions labelled $A$ and $E$ have electric field polarization parallel to the crystalline $c$ axis, which is aligned to our nanophotonic cavity mode. They experience Purcell enhancement (see Supplementary Information Section~\ref{devices}), are highly cyclic, and can be coherently driven via the cavity mode. We use the $A$ transition to generate single photons during entanglement heralding and for readout. Transitions $F_1$ and $F_2$ address the two $\ket{\text{aux}}$ states (with splitting between $2\pi\times20~$MHz and $2\pi\times60~$MHz), are polarized perpendicularly to the cavity mode and do not experience Purcell enhancement; we use them for initialization. We cannot coherently drive the $F$ transitions due to a combination of poor coherence (the $\ket{\text{aux}}$ states aren't protected from magnetic field noise) and weak $a$-directed electric field strength.

More detail on this energy level structure can be found in \cite{Kindem2018}.

\subsection{Devices and Cavity QED}
\label{devices}
Each device consists of a nanophotonic resonator fabricated via focused ion beam milling from YVO$_4$ as detailed in \cite{Zhong:16}, the cavity mode is TM polarized with electric field parallel to the crystalline $c$ axis. The main factor in determining device yield\footnote{Current device yields are approximately 50\%.} is the resonance frequency which must exceed the $^{171}$Yb optical transition (since nitrogen deposition tuning can only shift the resonances to lower frequencies). The YVO$_4$ chips used in this work were supplied by Gamdan Optics, they are nominally undoped, however, $^{171}$Yb is present at impurity levels of approximately 20 parts per billion leading to roughly 20 $^{171}$Yb ions within the cavity mode volume.

The parameters for the two devices used in this work are given in the table below:
\begin{center}
\begin{tabular}{|c|c|c|}
 \hline
   & Device 1 & Device 2 \\
 \hline\hline
  $Q$ & $(9.7\pm0.2)\times10^3$ & $(1.27\pm0.07)\times10^4$ \\
   \hline
  $\kappa$ & $2\pi\times(31.4\pm0.8)$~GHz & $2\pi\times(24.4\pm1.3)$~GHz \\
   \hline
  $\kappa_\text{in}/\kappa$ &$0.117\pm0.004$ & $0.065\pm0.004$ \\ 
 \hline
\end{tabular}
\end{center}
where $Q$ is the cavity quality factor, $\kappa$ is the cavity energy decay rate, and $\kappa_\text{in}/\kappa$ is the fraction of cavity decay into the waveguide mode. We estimate atom-cavity coupling rates of $g_1=2\pi\times(23.9\pm0.3)~$MHz, $g_2=2\pi\times(31.3\pm0.8)~$MHz, $g_3=2\pi\times(23.1\pm0.3)~$MHz and $g_4=2\pi\times(29.4\pm0.8)~$MHz for Ions 1, 2, 3 and 4, respectively. We note that based on previous estimations of the $A$ transition dipole moment \cite{Kindem2018,Kindem2020a}, there is a discrepancy in these predicted values of $g$ which appear to exceed the maximum values possible based on our cavity design by a factor of approximately 1.2. This doesn't affect any of our simulations or results, therefore, we leave a detailed spectroscopic analysis for future work. The free-space optical decay rate and lifetime of the $\ket{\text{e}}$ state are $\Gamma=3.7~$kHz and $267~\mu$s, respectively. This leads to effective Purcell enhancement factors \cite{Purcell1946} (defined as $4g^2/\kappa\Gamma$) of 122, 271, 114 and 239 with corresponding $\ket{\text{e}}$ state lifetimes of 2.17$~\mu$s, 0.985$~\mu$s, 2.33$~\mu$s and 1.115$~\mu$s for Ions 1, 2, 3 and 4, respectively  (note that Ions 1 and 3 are in Device 1, Ion 2 and 4 are in Device 2).

We also fabricate gold coplanar waveguides on the surface of these chips (see \cite{Bartholomew2020} for details), enabling microwave driving of spin transitions with magnetic field polarization parallel to the crystalline $c$ axis.

\subsection{Hong-Ou-Mandel Interference}

We experimentally investigate the indistinguishably of photons emitted by two remote $^{171}$Yb ions using Hong-Ou-Mandel (HOM) interference \cite{Hong1987}. This serves as a demonstration of mutual photonic coherence, a pre-requisite for the entanglement experiments presented in this work. Conventionally, perfectly indistinguishable photons impinging on two ports of a beamsplitter will always emerge from the same output port (bunch). Such experiments have been used to study photonic emission in a range of emitter platforms \cite{Maunz2007,Beugnon2006,bernien2012,flagg2010,lettow2010,Lukin2023}.

In our measurements, photons emitted by Ions 1 and 2, at frequencies $\omega_1$ and $\omega_2$ (where $\omega_2-\omega_1=\Delta\omega_{12}\approx 2\pi\times 33~$MHz\footnote{Note that the ions' optical frequencies exhibit slow drift on the timescale of several weeks, hence $\Delta\omega_{12}$ is slightly different here compared to the bipartite entanglement experiments in Fig.~2.}) are collected into the same spatial mode in orthogonal polarization states. The combination of a half waveplate and polarizing beamsplitter (PBS) acts to mix (or interfere) these two input modes. The two output modes are measured with single photon detectors (SNSPD 1 and SNSPD 2) as depicted in Extended Data Fig.~1a. A more detailed experimental setup description can be found in Methods Section~A and Extended Data Fig.~3.

To analyze this measurement we follow the approach in \cite{Kambs2018}. Our $^{171}$Yb ions generate photons with spatio-temporal wavepackets given by:
\begin{equation}
    \zeta_i(z,t)=\frac{1}{\sqrt{T_{1,o}^{(i)}}}\mathcal{H}(t)e^{-\frac{t}{2T_{1,o}^{(i)}}-i(\omega_i t -k_iz +\Phi_i)},
\end{equation}
where $i$ labels the ion number, $T_{1,o}^{(i)}$ is the optical lifetime, $\omega_i$ is the frequency, $k_i$ is the wavenumber, $\Phi_i$ is an optical phase associated with the ion's excitation, and $\mathcal{H}(t)$ is the Heaviside step function, i.e., the photon is emitted at $t=0$ \cite{Kambs2018}. We don't consider pure dephasing in this model as quasi-static fluctuations in the optical frequencies, $\omega_i$, are the dominant source of decoherence.

We consider field operators associated with the orthogonally polarized input modes populated by Ions 1 and 2, labelled $A$ and $B$ respectively. We define annihilation operators $\hat{A}$ and $\hat{B}$ and electric field operators $\hat{E}^+_A(z,t)=\zeta_1(z,t)\hat{A}$ and $\hat{E}^+_B(z,t)=\zeta_2(z,t)\hat{B}$. We note that the complete electric field operators consist of an infinite sum over spatio-temporal modes, however, by operating in a basis that includes $\hat{A}$ and $\hat{B}$ we can ignore all other orthogonal modes \cite{Raymer2020}.

Now, considering the measurement setup in Extended Data Fig.~1a, we define electric field operators for the two photodetection channels: $\hat{E}^+_C(z,t)$ and $\hat{E}^+_D(z,t)$ for SNSPD 1 and SNSPD 2, respectively. The half wave plate and PBS transform the input channels into the output channels according to:
\begin{equation}
    \begin{split} \hat{E}^+_C(z,t)&=\cos(2\theta)\hat{E}^+_A(z,t)+\sin(2\theta)\hat{E}^+_B(z,t),\\
        \hat{E}^+_D(z,t)&=-\cos(2\theta)\hat{E}^+_B(z,t)+\sin(2\theta)\hat{E}^+_A(z,t),
    \end{split}
\end{equation}
where $\theta$ is the rotation angle of the half wave plate. We set $\theta=\pi/8$, where the input mode contributions to the output modes are balanced, this is the condition for optimum HOM visibility.

The joint photodetection probability associated with a detection event at time $t_1$ in SNSPD 1 and $t_2$ in SNSPD 2 is given by:
\begin{equation}
\begin{split}
&P(t_1,t_2)=\bra{0}\hat{A}\hat{B}\hat{E}^-_C(t_1)\hat{E}^-_D(t_2)\hat{E}^+_C(t_1)\hat{E}^+_D(t_2)\hat{A}^\dag\hat{B}^\dag\ket{0}\\
&=\frac{1}{4}\left[|\zeta_1(t_1)\zeta_2(t_2)|^2+|\zeta_1(t_2)\zeta_2(t_1)|^2-2\mathfrak{Re}\{\zeta_1(t_1)\zeta_1^*(t_2)\zeta_2^*(t_1)\zeta_2(t_2)\}\right]\\
&\propto e^{-t_1/T_{1,o}^{(1)}}e^{-t_2/T_{1,o}^{(2)}}+e^{-t_2/T_{1,o}^{(1)}}e^{-t_1/T_{1,o}^{(2)}}-2e^{-\frac{t_1+t_2}{2\tau}}\cos(\Delta\omega_{12}(t_1-t_2)),
\end{split}
\end{equation}
where, for simplicity, we have dropped the positional coordinate and $\tau$ is defined according to $1/\tau=1/T_{1,o}^{(1)}+1/T_{1,o}^{(2)}$.

Finally, we define the photon arrival time difference $\Delta t=t_2-t_1$ and marginalize over the remaining temporal degree of freedom to arrive at the coincidence probability:
\begin{equation}
    P(\Delta t)=e^{-|\Delta t|/T_{1,o}^{(2)}}+e^{-|\Delta t|/T_{1,o}^{(1)}}-2e^{-|\Delta t|/2\tau}e^{-|\Delta t|^2(\gamma_1^{*2}+\gamma_2^{*2})/2}\cos(\langle\Delta\omega_{12}\rangle\Delta t).
\end{equation}
We have also taken an ensemble average over the ions' frequency distributions whereby $\gamma^*_i$ is the linewidth of Ion $i$ (defined as the standard deviation of the optical frequency fluctuation) and $\langle\Delta\omega_{12}\rangle$ is the average frequency difference between the two ions. Also note that we have assumed that the experiment repetition time is much longer than either of the two ions' lifetimes.

The HOM visibility is then given by the oscillation contrast of the two-photon coincidence probability at frequency $\langle\Delta\omega_{12}\rangle$. For $\Delta t>0$, this is given by\footnote{A similar expression can be derived for $\Delta t<0$.}:
\begin{equation}
\label{hom_vis}
    V(\Delta t)=\frac{2\sqrt{\alpha}e^{-\Delta t/2\tau}e^{-\Delta t^2(\gamma_1^{*2}+\gamma_2^{*2})/2}}{e^{-\Delta t/T_{1,o}^{(2)}}+\alpha e^{-\Delta t/T_{1,o}^{(1)}}}.
\end{equation}
Here, we have introduced an additional infidelity (not present in the previous expressions) associated with relative photodetection probabilities. This is characterized by the parameter $\alpha$, defined as:
\begin{equation}
\label{hom_balance}
    \alpha=\frac{\eta^{AD}\eta^{BC}}{\eta^{AC}\eta^{BD}},
\end{equation}
where $\eta^{AC}$ and $\eta^{AD}$ are probabilities that photons originating from Ion 1 are detected on SNSPD 1 and SNSPD 2, respectively. Similarly, $\eta^{BC}$ and $\eta^{BD}$ are probabilities that photons originating from Ion 2 are detected on SNSPD 1 and SNSPD 2, respectively. In cases where $\alpha\neq1$, small adjustments of the half waveplate angle, $\theta$, are used to balance these photodetection probabilities and maximize HOM contrast.

Experimentally, Extended Data Fig.~1b shows a normalized histogram of coincidences between the two detectors with a bin size of $180~$ns. We don't observe any correlation in photodetection statistics, this is because the bin size is much larger than $2\pi/\langle\Delta\omega_{12}\rangle\approx30~$ns so the photons appear distinguishable due to their optical frequency difference.

Extended Data Fig.~1c shows the same measurement with a $4~$ns bin size over the central region with $|\Delta t|<200~$ns. We observe an oscillation between bunching and antibunching at the optical frequency difference, $\langle\Delta\omega_{12}\rangle\approx2\pi\times33~$MHz, i.e. using a smaller bin size leads to erasure of frequency information \cite{Legero2003,Legero2004}. In Extended Data Fig.~1d we plot this result over a larger range of $\Delta t$, where the window positions are chosen to coincide with the trough of each oscillation: we recover a conventional HOM dip.

We can use the oscillation contrast to define an effective photon indistinguishability: quantifying the similarity of photon wavepackets in all aspects other than the static frequency difference, $\langle\Delta\omega_{12}\rangle$. In Extended Data Fig.~1e we plot this contrast when averaged over a window of width $W$ which extends from $-W/2<\Delta t<W/2$. For the smallest window size ($W=6~$ns) we obtain an effective photon indistinguishability of $96\pm2\%$.

In all figures, solid lines are obtained from a model based on equation~\eqref{hom_vis} which additionally includes the effect of dark counts and experiment repetition rate in relation to the ion lifetime. At $\Delta t=0$, the dominant limitation to HOM fringe contrast is the finite bin size relative to the inverse optical frequency difference. At larger values of $\Delta t$, the contrast exhibits Gaussian decay with a timescale of, $205\pm2~$ns, which is caused by the quasi-static variation of the two ions' optical transition frequencies.

\subsection{Single Ion Optical Properties}
\label{optprops}
\subsubsection{Second Order Photon Correlation}
We verify that peaks in our fluorescence spectra (Fig.~1b) originate from single emitters using pulse-wise second order photon correlation measurements (as detailed in \cite{Kindem2020a}). For Ions 1, 2, 3 and 4 used in this work we find value of $g^{(2)}(0)=(1.0\pm0.2)\times 10^{-2}$, $(2.5\pm0.2)\times 10^{-2}$, $(1.1\pm0.3)\times 10^{-2}$, and $(0.9\pm0.4)\times 10^{-2}$, respectively (Fig.~S1b and Fig.~S2b). To characterize the multiplexing capacity of these nodes we found 5 additional emitters in Device~1 and 6 additional emitters in Device~2, all with a value of  $g^{(2)}(0)<0.1$ (labelled Ions~5--15 in Fig.~S1a and Fig.~S2a). The corresponding photon correlation measurements are plotted in Fig.~S1c and Fig.~S2c along with the single photon detection efficiencies. These ions are all separated by $<1475~$MHz, therefore any combination could be robustly entangled using our current detectors with minimal infidelity due to timing jitter ($\mathcal{F}<0.96$), enabling a straightforward increase of our multiplexed entanglement scheme to 7 modes.

\subsubsection{Ion Linewidth and Ramsey Coherence}
The ion linewidths (defined as the standard deviation of the Gaussian spectral distribution) are derived according to $\gamma^*=\sqrt{2}/T_{2,o}^*$, where $T_{2,o}^*$ is the optical Ramsey coherence time. This yields values of $2\pi\times(0.73\pm0.05)~$MHz,  $2\pi\times(1.02\pm0.05)~$MHz, $2\pi\times(0.69\pm0.03)~$MHz and $2\pi\times(1.02\pm0.03)~$MHz for Ions 1, 2, 3 and 4, respectively.
 
\subsubsection{Cooperativities and Echo Coherence}
Optical echo coherence times measured on the $A$ transition ($\ket{1}\leftrightarrow\ket{\text{e}}$) yield values of $T_{2,o}=3.66\pm0.12~\mu$s, $1.75\pm0.04~\mu$s, $3.89\pm0.19~\mu$s and $1.89\pm0.11~\mu$s for Ions 1, 2, 3 and 4, respectively. In all four cases, the echo coherence times are close to the lifetime limit ($T_{2,o}=2T_{1,o}$) with corresponding values of $2T_{1,o}= 4.34\pm0.03~\mu$s, $1.970\pm0.006~\mu$s, $4.66\pm0.04~\mu$s and $2.23\pm0.02~\mu$s for Ions 1, 2, 3 and 4, respectively, i.e. the dominant source of decoherence is lifetime decay. We use these values to extract cooperativities of $5.1\pm1.1$, $7.7\pm1.6$, $4.8\pm1.4$ and $5\pm2$ for Ions 1, 2, 3 and 4, respectively.

We believe that residual Markovian dephasing on the optical $A$ transition may be caused by magnetic interaction of the spin states $\ket{1}$ and $\ket{\text{e}}$ with local fluctuating magnetic fields. Under this assumption, an applied magnetic field would cause the states $\ket{1}$ and $\ket{\text{e}}$ to undergo an opposite frequency shift, whereas states $\ket{0}$ and $\ket{\text{e}}$ will shift frequency in the same direction. Therefore, we expect that a superposition of $\ket{0}$ and $\ket{\text{e}}$ (as present during the entanglement protocol) will be more robust to magnetic perturbation with a correspondingly lower pure dephasing rate compared to a superposition of $\ket{1}$ and $\ket{\text{e}}$ (as probed here). In our simulations, we don't include pure dephasing and observe a close correspondence with the experimental data (Extended Data Fig.~6), thereby verifying that fast (Markovian) optical dephasing has negligible effect in these experiments. A more detailed analysis of fast optical dephasing processes is left for future work.

\subsubsection{Optical Correlation Timescales}
Utilizing rephasing methods in entanglement heralding protocols requires differences between ions' optical frequencies to be quasi-static on the timescale of a single entanglement attempt (i.e. $\Delta\omega_{12}$ shouldn't change between the heralding period and dynamic rephasing). To verify this, we experimentally characterize the dominant spectral diffusion correlation timescale, $\tau_c$.

A direct measurement of this timescale using an optical echo is precluded by the relatively short optical lifetime, i.e. decoherence from spectral diffusion is a small contribution that's difficult to measure. Several alternative approaches exist \cite{Santori2002,Bottger2006,Sallen2010a}, including the use of a delay line to interfere photons emitted at two separate times, thereby comparing their frequencies. We adopt an alternative approach, whereby we probe the optical transition frequency using phase accumulation of a superposition state. Subsequently, this frequency information is stored in the form of a quantum phase on our ground state qubit for a variable duration which can be much longer than the optical lifetime. Finally, we probe the optical transition frequency for a second time and compare the accumulated and stored quantum phases, which acts as a measure of the optical frequency difference. We term this measurement a `delayed echo'.

More specifically, we first generate an optical superposition state via a combination of a qubit $\pi/2$ pulse followed by an optical $\pi$ pulse, leading to  $\ket{\psi}=1/\sqrt{2}(\ket{0}+\ket{\text{e}})$ (Extended Data Fig. 2a). Next, we accumulate optical phase for a duration $\tau$, leading to $\ket{\psi}\approx1/\sqrt{2}(\ket{0}+\ket{\text{e}}e^{-i\omega(0)\tau})$, this serves as a probe of the optical frequency at the start of the sequence. Note that we explicitly label the time dependence of the optical frequency, $\omega(t)$, and assume that it varies on timescales longer than $\tau$. We map this coherence to the qubit state using a second optical $\pi$ pulse ($\ket{\psi}\approx1/\sqrt{2}(\ket{0}+\ket{1}e^{-i\omega(0)\tau})$) and store it using an XY-8 dynamical decoupling sequence \cite{Gullion1990a} for duration $T_W$. We keep the inter-pulse spacing fixed at $2\tau_s=5.8~\mu$s during dynamical decoupling and vary the delay time by changing the number of $\pi$ pulses, $2N+1$. Note that we use an odd number of qubit $\pi$ pulses, leading to: $\ket{\psi}\approx1/\sqrt{2}(\ket{1}+\ket{0}e^{-i\omega(0)\tau})$. Finally, we apply another pair of optical pulses separated by $\tau$, which probes the optical frequency for a second time, resulting in a state:
\begin{equation}
\ket{\psi}\approx\frac{1}{\sqrt{2}}\left(\ket{1}+\ket{0}e^{-i\left[\omega(0)-\omega(T_W)\right]\tau}\right).
\end{equation}
In cases where $T_W$ is much less than the optical correlation timescale ($T_W\ll\tau_c$), $\omega(0)\approx\omega(T_W)$ and the second period of duration $\tau$ acts to cancel dephasing from the first period (i.e. optical coherence is rephased). However, in cases where $T_W\gg\tau_c$,  $\omega(0)$ and $\omega(T_W)$ are uncorrelated and the two periods of duration $\tau$ independently act to dephase the qubit, leading to a reduction in coherence by a factor $e^{-2\tau^2/(T^*_{2,o})^2}$, where $T^*_{2,o}$ is the optical Ramsey coherence time.

We theoretically analyse the performance of this pulse sequence using the filter function formalism \cite{Degen2017}. We use a time domain filter function:
\begin{equation}
F(t)=
    \begin{cases}
        1 & \text{if } -\frac{1}{2}(T_W+\tau)<t<-\frac{1}{2}(T_W-\tau) \\
        -1 & \text{if } \frac{1}{2}(T_W-\tau)<t<\frac{1}{2}(T_W+\tau) \\
        0 & \text{otherwise}
    \end{cases}
\end{equation}
combined with an Ornstein-Uhlenbeck model for the optical frequency, with correlation function \cite{Sun2022}:
\begin{equation}
S(\omega)=\frac{4\tau_c}{\tau_c^2\omega^2+1}\frac{1}{(T^*_{2,o})^2}.
\end{equation} This leads to the following functional form for the qubit coherence, $C(\tau,T_W)$, at the end of the pulse sequence:
\begin{equation}
C(\tau,T_W)=e^{-\frac{\tau}{T_{1,o}}}e^{-\frac{T_W}{T_{2,s}}}e^{-\frac{2\tau^2}{(T^*_{2,o})^2}(1-e^{-T_W/\tau_c})},
\end{equation}
where $T_{2,s}$ is the XY-8 qubit coherence time, $T_{1,o}$ is the optical transition lifetime and all other terms were previously defined.

In our experiments, we apply this pulse sequence to Ion 3 with fixed $\tau=1.5~\mu$s, and vary $T_W$ from 0 to $9~$ms. In Extended Data Fig. 2b we plot the coherence on a logarithmic $y$ axis (markers), and fit this to a functional form (solid line):
\begin{equation}
\log(C)=a-\frac{T_W}{T_{2,s}}-b\left(1-e^{-\frac{T_W}{\tau_c}}\right),
\end{equation}
where $a$, $b$ and $\tau_c$ are free parameters and $T_{2,s}=27\pm2~$ms is extracted from independent measurements. We identify contributions from both spin decoherence (green region) and optical decorrelation (red region) and extract an optical correlation timescale of $\tau_c=1.42\pm0.04~$ms. We note that this is considerably longer than the delay time between entanglement heralding and dynamic rephasing ($\approx 20~\mu$s), thereby satisfying the requirement for quasi-static optical frequencies. We also note that the fitted value of $b$ predicts a considerably longer optical Ramsey coherence timescale of $3.38\pm0.04~\mu$s compared to the experimentally observed results presented earlier in this section. This indicates the presence of additional optical spectral diffusion processes with correlation timescales longer than the spin XY-8 coherence time (i.e. $\tau_c>27~$ms), we leave a detailed investigation of these for future work.

\subsection{System Parameters}
\label{ionprops}
The table below presents a summary of ion properties and system parameters which are used in modelling the entanglement protocols as detailed in Supplementary Information Section~\ref{simulation}.
\begin{center}\footnotesize
\begin{tabular}{|c|c|c|c|c|}
 \hline
  & Ion 1 & Ion 2 & Ion 3& Ion 4 \\ 
 \hline\hline
 Optical lifetime, $T_{1,o}$ & $2.17\pm0.017~\mu$s & $0.985\pm0.003~\mu$s\ & $2.33\pm0.019~\mu$s&$1.115\pm0.010~\mu$s\\ 
  \hline
 Single photon efficiency, $p_\text{det}$ & $(1.100\pm0.002)\!\times\!10^{-2}$ & $(4.60\pm0.05)\!\times\!10^{-3}$ & $(1.396\pm0.002)\!\times\!10^{-2}$&$(5.210\pm0.013)\!\times\!10^{-3}$\\ 
 \hline
Qubit initialization fidelity &  $0.9976\pm0.0003$& $0.9954\pm0.0008$ &$0.9979\pm0.0002$&$0.9962\pm0.0004$\\ 
 \hline
Resonant noise count rate & $8.3\pm0.5~$Hz & $7.5\pm0.5~$Hz &$1.8\pm0.2~$Hz& $1.2\pm0.2~$Hz\\ 
\hline
Optical  Ramsey coherence, $T^*_{2,o}$& $310\pm20~$ns & $220\pm10~$ns & $328\pm16~$ns&$221\pm7~$ns\\ 
\hline
Optical  linewidth (standard deviation), $\gamma^*$ & $2\pi\!\times\!(0.73\pm0.05)$MHz & $2\pi\!\times\!(1.02\pm0.05)$MHz & $2\pi\!\times\!(0.69\pm0.03)$MHz&$2\pi\!\times\!(1.02\pm0.03)$MHz \\ 
\hline
Optical echo coherence, $T_{2,o}$& $3.66\pm0.12~\mu$s & $1.75\pm0.04~\mu$s &$3.89\pm0.19~\mu$s&$1.89\pm0.11~\mu$s\\
\hline
Qubit lifetime, $T_{1,s}$ & $53\pm3~$ms & $23\pm1~$ms & $55\pm6~$ms&$17.0\pm1.1~$ms\\ 
\hline
Qubit XY8 coherence, $T_{2,s}$ & $21.2\pm0.7~$ms & $16.9\pm0.7~$ms &$27\pm2~$ms&$12.0\pm0.7~$ms\\ 
\hline
\end{tabular}
\end{center}

The optical properties associated with the $A$ transitions of all four ions listed in this table were measured according to the details provided in Supplementary Information Section~\ref{optprops}. During entanglement experiments, we use optical Rabi frequencies of $\Omega=2\pi\times2.5~$MHz, $2\pi\times10~$MHz and $2\pi\times3.5~$MHz  for entanglement heralding, rephasing and readout, respectively. Generally, larger Rabi frequencies minimize pulse error whereas smaller Rabi frequencies minimize noise counts from weakly coupled ions and device heating, these considerations inform our choice of these different values.

The qubit coherence times are measured using an XY-8 pulse sequence \cite{Gullion1990a} where the inter-pulse separation is fixed at $2\tau_s=5.8~\mu$s and the sequence duration is varied by changing the number of pulses, $N$. Qubit Rabi frequencies of $2\pi\times5~$MHz are used throughout this work.

Single photon efficiencies, noise count rates and initialization fidelities are independently measured for all four ions according to the following experimental protocol. We start by initializing a given ion in $\ket{1}$, applying a single optical $\pi$ pulse and collecting photons in a detection window of duration $\tau_M$. After $N$ experiment repetitions we detect a total of $N_{\ket{1}}$ photons and extract a single photon detection efficiency of\footnote{Note the quoted values for Ions 1 and 2 were measured using Detection setup 2 in Extended Data Fig.~3b, whereas Ion 3 was measured using Detection setup 3.} $N_{\ket{1}}/N$. We now repeat this measurement with the qubit initialized in $\ket{0}$. We temporally resolve the resulting photon count distribution and identify two components: one which decays exponentially with the predetermined ion optical lifetime with total number of counts $N_{\ket{0}}^{\text{exp}}$ and one with minimal temporal variation (i.e. a static contribution) with total counts $N_{\ket{0}}^{\text{DC}}$. The exponentially decaying contribution originates from qubit initialization error, we estimate the initialization fidelity as $N_{\ket{1}}/(N_{\ket{1}}+N_{\ket{0}}^{\text{exp}})$. The static component originates from two sources: weakly coupled ions and dark counts from room lights or laser leakage. To distinguish these sources, we repeat the experiment with all lasers detuned from the ions' optical transitions, leading to total counts $N_{\text{dark}}$. Since the weakly coupled ions have long optical lifetimes we can treat their contribution as approximately time-independent, we term this the resonant noise count rate which can be calculated as $(N_{\ket{0}}^{\text{DC}}-N_{\text{dark}})/(N\tau_M)$. The dark count rate from room lights and laser leakage is identified as  $N_{\text{dark}}/(N\tau_M)$, we measure a value of $8.0\pm0.2~$Hz when using Detection Setups 1 or 2 and a value of $6.3\pm0.12~$Hz when using Detection Setup 3. (see Methods Section~A and Extended Data Fig.~3b for detail on these detection setups).

\subsection{Coherence in Entanglement Heralding}
We use three different sequences to explore the effects of optical and spin coherence on entanglement heralding between Ions 1 and 2 (Fig.~S3).

\subsubsection{Ramsey Entanglement Heralding}
First, we perform a conventional Ramsey heralding measurement (Fig.~S3a). We initialize both qubits in $\ket{0}$, prepare a weak superposition state with probability $\alpha$ in $\ket{1}$, optically excite both ions and herald entanglement on a single photon detection which occurs at stochastic time $t_0$ in a window of size $\tau_h$. At this stage in the protocol the heralded state is given by:
\begin{equation}
\label{ramsey_state_SI}
    \ket{\psi(t_0)}=\frac{1}{\sqrt{2}}\left(\ket{10}+\ket{01}e^{-i(\Delta\omega_{12} t_0+\delta\omega_q\tau_h)}\right),
\end{equation}
where $\Delta\omega_{12}$ is the optical $A$ transition frequency difference between Ions 1 and 2. Note here that we are also considering the effect of spin detuning and are in a frame rotating at the qubit drive frequency for the $\ket{0}$ state. Specifically, $\delta\omega_q=\left(\omega^{(2)}_q-\widetilde{\omega}^{(2)}_q\right)-\left(\omega^{(1)}_q-\widetilde{\omega}^{(1)}_q\right)$ where $\omega^{(i)}_q$ and $\widetilde{\omega}^{(i)}_q$ are the qubit transition and drive frequencies for Ion $i$, respectively (i.e. transition $\ket{0}\leftrightarrow\ket{1}$). The entangled state coherence is defined as $C=2\bra{01}\hat{\rho}(t_0)\ket{10}$ and has real part given by $\real\{C\}=\cos(\Delta\omega_{12} t_0+\delta\omega_q\tau_h)$. Both optical and spin transitions are quasi-static, meaning that they have correlation timescales much longer than the duration of a single experiment. Hence, we regard $\Delta\omega_{12}$ and $\delta\omega_q$ as shot-to-shot random variables which are static during a single experiment, but are sampled from Gaussian probability distributions between consecutive experiment repetitions. Specifically, the optical transition frequency difference, $\Delta\omega_{12}$ has mean value $\langle\Delta\omega_{12}\rangle\approx2\pi\times31~$MHz and standard deviation $\sqrt{(\gamma^*_1)^2+(\gamma^*_2)^2}=2\pi\times1.25\pm0.07~$MHz where $\gamma^*_1$ and $\gamma^*_2$ are the optical linewidths of Ions 1 and 2, defined as the standard deviations of their respective optical frequency distributions, and we have assumed that their time-dependent spectral diffusion is uncorrelated. Similarly, $\delta\omega_q$ has mean value 0 and standard deviation given by $\sqrt{2\left(T_{2,s}^{*(1)}\right)^{-2}+2\left(T_{2,s}^{*(2)}\right)^{-2}}$ where $T_{2,s}^{*(i)}$ is the qubit Ramsey coherence time associated with Ion $i$ \cite{Kindem2020a}.

When ensemble-averaged over these optical and spin transition frequency distributions, the real part of the entangled state coherence is given by:
\begin{equation}
\label{ramsey_decay}
    \real\{C\}=\cos(\langle\Delta\omega_{12}\rangle t_0)e^{-\frac{1}{2}t_0^2(\gamma_1^{*2}+\gamma_2^{*2})}e^{-\tau_h^2\left[{\left(T_{2,s}^{*(1)}\right)^{-2}+\left(T_{2,s}^{*(2)}\right)^{-2}}\right]}.
\end{equation}
 Experimentally, we measure the qubits in the $X$ basis and correlate the parity expectation value, $\langle\hat{X}\hat{X}\rangle$ with the photon emission time, $t_0$. Assuming that coherences associated with $\ket{00}\bra{11}$ are negligible we find that $\langle\hat{X}\hat{X}\rangle=\real\{C\}$. In Fig.~S3a we plot these results (markers) and find a functional form that agrees well with equation~\eqref{ramsey_decay}, i.e., a Gaussian decay profile centred at $t_0=0$. The solid line corresponds to a simulation result with only one free parameter: the optical path phase difference, $\Delta\Phi$, which determines the phase of the sinusoidal parity oscillation, see  Supplementary Information Section~\ref{simulation} for more details.
 
\subsubsection{Heralding with Precompensated Phase Accumulation}

The second sequence is depicted in Fig.~S3b and starts, as before, with the preparation of a superposition state; however, this time $\alpha$ is chosen such that $(1-\alpha)\ll 1$. Under this condition we will treat the $\ket{00}$ component as negligible. Next we optically excite both ions with optical $\pi$ pulses, wait for a duration $\tau_0$ and coherently transfer them back to the ground state with another pair of optical $\pi$ pulses. The resulting state is given by:
\begin{equation}
\ket{\psi}=(1-\alpha)\ket{00}+\sqrt{\alpha(1-\alpha)}\left[\ket{10}+\ket{01} e^{-i\left[\Delta\omega_{12}-\Delta\widetilde{\omega}_{12}+\delta\omega_q\right]\tau_0}\right]+\alpha\ket{11}e^{-i\theta},
\end{equation}
where $\Delta\widetilde{\omega}_{12}$ is the difference between $A$ transition driving laser frequencies for Ions 1 and 2, and we don't give an explicit expression for $\theta$ as it is irrelevant for the subsequent discussion. 
Next, we apply a qubit $\pi$ pulse, optically excite both ions, and herald entanglement by detecting a single photon at stochastic time $t_0$ in a window of size $\tau_h$, leading to the heralded state:
\begin{equation}
\ket{\psi(t_0)}=\frac{1}{\sqrt{2}}\left(\ket{01}+\ket{10} e^{-i\left[\Delta\omega_{12}(\tau_0-t_0)+\delta\omega_q(\tau_0-\tau_h)-\Delta\widetilde{\omega}_{12}\tau_0\right]}\right).
\end{equation}
Taking an ensemble average over the optical and spin frequency distributions defined in the previous sub-section, we find that the entangled state coherence is given by:
\begin{equation}
    \real\{C\}=\cos(\langle\Delta\omega_{12}\rangle t_0-\phi)e^{-\frac{1}{2}(t_0-\tau_0)^2(\gamma_1^{*2}+\gamma_2^{*2})}e^{-(\tau_h-\tau_0)^2\left[{\left(T_{2,s}^{*(1)}\right)^{-2}+\left(T_{2,s}^{*(2)}\right)^{-2}}\right]},
\end{equation}
where $\phi=(\langle\Delta\omega_{12}\rangle-\Delta\widetilde{\omega}_{12})\tau_0$ and is equal to zero if the difference between optical driving frequencies is chosen to match the static ion optical frequency difference, i.e. $\langle\Delta\omega_{12}\rangle=\Delta\widetilde{\omega}_{12}$.

The coherence of the resulting entangled state is now maximized when the stochastic photon emission satisfies $t_0=\tau_0$. Under this condition, dephasing due to quasi-static optical frequency variation during the initial free evolution time $\tau_0$ is perfectly counteracted by phase accumulation in the heralding window prior to the photon emission event. We observe this experimentally in Fig.~S3b where we choose three different values of $\tau_0=150,~450$ and $750~$ns and observe a Gaussian decay profile in entangled state coherence centred at different values of $t_0$. The Gaussian decay timescale is still limited by the ion linewidths, however, by shifting the point of maximum coherence to $t_0>0$ we can now herald with a two-sided acceptance window (i.e., both before and after the point of maximal coherence), which boosts the entanglement rate by a factor of two compared to the previous protocol, whilst still maintaining the same fidelity. Solid lines correspond to simulation results as described previously.

\subsubsection{Heralding with Dynamic Rephasing}
The third sequence, presented in Fig.~S3c is the dynamic rephasing protocol that was described in the main text and Methods Section~C. The sequence starts with a Ramsey heralding sequence leading to a state given by equation~\eqref{ramsey_state_SI} with stochastic photon emission time, $t_0$. Subsequently, a qubit $\pi$ pulse is applied and a delay of $\tau_h-t_0$ rephases qubit decoherence. Next, both ions are optically excited with $\pi$ pulses, after waiting for a duration $t_0$ to rephase the optical coherence, optical $\pi$ pulses coherently transfer both ions back to the ground state. Subsequently, the quantum state is given by:
\begin{equation}
\ket{\psi(t_0)}=\frac{1}{\sqrt{2}}\left(\ket{01}+\ket{10} e^{-i\Delta\widetilde{\omega}_{12}t_0}\right),
\end{equation}
where $\Delta\widetilde{\omega}_{12}$ is the difference between $A$ transition driving laser frequencies for Ions 1 and 2. Note that this sequence rephases the effects of quasi-static frequency variation on both optical and spin transitions such that the entangled state coherence is no longer limited by the ions' optical linewidths, $\gamma^*_1$ and $\gamma^*_2$, or the spin Ramsey coherence (i.e. the parity oscillation only depends on the static laser frequency difference). The dominant source of decoherence is now due to undetected spontaneous optical emission during the optical rephasing period.  Specifically, during the delay time with duration $t_0$, an undetected emission event would randomize the entangled state phase, thereby leading to a mixed state $\hat{\rho}=1/2(\ket{01}\bra{01}+\ket{10}\bra{10})$. As $t_0$ increases, the probability of such a deleterious event occurring exponentially saturates, with a timescale determined by the ions' optical lifetimes, thereby leading to the following entangled state coherence:
\begin{equation}
    \real\{C\}=\cos(\Delta\widetilde{\omega}_{12}t_0)e^{-t_0\left(\frac{1}{2T^{(1)}_{1,o}}+\frac{1}{2T^{(2)}_{1,o}}\right)},
\end{equation}
where $T^{(i)}_{1,o}$ is the optical lifetime of Ion $i$. Note that pure dephasing of the optical transitions isn't included here as it has negligible contribution compared to the lifetime decay (see Supplementary Information Section~\ref{optprops} for more details). 

We experimentally investigate these entangled state coherence limitations using the dynamic rephasing protocol in Fig.~S3c where, in contrast to the main text, we probe later photon emission times. We observe a close correspondence between the experimental data (markers) and our simulation (solid line), thereby verifying that our entangled state coherence decay is indeed limited by the optical lifetimes of the two ions.

\subsubsection{Application to Long-range Entanglement Heralding}
Finally, we will consider limitations to the entangled state coherence when applying the dynamic rephasing protocol to herald entanglement over long distances. This is an especially important consideration for future long-range multiplexing experiments where the photon travel time between emitter and detector is non-negligible.

More concretely, the dynamic rephasing protocol requires the optical transition frequency to be constant between the heralding and rephasing portions of the experiment, which are separated by some delay time, $T$. For a node separation $L$ this delay time is lower-bounded by $T>L/c$ (where $c$ is the speed of light), which accommodates the time for photons to travel from each emitter to the central detector and for a classical heralding signal to be sent back to each emitter node. If $T$ is greater than the optical frequency correlation timescale, $\tau_c$, then the heralded coherence won't be dynamically rephased.

We can analytically derive the entangled state coherence caused by spectral diffusion during a delay time of duration $T=L/c$ by applying a similar filter function formalism as detailed in Supplementary Information Section~\ref{optprops}. For a photon acceptance window size of duration $\tau_A$, the entangled state coherence is given by:
\begin{equation}
\real\{C\}=\frac{1}{\tau_A}\int^{\tau_A}_0e^{-t_0^2(\gamma_1^{*2}+\gamma_2^{*2})(1-e^{-L/c\tau_c})}e^{-t_0\left(1/2T_{1,o}^{(1)}+1/2T_{1,o}^{(2)}\right)}dt_0
\end{equation}
where $\gamma_i^{*}$ and  $T_{1,o}^{(i)}$ are the optical linewidth and lifetime of Ion $i$, respectively, note that we have assumed that the spectral diffusion correlation timescale is the same for both ions. Evaluating this integral we obtain:
\begin{equation}
\real\{C\}=\frac{\sqrt{\pi}}{2\Gamma\tau_A}e^{\frac{1}{4\Gamma^2\overline{T}_{1,o}^2}}\left[\erf\left(\Gamma\tau_A+\frac{1}{2\Gamma\overline{T}_{1,o}}\right)-\erf\left(\frac{1}{2\Gamma\overline{T}_{1,o}}\right)\right],
\end{equation}
where $\Gamma^2=(\gamma_1^{*2}+\gamma_2^{*2})\left(1-e^{-L/c\tau_c}\right)$ and $\frac{1}{\overline{T}_{1,o}}=\frac{1}{2T_{1,o}^{(1)}}+\frac{1}{2T_{1,o}^{(2)}}$.

Evaluating this expression for our system is complicated by the presence of multiple spectral diffusion processes. As measured in Supplementary Information Section~\ref{optprops}, the fastest process has a correlation timescale of $\tau_c=1.42~$ms and a corresponding linewidth of $\gamma^*=2\pi\times67~$kHz. Additional spectral diffusion processes with correlation timescales longer than 27~ms broaden the linewidth of our ions further to $\approx2\pi\times1~$MHz. Considering only the fastest diffusion process, and assuming that $\tau_c$ and $\gamma^*$ are the same for both ions, we find that the entangled state coherence limitation due to optical spectral diffusion is only $0.97$ as $L\rightarrow\infty$, hence this contributes only a negligible amount to the infidelity (less than 1.5\%) even for long transmission distances. The slower spectral diffusion processes are likely to have a greater impact on fidelity due to their larger contribution to the optical transition linewidth, however, since their correlation timescales are lower-bounded to $\tau_c>27~$ms, they will only impact the entangled state fidelities when node separations, $L$, are greater than $\approx5400~$km. Hence, we conclude that our combined dynamic rephasing protocol and platform will be suitable for future long-range entanglement experiments.

\subsection{Modelling Entanglement}
\label{simulation}
\subsubsection{Formalism}
We model the entanglement measurements using a quantum stochastic master equation formalism whereby the process of photon detection to herald entangled states can be encoded via quantum jumps. We will derive the time evolution for the specific case of bipartite remote entanglement between Ions 1 and 2 (Fig.~2), however, this modelling approach can be straightforwardly extended to all other measurements in this work.

Initially, let us consider each ion (labelled $i$) coupled to its cavity with rate $g_i$ via the $A$ transition which has frequency $\omega_i$ and operators $\hat{\sigma}_z^{(i)}=\ket{\text{e}}\bra{\text{e}}-\ket{1}\bra{1}$, $\hat{\sigma}_i^+=\ket{\text{e}}\bra{1}$, and $\hat{\sigma}_i^-=\ket{1}\bra{\text{e}}$. The cavity has frequency $\omega_c^{(i)}$ and bosonic creation operator $\hat{a}^\dag_i$. Additionally, the cavity is coupled with rate $\kappa_i$ to a waveguide which supports an infinite set of 1D continuum mode operators,  $\hat{b}^\dag_i(\omega)$, that satisfy the commutation relation $[\hat{b}_i(\omega),\hat{b}^\dag_i(\omega')]=\delta(\omega-\omega')$ \cite{Blow1990}. We treat this as a closed system with Hamiltonian:
\begin{equation}
\begin{split}
    \hat{H}=\sum_{i=1}^2\biggl(\frac{\hbar\omega_i}{2}\hat{\sigma}_z^{(i)}+\hbar\omega_c^{(i)}\hat{a}^\dag_i\hat{a}_i&+\int d\omega\left(\hbar\omega\hat{b}^\dag_i(\omega)\hat{b}_i(\omega)\right)
    \\&+\hbar\sqrt{\frac{\kappa_i}{2\pi}}\int d\omega\left(\hat{a}^\dag_i \hat{b}_i(\omega )+\hat{a}_i \hat{b}^\dag_i (\omega )\right)+\hbar g_i\left(\hat{a}^\dag_i\hat{\sigma}^-_i + \hat{a}_i\hat{\sigma}^+_i \right)\biggr).
    \end{split}
\end{equation}
First, we assume that each cavity is resonant with its respective ion, $\omega_i=\omega_c^{(i)}$. Since we are operating in the bad cavity regime, we adiabatically eliminate the cavity modes, $\hat{a}_i$. We also move into an interaction picture and re-write the system Hamiltonian as \cite{Angerer2018a}:
\begin{equation}
    \hat{H}=i\hbar\sum_{i=1}^2\sqrt{\frac{2g_i^2}{\pi\kappa_i}}\int d\omega \left(\hat{\sigma}^+_i\hat{b}_i(\omega)e^{-i(\omega-\omega_i) t}-\hat{\sigma}^-_i\hat{b}^\dag_i(\omega)e^{i(\omega -\omega_i)t}\right).
\end{equation}
We simplify this Hamiltonian by defining slowly-varying temporal continuum operators:\\ $\hat{b}_i(t)=\frac{1}{\sqrt{2\pi}}\int d\omega\hat{b}_i(\omega)e^{-i(\omega-\omega_i) t}$ that have commutation relation $[\hat{b}_i(t),\hat{b}^\dag_i(t')]=\delta(t-t')$ \cite{Brecht2015} leading to:
\begin{equation}
    \hat{H}=i\hbar\sum_{i=1}^2\sqrt{\frac{4g_i^2}{\kappa_i}}\left(\hat{\sigma}^+_i \hat{b}_i (t)-\hat{\sigma}^-_i \hat{b}^\dag_i (t)\right).
\end{equation}
We integrate the continuum operators over small time intervals, $\delta t$,  yielding the quantum noise increments $d\hat{B}_t=\int_t^{t+\delta t}\hat{b}(s)ds$ \cite{Baragiola2017}. The operators $d\hat{B}_t$ and $d\hat{B}^\dag_t$ are infinitesimal annihilation and creation operators that act nontrivially only on the time interval $[t,t + \delta t]$. The resulting infinitesimal evolution operator is given by \cite{Baragiola2017,wiseman_milburn_2009}:
\begin{equation}
    \hat{U}(t+\delta t,t)=\hat{\mathbb{I}}+\sum_{i=1}^2\left(\sqrt{\frac{4g_i^2}{\kappa_i}}\left(\hat{\sigma}^+_i d\hat{B}_t^{(i)}-\hat{\sigma}^-_i d\hat{B}_t^{\dag(i)}\right)-\frac{2g_i^2}{\kappa_i}\hat{\sigma}^+_i\hat{\sigma}^-_i\delta t\right).
\end{equation}
We derive the time evolution of our density matrix according to:
\begin{multline}
\label{rhoevol_2}
    \frac{\delta\rho}{\delta t}=\frac{\hat{U}(t+\delta t,t)\hat{\rho}\hat{U}(t+\delta t,t)^\dag-\hat{\rho}}{\delta t}=-\sum_{i=1}^2\left(\frac{i}{\hbar}\left[\hat{H}_\text{int}^{(i)},\hat{\rho}\right]+\frac{2g_i^2}{\kappa_i}\left\{\hat{\sigma}^+_i\hat{\sigma}^-_i,\hat{\rho} \right\}\right)\\
    +\frac{1}{\delta t}\left(\sqrt{\frac{4g_1^2}{\kappa_1}}\hat{\sigma}^-_1d\hat{B}_t^{\dag(1)}+\sqrt{\frac{4g_2^2}{\kappa_2}}\hat{\sigma}^-_2d\hat{B}_t^{\dag(2)}\right)\hat{\rho}\left(\sqrt{\frac{4g_1^2}{\kappa_1}}\hat{\sigma}^+_1d\hat{B}_t^{(1)}+\sqrt{\frac{4g_2^2}{\kappa_2}}\hat{\sigma}^+_2d\hat{B}_t^{(2)}\right),
\end{multline}
where we have assumed that all bath modes are initially in the vacuum state and have ignored terms that are $\mathcal{O}(\sqrt{\delta t})$ and higher. Note that we have also included local Hamiltonians, $\hat{H}_\text{int}^{(i)}$, for each ion which include additional spin energy level structure, and both optical and microwave driving.

Next, we consider the process of photodetection to herald entanglement. For simplicity, here we assume that each device is connected directly to a single non-polarizing beamsplitter with a 1D waveguide of length $L_i$. A single photon detector measures a photon at time $t_0$ in a window of size $\delta t$ in one of the beamsplitter output ports, described by an infinitesimal photonic annihilation operator $d\hat{C}_{t_0}$. Assuming dispersionless photon propagation in the waveguide with phase velocity $c$ we can use standard linear quantum optics relations to write this annihilation operator as:
\begin{equation}
\label{det_jumper}
    d\hat{C}_{t_0}=d\hat{B}_{(t_0-L_2/c)}^{(2)}e^{-i\omega_2(t_0-L_2/c)}+d\hat{B}_{(t_0-L_1/c)}^{(1)}e^{-i\omega_1(t_0-L_1/c)},
\end{equation}
where $d\hat{B}_{(t_0-L_i/c)}$ annihilates a photon at the location of Device $i$ during the advanced time interval $[t_0-L_i/c,t_0-L_i/c+\delta t]$ (as defined previously). We can simplify this equation by removing a global phase:
\begin{equation}
\label{det_jumper_2}
    d\hat{C}_{t_0}=d\hat{B}_{(t_0-L_2/c)}^{(2)}e^{-i\left(\Delta\omega_{12}(t_0-\bar{L}/c)+\Delta\Phi\right)}+d\hat{B}_{(t_0-L_1/c)}^{(1)},
\end{equation}
where $\bar{L}=\frac{L_1+L_2}{2}$, $\Delta\Phi=\frac{(\omega_1+\omega_2)(L_1-L_2)}{2c}$, and $\Delta\omega_{12}=\omega_2-\omega_1$.

Experimentally, $L_1-L_2\approx3~$m, causing photons from different devices to have different times-of-flight and leading to collapse operators, $d\hat{B}_{(t_0-L_i/c)}$, that act at different times. This effect is mitigated with a calibrated timing delay in the application of pulse sequences to ions in the two separate devices, thereby ensuring that the photons arrive to the detection setup at the same time. In the subsequent analysis we will make the simplifying assumption that $L_1=L_2=L$, however, we will retain the phase difference, $\Delta\Phi$, since wavelength-scale variations in path length difference will affect our Bell state coherence. Hence, we define a simplified collapse operator:
\begin{equation}
\label{det_jump_2}
    d\hat{C}_{t_0}=d\hat{B}_{\widetilde{t}_0}^{(2)}e^{-i\left(\Delta\omega_{12}\widetilde{t}_0+\Delta\Phi\right)}+d\hat{B}_{\widetilde{t}_0}^{(1)}
\end{equation}
where $\widetilde{t}_0$ is the advanced time defined as $\widetilde{t}_0=t_0-L/c$.

Using equations~\eqref{rhoevol_2} and \eqref{det_jump_2}  we can derive a time evolution of the quantum state that is independent of the photonic modes and can be described in two parts. Firstly,  time evolution conditioned on the absence of a photon detection given by $\hat{\rho}(t)=e^{\mathcal{L}t}[\hat{\rho}(0)]$ where:
\begin{equation}
\label{nophoton_2}
\hat{\mathcal{L}}[\hat{\rho}]=-\sum_{i=1}^2\left(\frac{i}{\hbar}\left[\hat{H}_\text{int}^{(i)},\hat{\rho}\right]+\frac{2g_i^2}{\kappa_i}\left\{\hat{\sigma}^+_i\hat{\sigma}^-_i,\hat{\rho} \right\}\right).
\end{equation}
Secondly, a quantum jump associated with photon detection at time $t_0$ in a window of size $\delta t$ which can be described by application of the combined two-ion collapse operator:
\begin{equation}
\hat{S}=\sqrt{\frac{4g_1^2\delta t}{\kappa_1}}\hat{\sigma}^-_1+\sqrt{\frac{4g_2^2\delta t}{\kappa_2}}\hat{\sigma}^-_2e^{-i\left(\Delta\omega_{12}\widetilde{t}_0+\Delta\Phi\right)}\\
\end{equation}
at time $\widetilde{t}_0=t_0-\frac{L}{c}$.

Hence, the conditional density matrix at the end of an experiment of duration $t_f$, where a single photon was detected at time $t_0$, is given by:
\begin{equation}
\label{cond_ev}
\hat{\rho}(t_f|t_0)=e^{\hat{\mathcal{L}}(t_f-\widetilde{t}_0)}\left[\hat{S}\left(e^{\hat{\mathcal{L}}\widetilde{t}_0}\left[\hat{\rho}(0)\right]\right)\hat{S}^\dag\right].
\end{equation}
Note that this density matrix is unnormalized and has trace equal to the probability of photodetection in a window of size $\delta t$. The unconditional heralded density matrix is then obtained by summing over all photon detection times, $t_0$, in increments of $\delta t$ i.e. $\hat{\rho}(t_f)=\sum_{t_0=0}^{\tau_A}\hat{\rho}(t_f|t_0)$ where $\tau_A$ is the photon acceptance window size.

Other detection setup configurations (such as those depicted in Extended Data Fig.~3b) can be characterized using this formalism by modifying equation~\eqref{det_jumper}. In situations where multiple detection modes are required (for example, orthogonally polarized modes impinging on one detector, or the use of multiple detectors) the time evolution is evaluated independently for each detection mode. The result is then a classical average of the resulting time evolved density matrices, weighted according to the respective detection efficiencies.

We include various sources of experimental infidelity in this model, as detailed in the following subsections.

\subsubsection{Photon Inefficiencies}

So far, we have assumed that the entire cavity emission is concentrated in the waveguide mode. In reality, the cavity can have multiple loss paths: $\kappa_i=\kappa_{\text{wg}}^{(i)}+\kappa_{\text{s}}^{(i)}$ where $\kappa_{\text{wg}}^{(i)}$ is the energy loss rate into the waveguide and $\kappa_{\text{s}}^{(i)}$ is the rate of energy scatter into other modes. Furthermore, due to various sources of inefficiency (e.g. waveguide-fibre coupling) only a fraction ($\eta_i$) of light in the waveguide will be detected. The total probability to detect a photon emitted by Ion $i$ into the cavity is given by $p_\text{det}^{(i)}=\kappa_{\text{wg}}^{(i)}\eta_i/\kappa_i$. All scattered and undetected light can be treated using a Lindbladian master equation which traces over (ignores) the radiation modes. Only the detected fraction needs to be treated using the time evolution according to equation \eqref{rhoevol_2}.

Specifically, we redefine the jump operator according to:

\begin{equation}
\hat{S}=\sqrt{\frac{4g_1^2p_\text{det}^{(1)}\delta t}{\kappa_1}}\hat{\sigma}^-_1+\sqrt{\frac{4g_2^2p_\text{det}^{(2)}\delta t}{\kappa_2}}\hat{\sigma}^-_2e^{-i\left(\Delta\omega_{12}\widetilde{t}_0+\Delta\Phi\right)}.
\end{equation}

The no-jump evolution is also redefined according to:
\begin{equation}
\label{nophoton_3}
\hat{\mathcal{L}}[\hat{\rho}]=-\sum_{i=1}^2\left(\frac{i}{\hbar}\left[\hat{H}_\text{int}^{(i)},\hat{\rho}\right]+\frac{2g_i^2p_\text{det}^{(i)}}{\kappa_i}\left\{\hat{\sigma}^+_i\hat{\sigma}^-_i,\hat{\rho} \right\}-\hat{C}_i\hat{\rho}\hat{C}^\dag_i+\frac{1}{2}\left\{\hat{C}^\dag_i\hat{C}_i,\hat{\rho}\right\}\right),
\end{equation}
where $\hat{C}_i$ are collapse operators associated with undetected emission on the $A$ transition of Ion $i$ and have form:
\begin{equation}
\hat{C}_i=\ket{1}\bra{\text{e}}\sqrt{\frac{4g_i^2(1-p_\text{det}^{(i)})}{\kappa_i}+\Gamma_A^{(i)}},
\end{equation}
where $\Gamma_A^{(i)}$ is the un-enhanced optical decay rate for Ion $i$ via the $A$ transition.

\subsubsection{Noise Counts}
We also consider the possibility of noise counts occurring with uniform time distribution at rate $R$. Any experiments where $>1$ photons are detected will be rejected. Furthermore, we assume that the detector dead-time is sufficiently short relative to average photon measurement rates that we can ignore its effect (i.e. assume negligible probability that a second photon is detected during the dead time following the detection of a first photon). To account for these noise counts, we make a correction to the conditional density matrix defined in equation~\eqref{cond_ev}:
\begin{equation}
    \hat{\rho}(t_f|t_0)=R\delta t e^{\hat{\mathcal{L}}t_f}[\hat{\rho}(0)]+e^{\hat{\mathcal{L}}(t_f-\widetilde{t}_0)}\left[\hat{S}\left(e^{\hat{\mathcal{L}}\widetilde{t}_0}\left[\hat{\rho}(0)\right]\right)\hat{S}^\dag\right].
\end{equation}

\subsubsection{Fast (Markovian) Noise}

For Markovian processes we add additional terms with form $\hat{C}_i\hat{\rho}\hat{C}^\dag_i-\frac{1}{2}\left\{\hat{C}^\dag_i\hat{C}_i,\hat{\rho}\right\}$ into our no-jump time evolution, equation~\eqref{nophoton_3}. The corresponding collapse operators, $\hat{C}_i$, are listed below.
\begin{enumerate}
\item
Undetected optical emission on the $E$ transition is described by:
\begin{equation}
\hat{C}_i=\ket{0}\bra{1_e}\sqrt{\frac{4g_i^2}{\kappa_i}+\Gamma_E^{(i)}},
\end{equation}
where $\Gamma_E^{(i)}$ is the un-enhanced optical decay rate for Ion $i$ via the $E$ transition.
\item
Ground state spin relaxation is included with two processes:
\begin{equation}
\hat{C}_i=\ket{0}\bra{1}\sqrt{\frac{\Gamma_s^{(i)}}{2}},~\text{and}~\hat{C}_i=\ket{1}\bra{0}\sqrt{\frac{\Gamma_s^{(i)}}{2}},
\end{equation}
where $\Gamma_s^{(i)}$ is the spin relaxation rate for Ion $i$, note that we operate in the high temperature limit where the thermalized occupation of the two qubit states is approximately equal.
\item
Pure dephasing on the spin transition is included with:
\begin{equation}
\hat{C}_i=\left(\ket{0}\bra{0}-\ket{1}\bra{1}\right)\sqrt{\frac{\gamma_s^{(i)}}{2}},
\end{equation}
where $\gamma_s^{(i)}$ is the pure spin dephasing rate for Ion $i$ derived from the XY-8 coherence time and spin lifetime.
\end{enumerate}

\subsubsection{Slow (Non-Markovian) Noise}
We model spectral diffusion of the optical and spin transitions by performing 300 Monte-Carlo repetitions of this time-evolution and taking an ensemble average. The transition frequencies are assumed quasi-static  and are re-sampled for each repetition. Specifically, the spin transitions have a probability distribution given in \cite{Ruskuc2022}. The optical transitions are sampled from Gaussian distributions with standard deviations given by the optical linewidth of each ion, $\gamma^*_i$.

\subsection{Entanglement Fidelity Analysis}
\label{fid_analysis}
In this section we use the model described in Supplementary Information Section~\ref{simulation} to gain insight into fidelity limitations to our remote entangled Bell state between Ions 1 and 2. Input parameters for this simulation are derived from independent calibration experiments, detailed in Supplementary Information Section~\ref{ionprops}. We introduce one free parameter: a correction to the single-photon detection efficiency associated with Ion 1, leading to a reduction by a factor of 0.955 compared to the measured value. We attribute this correction to a systematic change in detection efficiency from optical alignment drift between the calibration and data taking. Errors in simulated results are obtained by performing 50 repetitions of the entire simulation, each with input parameters sampled from Gaussian distributions with mean and standard deviation given by the experimentally determined values and errors presented in Supplementary Information Section~\ref{ionprops}. This model accurately predicts a range of experimental results, within standard error, as presented in Fig.~2 and Extended Data Fig.~6.

First, we run our simulation with all sources of error included, and with a photon acceptance window size of $500~$ns, leading to a fidelity of $\mathcal{F}_\text{sim}=0.729\pm0.004$; this agrees well with the experimentally measured value of $\mathcal{F}=0.723\pm0.007$. Next, we examine how different sources of error impact the fidelity. This is quantified in two different ways: first by simulating the improvement in fidelity if a given source of error were removed and, second, the resulting fidelity if all other sources were neglected and the given source were exclusively present. Note that adding or removing certain sources of error (such as noise counts) will lead to different optimal values of $\alpha_1$ and $\alpha_2$. For these cases, the simulated fidelities are quoted after re-optimizing $\alpha_1$ and $\alpha_2$ according to the procedure outlined in Methods Section~C. The resulting fidelity analysis is presented in Extended Data Fig.~7a and is reproduced here for convenience:

\begin{center}
\begin{tabular}{|c|c|c|}
 \hline
  Source of error & $\mathcal{F_\text{sim}}$ if error removed & $\mathcal{F_\text{sim}}$ if error exclusively present \\ 
 \hline\hline
 Undetected emission during rephasing & $0.791\pm0.004$ & $0.9163\pm0.0004$ \\  
\hline
 Noise counts & $0.777\pm0.003$ & $0.9096\pm0.0009$ \\ 
  \hline
  Optical gate errors & $0.777\pm0.002$ & $0.929\pm0.006$ \\ 
\hline
 Imperfect qubit initialization & $0.746\pm0.004$ & $0.9173\pm0.0002$ \\ 
 \hline
 Qubit gate errors & $0.746\pm0.003$ & $0.9805\pm0.0015$ \\ 
\hline
Ion optical lifetime difference& $0.740\pm0.003$ & $0.9978\pm0.0003$ \\ 
\hline
Qubit $T_1$ and $T_2$ during free evolution & $0.734\pm0.003$ & $0.9951\pm0.0004$ \\ 
\hline
Optical path phase instability & $0.733\pm0.004$ & $0.9851\pm0.0002$ \\ 
\hline

\end{tabular}
\end{center}

We provide a brief explanation of the different parameters considered:
\begin{itemize}
\item Undetected spontaneous emission events during the dynamic rephasing portion of our pulse sequence destroy the entangled state coherence. Specifically, after entanglement heralding the dynamic rephasing sequence consists of optical re-excitation of the ions for a duration $t_0$ during which this deleterious process can occur.
\item Noise counts are defined as photons that are detected during the heralding stage of the sequence, but originate from sources other than the two $^{171}$Yb ions being entangled, predominantly other background $^{171}$Yb ions. This contribution is characterized in Supplementary Information Section~\ref{ionprops}.
\item Optical and qubit gate errors are modelled using the simulation detailed in Supplementary Information Section~\ref{simulation}. Specifically, these errors consist of a combination of quasi-static frequency shifts of the optical and spin transition frequencies and fast (Markovian) optical spontaneous emission, qubit relaxation, and qubit dephasing.
\item The qubit initialization fidelity is independently measured in Supplementary Information Section~\ref{ionprops}.
\item The optical lifetimes of the two ions are independently measured in Supplementary Information Section~\ref{ionprops}.
\item The qubit $T_1$ and $T_2$ times are independently measured in Supplementary Information Section~\ref{ionprops}.
\item The optical path length difference phase instability is characterized in Methods Section~B.
\end{itemize}

We identify three dominant sources of infidelity and propose approaches to mitigate their effect. Firstly, noise counts which primarily come from weakly coupled ions can be reduced by using crystals with lower concentrations of $^{171}$Yb ions or switching to alternative host materials with lower rare-earth impurities \cite{Stevenson2022,Gritsch:23,Uysal2024b}. Secondly, undetected spontaneous emission events during optical rephasing can be reduced by using a non-Purcell-enhanced transition. For instance, cavities with approximately $10\times$ narrower linewidth would selectively enhance the optical $A$ transition, the optical $E$ transition between states $\ket{0}\leftrightarrow\ket{1_e}$ could then be used to rephase coherence. This requires that optical spectral diffusion between these two transitions is identical. For electric field fluctuations (which we believe are the dominant source of spectral diffusion in our system) this is a justified assumption, as the spin sub-levels must be equally affected according to Kramers’ theorem. However, note that even if dominated by magnetic field fluctuations, the transition frequency shifts will still be correlated, albeit with a relative magnitude determined by the ratio of their g-factors (namely $g_A/g_E$). Under these circumstances, dynamic rephasing on the optical $E$ transition could still counteract phase accumulation on the $A$ transition via a re-scaled rephasing delay time (i.e. $t_0\rightarrow g_A/g_Et_0$). Alternatively, fast tuning of the cavity resonance \cite{Casabone2021,Xia:22,Yu2023,Yang2023} could  be used to temporarily extend the optical lifetime of the $A$ transition during the dynamic rephasing period. Note that these approaches would eliminate the lifetime-limitation in entangled state coherence, we could then also boost the entanglement rate by heralding with the entire photonic emission without suffering any reduction in fidelity. Finally, optical gate errors, which predominantly come from $T_1$ decay (spontaneous emission) during application of the optical pulse can be mitigated by using larger optical Rabi frequencies with correspondingly shorter gate times. 

If all three of these error sources were eliminated, we could achieve entangled state fidelities of $\mathcal{F_\text{sim}}=0.883\pm0.004$.

\subsection{Entanglement Rate Analysis}
\label{rate_anal}
In the main text we demonstrated entangled state generation between Ions 1 and 2 with a heralding rate of $\mathcal{R}=3.1~$Hz using a photon acceptance window size of $500~$ns (Fig.~2c). In Extended Data Fig.~7b we provided a brief summary of factors limiting this rate. In this section, we provide a more detailed analysis and provide approaches to improve this rate.

Given the low photodetection probabilities, the entanglement heralding rate in the single photon protocol can be approximated using the following equation:
\begin{equation}
\mathcal{R}=R_\text{exp}\times\left[\alpha_1p_\text{det}^{(1)}\left(1-e^{-\tau_A/T^{(1)}_{1,o}}\right)+\alpha_2p_\text{det}^{(2)}\left(1-e^{-\tau_A/T^{(2)}_{1,o}}\right)\right],
\end{equation}
where $R_\text{exp}=12.3~$kHz is the experiment repetition rate, $\tau_A=500~$ns is the acceptance window size, $T^{(1)}_{1,o}=2.17\pm0.017~\mu$s and $T^{(2)}_{1,o}=0.985\pm0.003~\mu$s are the optical lifetimes, $p_\text{det}^{(1)}=(1.100\pm0.002)\times10^{-2}$ and $p_\text{det}^{(2)}=(4.60\pm0.05)\times10^{-3}$ are the photon detection efficiencies (measured in Supplementary Information Section~\ref{ionprops}), and $\alpha_1=0.062$ and $\alpha_2=0.078$ are the probabilities of occupying $\ket{1}$ in the initial weak superposition states (in all cases, the numbered subscripts or superscripts indicate whether the value relates to Ion 1 or Ion 2). This leads to a predicted entanglement rate of $3.48\pm0.03$~Hz which is close to the experimentally observed value of 3.1~Hz, albeit outwith the measured error margin. We attribute this deviation to systematic errors associated with slow drift in device coupling and optical setup alignment between data taking and calibration.

The experiment repetition rate corresponds to an average time of $81~\mu$s per entanglement attempt which is attributed to the following segments of the pulse sequence (reproduced from Extended Data Fig.~7b for convenience):
\begin{center}
\begin{tabular}{|c|c|}
\hline
Operation & Time per entanglement attempt \\ 
\hline\hline
Emptying $\ket{\text{aux}}$ & $24~\mu$s \\ 
\hline
Qubit initialization & $33~\mu$s \\ 
\hline
Path phase difference measurement & $14~\mu$s \\ 
\hline
Heralding pulse sequence & $10~\mu$s \\
\hline
\end{tabular}
\end{center}
The most significant improvement in experiment repetition rate can be achieved by reducing the initialization time per entanglement attempt. This could be accomplished by either enhancing the cavity quality factor, which increases the $A$ transition decay rate, or by implementing entanglement multiplexing. Both approaches will be discussed later in this section.

The values of $\alpha_1$ and $\alpha_2$ cannot be increased to improve $\mathcal{R}$ without reducing the entangled state fidelity, this trade-off is inherent to single photon protocols (see discussion in Methods Section~C and \cite{Hermans2023}). Moving to a two-photon protocol (as in Supplementary Information Section~\ref{bk}) removes this limitation, however, it ultimately leads to a reduction in the over all heralding rate, which depends on the photon detection efficiencies squared.

The photon detection efficiencies are limited by the following factors:
\begin{center}
\begin{tabular}{|c|c|c|}
\hline
Experimental  contribution & Device 1 efficiency & Device 2 efficiency \\ 
\hline\hline
Cavity--waveguide coupling  & $0.117\pm0.004$ & $0.065\pm0.004$\\ 
\hline
Chip--fibre coupling & 0.27& 0.12\\ 
\hline
Optical setup & 0.66 &  0.65\\ 
\hline
SNSPD & 0.85 & 0.85\\
\hline
\end{tabular}
\end{center}

Parameters without error bars were obtained using an optical power meter resulting in negligible random error. The cavity-waveguide coupling and associated error bar were obtained by fitting the cavity reflection spectrum. The resulting predicted single-photon efficiencies are $(1.77\pm0.06)\times10^{-2}$ and $(4.3\pm0.3)\times10^{-3}$ for Device~1 and Device~2, respectively. These show considerable deviation from the independently measured single photon detection efficiencies (see Supplementary Information Section~\ref{ionprops}). We attribute this difference to a large systematic error due to the long period of time between data taking and calibration ($\sim1$~month).  Nevertheless, we see that photon detection efficiencies could be substantially improved via an increase in the cavity-waveguide coupling rate: similar devices in YVO$_4$ have demonstrated critical coupling with $\kappa_\text{in}/\kappa=0.5$ \cite{Zhong2017d}. Furthermore, the waveguide--fibre coupling could be considerably improved by using alternative approaches such as adiabatic mode transfer to a tapered fibre \cite{Tiecke:15}. With these improvements we estimate a $15\times$ improvement to the entanglement rate.

The effect of a finite heralding window size ($\tau_A$) can mitigated by rephasing optical coherence on non-Purcell-enhanced transitions as discussed in Supplementary Information Section~\ref{fid_analysis}. This would lead to an entangled state coherence which is no longer limited by the optical lifetime, thereby enabling an increase in the acceptance window size without reducing the heralded fidelity. With this improvement we estimate a $5\times$ increase in the entanglement rate.

The entanglement rate could be further boosted by multiplexing using a larger number of ions: with the current devices, expanding to 7 ion pairs is feasible (as discussed in Supplementary Information Section~\ref{optprops}) which would lead to a $4\times$ increase over single pair rates. Combined with the previously discussed improvements, an entanglement rate of $\sim1$~kHz should be achievable.

In the future, we will improve the multiplexing capacity by increasing the $^{171}$Yb concentration and tailoring the optical inhomogeneous distribution in a similar fashion to ref.~\cite{Ulanowski2024} where $>360$ spectrally resolved ions were measured. Furthermore, by fabricating devices from suspended YVO$_4$ membranes, we hope that an order of magnitude increase in device $Q$ factors will be achievable, as has recently been demonstrated with diamond thin-film photonics \cite{Ding2024}. With these advancements entanglement rates exceeding 10~kHz will be possible. 

\subsection{Estimation of $\ket{\text{aux}}$ Infidelity}
\label{aux_sec}
For the measurements presented in Fig.~2, Fig.~3 and Fig.~4, imperfect initialization into the qubit manifold (i.e. residual population in the $\ket{\text{aux}}$ states) would lead to an over-estimate of our entangled state fidelities. 

To understand this effect, consider the application of the single-photon entanglement sequence to the erroneous initial state $\ket{0}\ket{\text{aux}}$ where Ion 1 has been correctly initialized, however Ion 2 is outside the qubit manifold. Under these circumstances, optical and microwave driving will only affect Ion 1 which can then generate a heralding photon, however Ion 2 is unaffected. Therefore, a single photon detection heralds the incorrect state, $\ket{1}\ket{\text{aux}}$. Next, consider our readout sequence (described in Methods Section~C and depicted in Extended Data Fig.~4f). For each ion, it operates by separating two read periods (each consisting of 100 consecutive optical excitations) by a qubit $\pi$ pulse. A photon detection in either the first or second period corresponds to a successful measurement of $\ket{1}$ or $\ket{0}$, respectively. However, an absence of photodetections causes the measurement to fail (i.e. no state is ascribed to the qubit), which mitigates errors due to photon loss. However, a measurement failure could also be caused by occupation of the $\ket{\text{aux}}$ state, in other words, the measurement result is post-selected on the ion occupying the qubit manifold. Hence, the measurement of an incorrectly heralded state, $\ket{1}\ket{\text{aux}}$, would be discarded,
leading to an overestimate of the entangled state fidelity. We derive the relation $\mathcal{F}_\text{herald}=(1-P_\text{aux})\mathcal{F}$ where $\mathcal{F}$ is the measured fidelity (post-selected on occupation of the qubit manifold), $\mathcal{F}_\text{herald}$ is the non-post-selected fidelity, and $P_\text{aux}$ is the probability of occupying the $\ket{\text{aux}}$ state, assumed equal for both ions.

In order to estimate this source of infidelity we measure $P_\text{aux}$ of Ion 1 using the following experimental procedure. First, we prepare the ion into $\ket{0}$ and perform 600 optical reads (consisting of an optical $\pi$ pulse followed by photodetection), each separated by a qubit $\pi$ pulse. The total number of photons detected ($N_\text{init}$) measures the combined population of the $\ket{0}$ and $\ket{1}$ states. We then wait for a duration of 15~s, during which the ion's states thermalize to a Boltzmann distribution before repeating the readout procedure to obtain a number of photons, $N_\text{therm}$. 
We use these values to extract an estimate for the $\ket{\text{aux}}$ population according to:
\begin{equation}
P_\text{aux}=1-\frac{N_\text{init}}{N_\text{therm}}\left(\frac{e^{-\frac{\hbar\omega_{0\leftrightarrow\text{aux}}}{k_bT}}+e^{-\frac{\hbar\omega_{1\leftrightarrow\text{aux}}}{k_bT}}}{2+e^{-\frac{\hbar\omega_{0\leftrightarrow\text{aux}}}{k_bT}}+e^{-\frac{\hbar\omega_{1\leftrightarrow\text{aux}}}{k_bT}}}\right)=0.023\pm0.015,
\end{equation}
where $\omega_{1\leftrightarrow\text{aux}}$ and $\omega_{0\leftrightarrow\text{aux}}$ are the frequencies of the $\ket{1}\leftrightarrow\ket{\text{aux}}$ and $\ket{0}\leftrightarrow\ket{\text{aux}}$ transitions, respectively, and $T=0.64\pm0.07~$K is the device temperature, measured in a separate experiment.

During entanglement experiments we empty the $\ket{\text{aux}}$ state into the qubit manifold after every 50 failed heralding attempts (Extended Data Fig.~4b). During each attempt, optical pulses applied to the ion can lead to population shelving into $\ket{\text{aux}}$, thereby slightly increasing $P_\text{aux}$. When considering all 50 attempts, we predict an average $\ket{\text{aux}}$ state population of $\langle P_\text{aux}\rangle=0.060\pm0.017$. We note that the calibration experiments performed here likely overestimate the value of $\langle P_\text{aux}\rangle$ since they involve more population transfer to $\ket{\text{aux}}$ and less frequent initialization compared to the entanglement experiments. Therefore, we use this as a lower bound estimate for our entangled state fidelity given by $\mathcal{F}_\text{herald}>(0.940\pm0.017)\times\mathcal{F}$.

We note that the heralding and teleportation protocols presented in Supplementary Information Sections~\ref{bk} and \ref{teleportsec} are not susceptible to this issue since they rely on two photon detections: one in each of two photodetection periods, separated by a qubit $\pi$ pulse, thereby eliminating error states involving $\ket{\text{aux}}$ at the heralding stage (i.e. prior to measurement). In these cases, the measured fidelities will equal the heralded fidelities, i.e. $\mathcal{F}=\mathcal{F}_\text{herald}$. We note that detection of a dark count in one of the heralding periods could still lead to a heralded error state, however, the combined probability of $\ket{\text{aux}}$ state occupation and noise count detection is sufficiently small to be neglected.

\subsection{Two Photon Entanglement Protocols}
\label{bk}
We use a two photon protocol based on the Barrett-Kok scheme \cite{Barrett2005} to herald entangled states between two remote ions (Ions 1 and 2).

The experiment sequence is presented in Extended Data Fig.~8a. We start with both ions prepared in an equal superposition: $\ket{\psi}=1/2(\ket{0}+\ket{1})\otimes(\ket{0}+\ket{1})$. We then perform two successive rounds of optical excitation, separated by a qubit $\pi$ pulse, each of which is followed by a photodetection period with duration $\tau_h=\tau_s=2.9~\mu$s. An entangled state is heralded when a single photon is measured in each photodetection period with stochastic emission times $t_0$ and $t_1$, respectively (each measured relative to the start of the respective heralding period). At this stage, we carve out both $\ket{00}$ and $\ket{11}$ components yielding an entangled state:
\begin{equation}
\ket{\psi(t_0,t_1)}= \frac{1}{\sqrt{2}}\left(\ket{10}+\ket{01}e^{i\Delta\omega_{12}(t_0-t_1)}\right).
\end{equation}
Unlike the single-photon protocol, this scheme carves out $\ket{11}$, avoiding the need for weak superposition states. This not only eliminates infidelity associated with $\ket{11}$, but it also increases the ion brightness, leading to a reduced contribution from noise counts (see Supplementary Information Section~\ref{ionprops}). Next, we apply a dynamical decoupling sequence consisting of 5 $\pi$ pulses on both the qubit transition and excited state spin transition, $f_e$, with separation $2\tau_s=5.8~\mu$s. The excited state $\pi$ pulses help to mitigate qubit population-basis errors due to imperfect optical $\pi$ pulses (Methods Section~C).

During the dynamical decoupling sequence, we apply a variant of the dynamic rephasing protocol, however there are two distinct cases to consider. If $t_1>t_0$ then we rephase spin and optical coherences for durations $\tau_h-t_1+t_0$ and $t_1-t_0$, respectively, after the third $\pi$ pulse. If, however, $t_1<t_0$, we rephase spin and optical coherences for durations $\tau_h-t_0+t_1$ and $t_0-t_1$, respectively, after the second $\pi$ pulse. Finally, we apply a qubit $Z$ rotation with angle $\Delta\widetilde{\omega}_{12}(t_0-t_1)$ to Ion 2 (note, this gate isn't shown in the pulse sequence since it isn't applied globally) leading to a heralded state:
\begin{equation}
\ket{\psi}= \frac{1}{\sqrt{2}}\left(\ket{01}+\ket{10}\right).
\end{equation}

In contrast to the single photon heralding scheme, the second photon detection at time $t_1$ erases the phase $\Delta\Phi$ associated with the optical path length difference caused by the first photon detection at time $t_0$. Therefore, stabilization of the optical path phase difference is not required for this measurement.

The entangled state is verified by measuring populations in nine cardinal two-qubit Pauli bases along $X$, $Y$ and $Z$ and performing maximum likelihood quantum state tomography to reconstruct the density matrix (\cite{James2001} and Supplementary Information Section~P). Extended Data Fig.~8b plots the resulting entangled state fidelity and rate against the two-sided acceptance window size $W$, where heralding events are accepted if $-W/2<t_0-t_1<W/2$. The fidelity and rate range from $\mathcal{F}=0.84\pm0.03$ and $\mathcal{R}=0.017$~Hz to  $\mathcal{F}=0.75\pm0.02$ and $\mathcal{R}=0.091$~Hz, respectively, as the window size is varied from $W=200$~ns to $W=1.6~\mu$s. In Extended Data Fig.~8c we plot the density matrix corresponding to a window size $W=0.6~\mu$s, this state has an entanglement fidelity and rate of $\mathcal{F}=0.81\pm0.02$ and $\mathcal{R}=0.049$~Hz, respectively.

Also note that this entanglement protocol will carve out error states that occur when either of the two ions occupies the $\ket{\text{aux}}$ state. This is discussed in more detail in Supplementary Information Section~\ref{aux_sec}.

\subsection{Probabilistic Quantum State Teleportation}
\label{teleportsec}
We implement a protocol that extends this two-photon entanglement scheme to probabilistically teleport quantum states between the two remote network nodes: from Ion 2 to Ion 1 \cite{Hu2023}. We note that even with perfect photodetection efficiency, this protocol's reliance on a linear optics Bell state measurement will limit its success probability to $50\%$; nevertheless, it represents an advantage over remote state preparation \cite{Lo2000} by enabling transmission of quantum states that are unknown to the sender.

We prepare Ion 2's qubit in the target state:
\begin{equation}
\ket{\psi_T}=\cos(\theta_T/2)\ket{0}+e^{i\phi_T}\sin(\theta_T/2)\ket{1}
\end{equation}
by application of consecutive $Y$ and $Z$ rotations through angles $\theta_T\in[0,\pi]$ and $\phi_T\in[0,2\pi]$, respectively, to the initial state, $\ket{0}$ (Fig.~S4a). These angles encode all possible single qubit pure states. While this method of state preparation relies on classical knowledge of the target state by the sender, we emphasize that this is not a requirement of the protocol. Namely, Ion 2 could be provided with an unknown quantum state (for instance, by means of a SWAP gate applied to an auxiliary qubit) and the protocol would operate in the same manner. Ion 1 is prepared in the initial state $1/\sqrt{2}(\ket{0}+\ket{1})$.

 Subsequently, both ions are optically excited in two successive rounds, separated by a qubit $\pi$ pulse. Single photon detections in each round, at times $t_0$ and $t_1$ (measured relative to the start of each heralding window), carve out $\ket{00}$ and $\ket{11}$ leading to an entangled two qubit state:
\begin{equation}
\ket{\psi(t_0,t_1)}= \cos(\theta_T/2)\ket{10}\\+e^{i(\phi_T+\phi(t_0,t_1))}\sin(\theta_T/2)\ket{01},
\end{equation}
where $\phi(t_0,t_1)=\Delta\omega_{12}(t_0-t_1)$ is the undesired, random phase shift resulting from the combination of stochastic photon heralding and a fluctuating optical frequency difference. We note that optical path phase, $\Delta\Phi$, accumulated by the first photon is cancelled by the second; hence, active phase stabilization is not required for these measurements.

The stochastic phase, $\phi(t_0,t_1)$, is corrected using the pulse sequences presented in the right panels of Extended Data Fig.~8a which contain dynamically adapted parameters that depend on both photon detection times. Specifically, after the heralding windows we apply a dynamical decoupling sequence to the qubits consisting of five $\pi$ pulses. In cases where $t_1 < t_0$, we correct the stochastic phase after the second qubit $\pi$ pulse by waiting for a duration $\tau_h - t_0 + t_1$ before applying two optical $\pi$ pulses separated by a delay time of $t_0 - t_1$ to both ions.  These two delays correct qubit and optical dephasing, respectively, that occurred during the preceding heralding windows. In cases where $t_1 > t_0$ the correction happens after the third qubit $\pi$ pulse, the spin and optical coherences are rephased for durations of $\tau_h - t_1 + t_0$ and $t_1 - t_0$, respectively. At this stage, the two-qubit quantum state is given by:
\begin{equation}
\ket{\psi(t_0,t_1)}= \cos(\theta_T/2)\ket{01}\\+\sin(\theta_T/2) e^{i(\phi_T+\Delta\widetilde{\omega}_{12}(t_0-t_1))}\ket{10}.
\end{equation}
Finally, we apply a $Z$ rotation to Ion 1’s qubit by an angle $-\Delta\widetilde{\omega}_{12}(t_0-t_1)$, thereby yielding the state:
\begin{equation}
\ket{\psi}= \cos(\theta_T/2)\ket{01}+\sin(\theta_T/2) e^{i\phi_T}\ket{10}.
\end{equation}
We can rewrite this equation as:
\begin{equation}
\label{teleport}
\ket{\psi}=\frac{1}{\sqrt{2}}\left(\ket{\psi_T}\otimes\ket{+X}-\hat{Z}_1\ket{\psi_T}\otimes\ket{-X}\right),
\end{equation}
where $\hat{Z}_1=(\ket{0}\bra{0}-\ket{1}\bra{1})\otimes\mathbb{I}$, $\ket{\pm X} = (\ket{0} \pm \ket{1})/\sqrt{2}$ are $X$-basis eigenstates, and $\mathbb{I}$ is the identity operator.

Next, we read out Ion 2's qubit in the $X$ basis whilst simultaneously applying dynamical decoupling to preserve Ion 1's state. From equation~\eqref{teleport} we see that projection of Ion 2 onto $\ket{+X}$ leads to a collapse of Ion 1 onto $\ket{\psi_T}$, thereby achieving state teleportation. However, when Ion 2 is projected onto $\ket{-X}$, Ion 1 is collapsed onto the phase-flipped target state, $\hat{Z}_1 \ket{\psi_T}$; hence, an additional Pauli-$Z$ gate is applied to Ion 1's qubit in order to complete the quantum state transfer.

In Fig.~S4b we apply this protocol to a target superposition state with $\theta_T=\pi/2$. As we vary the azimuthal angle, $\phi_T$, between $0$ and $2\pi$ we observe an oscillation of Ion 1's $X$-basis population, $\langle\hat{X}\rangle$, verifying the teleportation of quantum coherence. To quantify the teleportation fidelity, in Fig.~S4c, we teleport the six cardinal Bloch states ($\ket{\pm X}$, $\ket{\pm Y}$, $\ket{0}$, $\ket{1}$); in each case, Ion 1's state is tomographically reconstructed  by measuring three Pauli expectation values along the $X$-, $Y$-, and $Z$-axes (here, $\ket{\pm Y} = (\ket{0} \pm i \ket{1})/\sqrt{2}$). The resulting six state transfer fidelities and Bloch vectors are plotted in the left and right panels, respectively, for a two-sided photon acceptance window size of $W=1~\mu$s. The average fidelity is $0.834\pm0.011$, which is above the classical limit of 2/3. The teleportation success probability is $1.9\times10^{-6}$ leading to a teleportation rate of 0.027~Hz.

\subsection{Multiplexed Entanglement Rate Enhancement}
\subsubsection{Additional Protocol Detail}
In this section we provide more detail regarding the use of multiplexing to boost the remote entanglement distribution rate. Since this rate is limited by the ion initialization time (Supplementary Information Section~I), we achieve an increase by parallelized initialization of multiple remote ion pairs, followed by sequential entanglement attempts.

Specifically, we use Ions 1 and 2 separated by $\Delta\omega_{12}\approx2\pi\times36.5~$MHz (Pair 1) and Ions 3 and 4 separated by $\Delta\omega_{34}\approx2\pi\times104~$MHz (Pair 2). Ions 1 and 3 are co-located in Device 1. Ions 2 and 4 are co-located in Device 2. Note that the ions' optical frequencies exhibit slow drift on the timescale of several weeks, hence $\Delta\omega_{12}$ is slightly different here compared to the entanglement experiments presented in Fig.~2. Optical pulses at each ion's $A$ transition frequency enable individual optical control. However, due to the narrow distribution of spin frequencies, all microwave control pulses are globally applied to qubits within the same device. Therefore, within each device we choose a microwave driving frequency that is intermediate between the co-located qubits.

The detailed entanglement sequence is presented in Extended Data Fig.~9a. We initialize all four qubits in parallel by applying optical $\pi$ pulses to all four ions, followed by excited state microwave $\pi$ pulses and decay into $\ket{0}$. Then we prepare each qubit in a weak superposition with probability $\alpha_i$ in the $\ket{1}$ state. Subsequently, we perform an entanglement attempt on Pair 2 comprising a pair of optical $\pi$ pulses applied to Ions 3 and 4, followed by a photon heralding window. We then apply two qubit dynamical decoupling periods before performing a consecutive entanglement attempt on Pair 1, comprising a pair of optical $\pi$ pulses applied to Ions 1 and 2, and a second photon heralding window. Pulse spacings are chosen carefully so that all qubit coherences are dynamically decoupled throughout both entanglement attempts. When both entanglement attempts fail, we re-start the sequence (i.e. return to qubit initialization). However, if a heralding photon is detected during either entanglement attempt we proceed exclusively with the corresponding ion pair. That is, we perform dynamic rephasing on the pair of ions associated with the successful heralding attempt, followed by real-time phase compensation, thereby retrieving a deterministic entangled state. Finally, we read out two-qubit populations of the successful pair in the $XX$, $YY$ or $ZZ$ bases. In the subsequent discussion, we will label the fidelity and rate derived from this measurement as $\mathcal{F}_\text{mult}$ and $\mathcal{R}_\text{mult}$.

To quantify the benefits of this multiplexed entanglement experiment we perform additional independent measurements that exclusively entangle either Pair 1 or Pair 2 (in a non-multiplexed fashion) as detailed in Fig.~2. Since these measurements involve only a single qubit per device, resonant microwave driving frequencies are used. In the subsequent discussions we will label the fidelities and rates associated with the Pair 1 entanglement experiment as $\mathcal{F}_1$ and $\mathcal{R}_1$, respectively. Results from the independent Pair 2 entanglement experiment are labelled $\mathcal{F}_2$ and $\mathcal{R}_2$.

\subsubsection{Multiplexed Rate Analysis}

In this section (and the subsequent fidelity analysis) we will exclusively utilize a 500~ns photon acceptance window for all quoted measurement results. The multiplexed experiment yields an entanglement rate of $\mathcal{R}_\text{mult}=7.1$~Hz. We compare this to the independent, non-multiplexed entanglement experiments which yield $\mathcal{R}_1=3.8~$Hz and $\mathcal{R}_2=3.7~$Hz for Pair 1 and Pair 2, respectively. Specifically, we calculate the ratio between the multiplexed rate and the average single-pair rate ($2\mathcal{R}_\text{mult}/(\mathcal{R}_1+\mathcal{R}_2)$) yielding a multiplexed entanglement rate enhancement factor of 1.9.

Note that multiplexed experiments use a repetition rate ($\mathcal{R}_\text{exp}^\text{mult}=10.8$~kHz) that is slightly lower than the independent experiment repetition rate ($\mathcal{R}_\text{exp}^\text{ind}=12.3$~kHz). This arises from the two sequential heralding periods in the multiplexed experiment which add $12~\mu$s per attempt (see Extended Data Fig.~9a). Therefore, the expected multiplexed rate enhancement factor is given by $2\mathcal{R}_\text{exp}^\text{mult}/\mathcal{R}_\text{exp}^\text{ind}=1.7$. We see that our measured entanglement rate slightly exceeds this prediction, the rest of this section will be devoted to analyzing this difference.

First, we can break down the multiplexed entanglement rate into contributions arising from Pair 1 or Pair 2, which are found to be $\mathcal{R}_\text{mult}^{(1)}=4.1$~Hz and $\mathcal{R}_\text{mult}^{(2)}=3.0$~Hz, respectively (note here that $\mathcal{R}_\text{mult}=\mathcal{R}_\text{mult}^{(1)}+\mathcal{R}_\text{mult}^{(2)}$). We compare these to predicted rates based on the independent measurements which yield $\mathcal{R}_1\times\mathcal{R}_\text{exp}^\text{mult}/\mathcal{R}_\text{exp}^\text{ind}=3.3$~Hz for Pair 1, and $\mathcal{R}_2\times\mathcal{R}_\text{exp}^\text{mult}/\mathcal{R}_\text{exp}^\text{ind}=3.2$~Hz for Pair 2. We find a small  deviation of $6\%$ for Pair 2 entanglement contributions, this can be explained by a drift in photon detection efficiencies between the multiplexed and independent measurements, which were separated by several days. However, for the Pair 1 entanglement results, there is a much larger $21\%$ discrepancy.

To understand this, we need to consider the effect of detuned qubit pulses. We will introduce the notation $P_{\ket{1}}^{(i)}$ to indicate the $\ket{1}$ state population of Ion $i$ just prior to an entanglement attempt in this multiplexed measurement. Specifically, during the first entanglement attempt (using Pair 2), the $\ket{1}$ state populations $P_{\ket{1}}^{(3)}$ and $P_{\ket{1}}^{(4)}$ are determined exclusively by the weak microwave pulses applied prior to optical excitation. From the simulation result in Fig.~S5a we see that pulse detuning has minimal effect on this population (i.e. $P_{\ket{1}}^{(i)}\approx\alpha_i$). However, for the subsequent Pair 1 entanglement attempt we observe considerably different behaviour. Specifically, the $\ket{1}$ state populations are now determined by a combination of the weak microwave pulses and two dynamical decoupling periods, leading to $P_{\ket{1}}^{(1)}$ and $P_{\ket{1}}^{(2)}$ considerably larger than $\alpha_1$ and $\alpha_2$ for relatively small qubit detuning as shown in Fig.~S5b. For reference, the microwave pulse detuning for Ion~1, 2, 3 and 4's qubits are approximately $0.76~$MHz, $0.69~$MHz, $-0.36~$MHz, and $-0.29~$MHz, respectively. The photon emission rate of each each ion scales with $P_{\ket{1}}^{(i)}$, leading to a higher entanglement rate as $P_{\ket{1}}^{(i)}$ increases (see Supplementary Information Section~\ref{rate_anal}). This mechanism explains the considerable increase in Pair~1 entanglement rate, and the resulting discrepancy between the predicted and measured rate enhancement factors.

\subsubsection{Multiplexed Fidelity Analysis}
Entangled state fidelities are calculated from the two-qubit population measurements according to:
\begin{equation}
\mathcal{F}=\left(2P_{01}+2P_{10}+\langle \hat{X}\hat{X}\rangle +\langle \hat{Y}\hat{Y}\rangle\right)/4,
\end{equation}
where $P_{01}$ and $P_{10}$ are defined as expectation values of the two-qubit projection operators $\ket{01}\bra{01}$ and $\ket{10}\bra{10}$, respectively. The multiplexed fidelity is averaged over all Bell pairs, regardless of whether Pair 1 or Pair 2 was delivered and has a value of $\mathcal{F}_\text{mult}=0.682\pm0.004$. We can also segregate the multiplexed result into two separate data sets for Pair 1 and Pair 2, based on which heralding attempt was successful. This yields fidelities of $\mathcal{F}_\text{mult}^{(1)}=0.700\pm0.005$ and $\mathcal{F}_\text{mult}^{(2)}=0.668\pm0.006$ for Pair 1 and Pair 2, respectively (population histograms in the $XX$, $YY$ and $ZZ$ measurement bases are plotted in Extended Data Fig.~9b). These can be compared to the independent (non multiplexed) entanglement measurement results where we obtain fidelities of $\mathcal{F}_1=0.745\pm0.006$ for Pair 1\footnote{We note that this value is slightly higher compared to the value of $\mathcal{F}_1=0.723\pm0.007$ measured in Fig.~2. This is attributed to slow drift of the ion and optical setup properties between taking these two data sets.}, and $\mathcal{F}_2=0.656\pm0.007$ for Pair 2. The reduced value of $\mathcal{F}_2$ compared to $\mathcal{F}_1$ is attributed to an increased noise count rate for Ions 3 and 4 (i.e. there is more background emission coming from weakly coupled $^{171}$Yb ions at these frequencies).

We see that Pair 2 entanglement results in the multiplexed experiment agree with the independent measurement within standard error (i.e. $\mathcal{F}_\text{mult}^{(2)}\approx\mathcal{F}_2$). However, there is a non-negligible decrease in Pair 1 entanglement fidelity for the multiplexed measurements (i.e. $\mathcal{F}_\text{mult}^{(1)}<\mathcal{F}_1$). To understand this reduction we will consider two possible sources of error. The first is optical control cross-talk between the two qubit pairs, namely, optical pulses applied to Ions 3 and 4 during the first heralding period could potentially induce excess qubit decoherence on Ions 1 and 2. To quantify this effect we perform an XY-8 dynamical decoupling sequence on Ion 1's qubit with fixed inter-pulse separation ($2\tau_s=5.9~\mu$s) and variable pulse number, $N$. At the centre of every free evolution period we apply an optical $\pi$ pulse to Ion 3, thereby emulating the entanglement heralding attempt on Pair 2 (Fig.~S6a). In Fig.~S6b we plot Ion 1's coherence as a function of the number of dynamical decoupling periods both with and without the optical pulses applied leading to exponential coherence decays with $1/e$ constants of $N_{1/e}=2890\pm60$ and $N_{1/e}=3170\pm50$, respectively. We repeat this measurement on ions in Device~2, namely, we measure the coherence on Ion~2 both with and without optical pulses applied to Ion~4 leading to exponential coherence decays that have $1/e$ constants of $N_{1/e}=2290\pm70$ and $N_{1/e}=2800\pm110$, respectively. For both Device~1 and Device~2 measurements we observe a small reduction in coherence when optical pulses are applied, this corresponds to an infidelity of $1-\mathcal{F}=(5.5\pm1.1)\times 10^{-5}$ for Pair~1 Bell states which is a negligible contribution. 

The second source of error arises from the same detuned qubit control that was discussed in the previous section. Specifically, with increased $P_{\ket{1}}^{(i)}$ the probability of $\ket{11}$ state occupation increases, and the entangled state fidelity decreases\footnote{For the ideal case of two identical emitters with the same $\ket{1}$ state population, $P_{\ket{1}}$, the fidelity will be upper bounded by $\mathcal{F}<1-P_{\ket{1}}$.} (see Methods Section~C for more details). Based on the simulation results in Fig.~S5 this would explain why the fidelity of Pair~2 Bell pairs is preserved, whereas the fidelity of Pair~1 decreases. However, based on the observed increase in entanglement rate for Pair 1 Bell states, this effect can only explain approximately half of the measured fidelity reduction. We attribute the residual discrepancy to drift in experimental setup parameters or ion properties between performing the multiplexed and independent measurements.

For future experiments we will investigate the use of more advanced qubit control pulses to counteract the effect of detuning. Specifically, Hermite pulses have a broader frequency response and could drive both qubits with lower error \cite{reiserer2016}. Alternatively, we could utilize ROTTEN composite pulses \cite{cummins2001} that are optimized to operate with maximum fidelity at two discrete qubit frequencies.

\subsection{Entangling Ions in the Same Device}
We can use the single photon protocol, described in the main text and Fig.~2 to generate entangled states of ions within the same nanophotonic cavity. We perform this measurement on Ions 1 and 3 within Device 1, which have an optical frequency difference of $\approx2\pi\times475~$MHz. There are three main differences with this measurement, summarized here and described in detail in Methods Section~D. First, we use AOM~5 in Extended Data Fig.~3a setup to generate two optical tones that resonantly address the ions in a manner that's passively phase stable. Second, photons from both ions are emitted into the same optical mode, hence, we use Detection Setup 1 in Extended Data Fig.~3b to herald entanglement and don't perform any optical path phase stabilization. Finally, since global microwave control precludes single qubit addressability, we use a differential AC Stark shift of the qubit transitions, generated by a detuned optical drive, to perform single-qubit $Z$ rotations \cite{chen2020}. By tuning the intensity of the AC Stark drive we can control the imparted phase shift.

Experimentally, we demonstrate the ability to control the phase of heralded Bell states using the AC Stark drive in Extended Data Fig.~10c. We herald entanglement and apply the dynamic rephasing protocol, then, we apply an AC Stark sequence (described in Methods Section~D and ref. \cite{chen2020}) with a Stark drive Rabi frequency $\Omega$, prior to measuring both ions in the $X$ basis. We correlate $\langle\hat{X}\hat{X}\rangle$ with the photon detection time, $t_0$, and observe a shift in the parity oscillation phase for different values of $\Omega^2$. Note, due to the large optical frequency difference, binning data in sufficiently fine increments to resolve the parity oscillation at $\approx2\pi\times475~$MHz would require prohibitively long experiment durations. Therefore,  we bin experimental results according to $t_0\pmod{6\pi/\Delta\widetilde{\omega}_{13}}$, where $\Delta\widetilde{\omega}_{13}$ is the difference between optical driving frequencies used to address Ions 1 and 3. In Extended Data Fig.~10b we compare these measured Bell state phases (markers) to single-qubit phase shift measurements (solid line, described in Methods Section~D) and find a close correspondence.

Having calibrated the AC Stark response, we repeat this measurement with a dynamically adaptive AC Stark intensity such that the resulting phase shift, $\Phi_\text{AC}$, counteracts the stochastic Bell State phase, i.e., $\Phi_\text{AC}=-\Delta\widetilde{\omega}_{13}t_0$. The resulting entangled state phase is now independent of photon detection time. We perform maximum likelihood tomography on the two-qubit state as detailed in Supplementary Information Section~P with a photon acceptance window size of $500~$ns.  The resulting density matrix is plotted in Extended Data Fig.~10d, with a corresponding heralding rate of $\mathcal{R}=4.5$~Hz and a fidelity of $\mathcal{F}=0.763\pm0.005$.

\subsection{Additional Tripartite W State Results}
In this section we provide additional experimental results related to the tripartite W state measurements presented in Fig.~4.

We first probe the time-dependent coherences of the tripartite W state prior to the dynamic rephasing part of our sequence. Specifically, as described in Methods Section~E, we prepare a weak superposition state on all three qubits followed by resonant optical excitation of all ions. Detection of a single photon at stochastic time $t_0$ heralds the following entangled state:
\begin{equation*}
\ket{\psi(t_0)}=\\\frac{1}{\sqrt{3}}\left(\ket{100}e^{-i\omega_1t_0}+\ket{010}e^{-i\omega_2t_0}+\ket{001}e^{-i\omega_3t_0}\right).
\end{equation*}
To measure the W state coherences of this state we read out all three qubits in the $X$ basis (Fig.~S7a). The W state has six quantum coherences, $C_{ij}=\Trace\{\hat{\rho}\ket{01}_{ij}\bra{10}_{ij}\otimes\ket{0}_k\bra{0}_k\}$, where $i,j,k$ correspond to permutations of qubits 1, 2 and 3. We note that the real part of the coherence term $\real\{C_{ij}\}$ will exhibit an oscillation at the optical frequency difference between Ions $i$ and $j$ with respect to photon measurement time, and, for the ideal W state, will be proportional to the two-qubit $X$-basis parity expectation value, i.e. $\real\{C_{ij}(t_0)\}\propto\cos(\Delta\omega_{ij}t_0)\propto\langle\hat{X}_i\hat{X}_j\rangle$. We plot the Fourier transform of each parity measurement in Fig.~S7b where the three panels correspond to $C_{23},~C_{13}$ and $C_{12}$ from top to bottom. We observe parity oscillations at $\approx2\pi\times508$~MHz, $\approx2\pi\times475$~MHz and $\approx2\pi\times34$~MHz, respectively which match the ion frequency differences. Note that the ions' optical frequencies exhibit slow drift on the timescale of several weeks, hence $\Delta\omega_{12}$ is slightly different here compared to the bipartite entanglement experiments in Fig.~2.

We also provide additional results related to the fully phase-corrected W state prepared in Fig.~4, which has the following form:
\begin{equation*}
\ket{\psi}=\\\frac{1}{\sqrt{3}}\left(\ket{100}+\ket{010}+\ket{001}\right).
\end{equation*}
Specifically, in Fig.~S7c we plot the entanglement fidelity and rate as a function of photon acceptance window size which range from $\mathcal{F}=0.59\pm0.01$ and $\mathcal{R}=1.0~$Hz for a window size of $150~$ns to $\mathcal{F}=0.458\pm0.003$ and $\mathcal{R}=10$~Hz for a window size of $2850~$ns.

\subsection{Quantum State Tomography Data and Results}
In this section we describe maximum likelihood tomography which is used to characterize the entangled and teleported states throughout this work. This involves finding a predicted density matrix that best describes the experimentally observed results \cite{James2001,Hradil1997}.

For bipartite entangled states we perform readout in 9 cardinal two-qubit Pauli bases along $X$, $Y$ and $Z$. Each measurement basis (denoted $(ij)$, with $i,j\in\{X,Y,Z\}$) yields four photon count numbers $N^{(ij)}_{00},~N^{(ij)}_{01},~N^{(ij)}_{10},~N^{(ij)}_{11}$, corresponding to readout of the four 2-qubit measurement basis eigenstates, $\ket{00},~\ket{01},~\ket{10},~\ket{11}$, respectively. Figure~2c shows the resulting populations for the $XX$, $YY$ and $ZZ$ measurement bases, the other 6 bases ($XY,~XZ,~YX,~YZ,~ZX,~ZY$) are not shown but are included in the maximum likelihood reconstruction that follows.

We assume that the results of each measurement basis are multinomially distributed such that the likelihood of a model density matrix, $\hat{\rho}$, is given by \cite{Bernien2013}:
\begin{equation}
\label{likelihood_func}
    \mathcal{L}^{(ij)}(\hat{\rho})=\frac{N^{(ij)}!}{N^{(ij)}_{00}!N^{(ij)}_{01}!N^{(ij)}_{10}!N^{(ij)}_{11}!}\times\left[n^{(ij)}_{00}\right]^{N^{(ij)}_{00}}\left[n^{(ij)}_{01}\right]^{N^{(ij)}_{01}}\left[n^{(ij)}_{10}\right]^{N^{(ij)}_{10}}\left[n^{(ij)}_{11}\right]^{N^{(ij)}_{11}},
\end{equation}
where $N^{(ij)}=\sum_{a\in\{00,01,10,11\}}N^{(ij)}_{a}$, and $\{n^{(ij)}_{a}\}$ are the predicted populations of each measurement basis eigenstate according to the model density matrix. For example, if we were to consider the $\ket{00}$ eigenstate in the $XX$ basis, the value of $n^{(XX)}_{00}=\Tr\{\ket{00}\bra{00}\hat{U}^{\dag}_{XX}\hat{\rho} \hat{U}_{XX}\}$ where $\hat{U}_{XX}$ rotates both qubits into the $X$ basis.

In order to account for readout infidelity, we need to define predicted population measurements $\{p^{(ij)}_{a}\}$ which are related to the ideal populations $\{n^{(ij)}_{a}\}$ by a readout transformation matrix $R_{a,b}$ which satisfies: \nopagebreak$p^{(ij)}_{a}=\sum_{b\in\{00,01,10,11\}}R_{a,b}n^{(ij)}_{b}$. Here, $R_{a,b}$ is the probability of measuring state $\ket{a}$, given that the two qubits were perfectly prepared in state $\ket{b}$, and $a,b\in\{00,01,10,11\}$ \cite{Nguyen2019}. With this readout correction, the predicted $\ket{00}$ measured population in the $XX$ basis becomes:
\begin{equation*}
\begin{split}
    p^{(XX)}_{00}=&\Trace\{\ket{00}\bra{00}\hat{U}^{\dag}_{XX}\hat{\rho} \hat{U}_{XX}\}R_{00,00}+\Trace\{\ket{01}\bra{01}\hat{U}^{\dag}_{XX}\hat{\rho} \hat{U}_{XX}\}R_{00,01}+\\
    &\Trace\{\ket{10}\bra{10}\hat{U}^{\dag}_{XX}\hat{\rho} \hat{U}_{XX}\}R_{00,10}+\Trace\{\ket{11}\bra{11}\hat{U}^{\dag}_{XX}\hat{\rho} \hat{U}_{XX}\}R_{00,11},
\end{split}
\end{equation*}
similar expressions are used for the other populations and other readout bases, but are not listed here for brevity. We use a redefined likelihood function where $\{n_a^{(ij)}\}$ in equation~\eqref{likelihood_func} are replaced with $\{p_a^{(ij)}\}$. We assume that $N^{(ij)}$ is sufficiently large that we can approximate the likelihood function with a normal distribution:
\begin{equation*}
    \mathcal{L}^{(ij)}(\hat{\rho})\propto \prod_{a\in\{00,01,10,11\}}e^{-\left[\frac{\left(\tilde{n}_{a}^{(ij)}-p_{a}^{(ij)}\right)^2N^{(ij)}}{2\tilde{n}_{a}^{(ij)}\left(1-\tilde{n}_{a}^{(ij)}\right)}\right]},
\end{equation*}
where $\tilde{n}_{a}^{(ij)}$ are normalized measured populations, i.e., $\tilde{n}_{a}^{(ij)}=N^{(ij)}_{a}/N^{(ij)}$. The likelihood function for all 9 measurement bases is simply the product of these individual basis likelihood functions:
\begin{equation*}
    \mathcal{L}(\hat{\rho})=\prod_{i,j\in\{X,Y,Z\}}\mathcal{L}^{(ij)}(\hat{\rho}).
\end{equation*}
Finally, we take the log-likelihood, leading to:
\begin{equation*}
    \mathcal{L}(\hat{\rho})=\!\!\!\!\!\!\!\!\!\sum_{i,j\in\{X,Y,Z\}}\sum_{a\in\{00,01,10,11\}}\left[\frac{-\left[\tilde{n}_{a}^{(ij)}-p_{a}^{(ij)}\right]^2N^{(ij)}}{2\tilde{n}_{a}^{(ij)}\left(1-\tilde{n}_{a}^{(ij)}\right)}\right].
\end{equation*}

We use the convex optimization package CVX \cite{gb08,cvx} in Matlab to find the density matrix, $\hat{\rho}$, which maximizes the log-likelihood. Error estimates are obtained using a bootstrapping protocol. This involves using our predicted density matrix to generate 1000 simulated sets of experimental results, each of which is used to predict a new density matrix via the maximum likelihood analysis. Errors are quoted as standard deviations of observables derived from this set of density matrices.

A similar approach is used for estimating the three-qubit density matrix of tripartite entangled states. In this case, measurements are performed in all 27 cardinal three-qubit Pauli bases along $X$, $Y$ and $Z$, and the readout matrix $R_{a,b}$ now transforms three-qubit populations  i.e. $a,b\in\{000,001,010,011,100,101,110,111\}$.

Single qubit tomography, used in the teleportation experiments, is performed via a direct reconstruction from measurements in the $X,~Y$ and $Z$ bases. Specifically, the predicted density matrix is given by:
\begin{equation*}
    \hat{\rho}=\frac{1}{2}\left[\mathbb{I}+\langle\hat{X}\rangle\hat{X}+\langle\hat{Y}\rangle\hat{Y}+\langle\hat{Z}\rangle\hat{Z}\right],
\end{equation*}
where $\hat{X}$, $\hat{Y}$, and $\hat{Z}$ are the Pauli matrices, and $\langle\hat{X}\rangle$, $\langle\hat{Y}\rangle$, and $\langle\hat{Z}\rangle$ are measured populations which have been corrected for readout infidelity.

Raw data for all tomography measurements in this work are presented in subsequent subsections, along with the corresponding readout transformation matrices ($R_{a,b}$) and resulting complex-valued predicted density matrices, $\hat{\rho}$.

\subsubsection{Entangling Two Remote Ions Using a Single Photon Protocol}
Data presented in this subsection is used to derive the results presented in Fig.~2c. The table below presents raw (uncorrected) Pauli-basis measurement results with a $500~$ns photon acceptance window, where $N_{ab}$ corresponds to the number of measurement outcomes where Ion 1 was in the $\ket{a}$ state and Ion 2 was in the $\ket{b}$ state.

\begin{center}
\begin{tabular}{|c|c|c|c|c|c|}
 \hline
 Ion 1 Basis & Ion 2 Basis & $N_{00}$ & $N_{01}$ & $N_{10}$ & $N_{11}$\\
 \hline\hline
  $X$ & $X$ & 1493 & 489 & 492 & 1385 \\ 
 \hline
  $X$ & $Y$ & 558 & 739 & 715 &  534\\ 
 \hline
  $X$ & $Z$ & 660 & 718 & 648 & 576 \\ 
 \hline
  $Y$ & $X$ & 693 & 579 & 541 & 683 \\ 
 \hline
  $Y$ & $Y$ & 1472 & 492 & 486 & 1442 \\ 
 \hline
  $Y$ & $Z$ & 593 & 633 & 614 & 677 \\ 
 \hline
  $Z$ & $X$ & 564 & 672 & 643 & 599 \\ 
 \hline
  $Z$ & $Y$ & 633 & 679 & 624 & 692 \\ 
 \hline
 $Z$ & $Z$ & 356 & 1646 & 1673 & 349 \\ 
 \hline
\end{tabular}
\end{center}
The readout transformation matrix is measured to be:
\begin{equation*}
\begin{pmatrix}
p_{00}\\p_{01}\\p_{10}\\p_{11}
\end{pmatrix}=
\begin{pmatrix}
0.959 & 0.052 & 0.044 & 0.002 \\
0.018 & 0.925 & 0.001 & 0.042 \\
0.023 & 0.001 & 0.937 & 0.051 \\
0.000 & 0.022 & 0.018 & 0.905 \\
\end{pmatrix}
\begin{pmatrix}
n_{00}\\n_{01}\\n_{10}\\n_{11}
\end{pmatrix},
\end{equation*}
where $\{p_{ab}\}$ and $\{n_{ab}\}$ are the measured and actual populations, respectively, associated with Ion 1 in the $\ket{a}$ state and Ion 2 in the $\ket{b}$ state. After performing maximum likelihood tomography, the resulting density matrix for our system is:
\begin{equation*}
\hat{\rho}=
\begingroup
\setlength\arraycolsep{10pt}
\begin{pmatrix}
0.046+0.000i & 0.026+0.009i & 0.013+0.003i & -0.002-0.011i \\ 
0.026-0.009i & 0.442+0.000i & 0.286+0.071i & -0.017+0.008i \\ 
0.013-0.003i & 0.286-0.071i & 0.433+0.000i & -0.010+0.015i \\ 
-0.002+0.011i & -0.017-0.008i & -0.010-0.015i & 0.079+0.000i \\ 

\end{pmatrix}
\endgroup.
\end{equation*}
This matrix is ordered according to the state vector $\begin{pmatrix}\ket{00}&\ket{01}&\ket{10}&\ket{11}\end{pmatrix}^T$ where we adopt a convention of $\ket{\text{Ion~1},\text{Ion~2}}$ for labelling. The absolute values of this matrix are graphically depicted in Fig.~2c. 

\subsubsection{Entangling Two Remote Ions Using a Two Photon Protocol}

Data presented in this subsection is used to derive the results presented in Extended Data Fig.~8. The table below presents raw (uncorrected) Pauli-basis measurement results with a two-sided acceptance window size $W=600~$ns, where $N_{ab}$ corresponds to the number of measurement outcomes where Ion 1 was in the $\ket{a}$ state and Ion 2 was in the $\ket{b}$ state.

\begin{center}
\begin{tabular}{|c|c|c|c|c|c|}
 \hline
 Ion 1 Basis & Ion 2 Basis & $N_{00}$ & $N_{01}$ & $N_{10}$ & $N_{11}$\\
 \hline\hline
  $X$ & $X$ & 134 & 29 & 35 & 129 \\ 
 \hline
  $X$ & $Y$ & 69 & 67 & 72 & 66\\ 
 \hline
  $X$ & $Z$ & 79 & 77 & 82 & 76 \\ 
 \hline
  $Y$ & $X$ & 65 & 89 & 70 & 70 \\ 
 \hline
  $Y$ & $Y$ & 119 & 36 & 35 & 117 \\ 
 \hline
  $Y$ & $Z$ &67 & 69 & 77 & 64\\ 
 \hline
  $Z$ & $X$ & 69 & 63 & 82 & 85 \\ 
 \hline
  $Z$ & $Y$ & 78 & 70 & 75 & 67 \\ 
 \hline
 $Z$ & $Z$ & 15 & 122 & 102 & 19\\ 
 \hline
\end{tabular}
\end{center}
The readout transformation matrix is measured to be:
\begin{equation*}
\begin{pmatrix}
p_{00}\\p_{01}\\p_{10}\\p_{11}
\end{pmatrix}=
\begin{pmatrix}
0.936 & 0.052 & 0.049 & 0.003 \\
0.027 & 0.911 & 0.001 & 0.048 \\
0.035 & 0.002 & 0.923 & 0.051 \\
0.001 & 0.034 & 0.027 & 0.898 \\
\end{pmatrix}
\begin{pmatrix}
n_{00}\\n_{01}\\n_{10}\\n_{11}
\end{pmatrix},
\end{equation*}
where $\{p_{ab}\}$ and $\{n_{ab}\}$ are the measured and actual populations, respectively, associated with Ion 1 in the $\ket{a}$ state and Ion 2 in the $\ket{b}$ state. After performing maximum likelihood tomography, the resulting density matrix for our system is:
\begin{equation*}
\hat{\rho}=
\begingroup
\setlength\arraycolsep{10pt}
\begin{pmatrix}
0.015+0.000i & 0.011+0.014i & -0.001+0.001i & 0.019-0.015i \\ 
0.011-0.014i & 0.470+0.000i & 0.343+0.028i & 0.018-0.001i \\ 
-0.001-0.001i & 0.343-0.028i & 0.465+0.000i & 0.004-0.016i \\ 
0.019+0.015i & 0.018+0.001i & 0.004+0.016i & 0.050+0.000i \\ 

\end{pmatrix}
\endgroup.
\end{equation*}
This matrix is ordered according to the state vector $\begin{pmatrix}\ket{00}&\ket{01}&\ket{10}&\ket{11}\end{pmatrix}^T$ where we adopt a convention of $\ket{\text{Ion~1},\text{Ion~2}}$ for labelling. The absolute values of this matrix are graphically depicted in Extended Data Fig.~8c.

\subsubsection{Entangling Three Remote Ions Using a Single Photon Protocol}
Data presented in this subsection is used to derive the results presented in Fig.~4. The table below presents raw (uncorrected) Pauli-basis measurement results with a $300~$ns photon acceptance window, where $N_{abc}$ corresponds to the number of measurement outcomes where Ion 1 was in the $\ket{a}$ state, Ion 2 was in the $\ket{b}$ state and Ion 3 was in the $\ket{c}$ state.

\begin{center}
\begin{tabular}{|c|c|c|c|c|c|c|c|c|c|c|}
 \hline
 Ion 1 Basis & Ion 2 Basis& Ion 3 Basis  & $N_{000}$ & $N_{001}$ & $N_{010}$ & $N_{011}$& $N_{100}$ & $N_{101}$ & $N_{110}$ & $N_{111}$\\
 \hline\hline
  $X$ & $X$ & $X$ & 513 & 164 & 166 & 196 & 208 & 179 & 233 & 475 \\ 
 \hline
  $X$ & $X$ & $Y$ & 316 & 464 & 188 & 185 & 231 & 185 & 398 & 329\\ 
 \hline
  $X$ & $X$ & $Z$ & 516 & 235 & 199 & 173 & 215 & 179 & 483 & 209 \\ 
 \hline
  $X$ & $Y$ & $X$ & 393 & 167 & 405 & 243 & 235 & 379 & 182 & 327\\ 
 \hline
  $X$ & $Y$ & $Y$ & 335 & 213 & 185 & 443 & 394 & 187 & 203 & 333 \\ 
 \hline
  $X$ & $Y$ & $Z$ & 349 & 195 & 358 & 211 & 358 & 196 & 337 & 175\\ 
 \hline
  $X$ & $Z$ & $X$ & 517 & 192 & 252 & 163 & 209 & 492 & 196 & 231 \\ 
 \hline
  $X$ & $Z$ & $Y$ & 338 & 384 & 230 & 237 & 392 & 295 & 221 & 229\\ 
 \hline
 $X$ & $Z$ & $Z$ & 337 & 310 & 349 & 65 & 362 & 312 & 372 & 67 \\ 
 \hline
  $Y$ & $X$ & $X$ & 345 & 168 & 186 & 359 & 414 & 200 & 192 & 301 \\ 
 \hline
  $Y$ & $X$ & $Y$ & 374 & 185 & 363 & 153 & 147 & 341 & 222 & 368\\ 
 \hline
  $Y$ & $X$ & $Z$ & 378 & 204 & 343 & 193 & 307 & 215 & 330 & 225 \\ 
 \hline
  $Y$ & $Y$ & $X$ & 205 & 189 & 374 & 412 & 375 & 333 & 205 & 178 \\ 
 \hline
  $Y$ & $Y$ & $Y$ & 514 & 187 & 202 & 213 & 185 & 186 & 190 & 529 \\ 
 \hline
  $Y$ & $Y$ & $Z$ & 506 & 209 & 210 & 169 & 178 & 188 & 511 & 251 \\ 
 \hline
  $Y$ & $Z$ & $X$ & 344 & 357 & 220 & 226 & 394 & 308 & 217 & 209 \\ 
 \hline
  $Y$ & $Z$ & $Y$ & 485 & 183 & 216 & 207 & 200 & 464 & 187 & 225 \\ 
 \hline
 $Y$ & $Z$ & $Z$ & 344 & 280 & 323 & 71 & 409 & 323 & 348 & 64 \\ 
  \hline
  $Z$ & $X$ & $X$ & 437 & 181 & 201 & 443 & 241 & 158 & 185 & 219 \\ 
 \hline
  $Z$ & $X$ & $Y$ & 362 & 329 & 365 & 318 & 215 & 214 & 211 & 201\\ 
 \hline
  $Z$ & $X$ & $Z$ & 378 & 349 & 375 & 316 & 343 & 60 & 319 & 73 \\ 
 \hline
  $Z$ & $Y$ & $X$ & 348 & 294 & 350 & 336 & 233 & 209 & 218 & 202 \\ 
 \hline
  $Z$ & $Y$ & $Y$ & 484 & 225 & 189 & 476 & 240 & 210 & 173 & 232 \\ 
 \hline
  $Z$ & $Y$ & $Z$ & 409 & 347 & 369 & 306 & 314 & 92 & 355 & 77 \\ 
 \hline
  $Z$ & $Z$ & $X$ &390 & 311 & 354 & 352 & 343 & 356 & 99 & 69 \\ 
 \hline
  $Z$ & $Z$ & $Y$ & 358 & 310 & 324 & 335 & 324 & 319 & 87 & 75 \\ 
 \hline
 $Z$ & $Z$ & $Z$ & 221 & 501 & 533 & 111 & 594 & 113 & 136 & 26 \\ 
 \hline
\end{tabular}
\end{center}
The readout transformation matrix is measured to be:
\begin{equation*}
\begin{pmatrix}
p_{000}\\p_{001}\\p_{010}\\p_{011}\\p_{100}\\p_{101}\\p_{110}\\p_{111}
\end{pmatrix}=
\begin{pmatrix}
0.910 & 0.059 & 0.052 & 0.003 & 0.058 & 0.004 & 0.003 & 0.000 \\
0.037 & 0.889 & 0.002 & 0.051 & 0.002 & 0.057 & 0.000 & 0.003 \\
0.016 & 0.001 & 0.874 & 0.057 & 0.001 & 0.000 & 0.056 & 0.004 \\
0.001 & 0.015 & 0.036 & 0.854 & 0.000 & 0.001 & 0.002 & 0.054 \\
0.034 & 0.002 & 0.002 & 0.000 & 0.886 & 0.058 & 0.050 & 0.003 \\
0.001 & 0.033 & 0.000 & 0.002 & 0.036 & 0.865 & 0.002 & 0.049 \\
0.001 & 0.000 & 0.032 & 0.002 & 0.015 & 0.001 & 0.851 & 0.055 \\
0.000 & 0.001 & 0.001 & 0.032 & 0.001 & 0.015 & 0.035 & 0.831 \\
\end{pmatrix}
\begin{pmatrix}
n_{000}\\n_{001}\\n_{010}\\n_{011}\\n_{100}\\n_{101}\\n_{110}\\n_{111}
\end{pmatrix},
\end{equation*}
where $\{p_{abc}\}$ and $\{n_{abc}\}$ are the measured and actual populations, respectively, associated with Ion 1 in the $\ket{a}$ state, Ion 2 in the $\ket{b}$ state and Ion 3 in the $\ket{c}$ state. After performing maximum likelihood tomography, the resulting density matrix for our system is $\hat{\rho}=\hat{\rho}_\text{real}+i\hat{\rho}_\text{imag}$, where:

\begin{equation*}
\hat{\rho}_\text{real}=
\begingroup
\setlength\arraycolsep{3pt}
\begin{pmatrix}
0.044 & -0.013 & 0.009 & -0.003 & 0.005 & -0.009 & -0.003 & -0.005 \\
-0.013 & 0.252 & 0.156 & -0.000 & 0.164 & -0.001 & -0.011 & 0.002 \\
0.009 & 0.156 & 0.278 & 0.007 & 0.168 & -0.002 & 0.006 & 0.003 \\
-0.003 & -0.000 & 0.007 & 0.046 & -0.002 & 0.017 & 0.021 & 0.000 \\
0.005 & 0.164 & 0.168 & -0.002 & 0.270 & 0.006 & -0.000 & 0.004 \\
-0.009 & -0.001 & -0.002 & 0.017 & 0.006 & 0.043 & 0.028 & 0.005 \\
-0.003 & -0.011 & 0.006 & 0.021 & -0.000 & 0.028 & 0.057 & -0.009 \\
-0.005 & 0.002 & 0.003 & 0.000 & 0.004 & 0.005 & -0.009 & 0.009 \\

\end{pmatrix}
\endgroup,
\end{equation*}

\begin{equation*}
\hat{\rho}_\text{imag}=
\begingroup
\setlength\arraycolsep{3pt}
\begin{pmatrix}
0.000 & -0.008 & 0.004 & 0.000 & -0.003 & -0.011 & 0.003 & 0.006 \\
0.008 & 0.000 & 0.014 & 0.008 & 0.037 & 0.011 & -0.003 & -0.003 \\
-0.004 & -0.014 & 0.000 & 0.012 & 0.021 & 0.018 & -0.000 & -0.002 \\
-0.000 & -0.008 & -0.012 & 0.000 & -0.006 & 0.013 & 0.004 & -0.000 \\
0.003 & -0.037 & -0.021 & 0.006 & 0.000 & 0.007 & 0.013 & -0.003 \\
0.011 & -0.011 & -0.018 & -0.013 & -0.007 & 0.000 & 0.006 & -0.003 \\
-0.003 & 0.003 & 0.000 & -0.004 & -0.013 & -0.006 & 0.000 & -0.001 \\
-0.006 & 0.003 & 0.002 & 0.000 & 0.003 & 0.003 & 0.001 & 0.000 \\

\end{pmatrix}
\endgroup.
\end{equation*}

These matrices are ordered according to the state vector $\begin{pmatrix}\ket{000}&\ket{001}&\ket{010}&\ket{011}&\ket{100}&\ket{101}&\ket{110}&\ket{111}\end{pmatrix}^T$ where we adopt a convention of $\ket{\text{Ion~1},\text{Ion~2},\text{Ion~3}}$ for labelling. The absolute values of this matrix are graphically depicted in Fig.~4c. We also plot the tripartite W state entanglement fidelities and rates for varying photon acceptance window sizes in Fig.~S7c ranging from 150~ns to 2850~ns.

\subsubsection{Entangling Two Ions In the Same Device Using a Single Photon Protocol}
Data presented in this subsection is used to derive the results presented in Extended Data Fig.~10. The table below presents raw (uncorrected) Pauli-basis measurement results with a $500~$ns photon acceptance window, where $N_{ab}$ corresponds to the number of measurement outcomes where Ion 1 was in the $\ket{a}$ state and Ion 3 was in the $\ket{b}$ state.

\begin{center}
\begin{tabular}{|c|c|c|c|c|c|}
 \hline
 Ion 1 Basis & Ion 3 Basis & $N_{00}$ & $N_{01}$ & $N_{10}$ & $N_{11}$\\
 \hline\hline
  $X$ & $X$ & 2767 & 795 & 773 & 2734 \\ 
 \hline
  $X$ & $Y$ & 1727 & 1861 & 1718 & 1699\\ 
 \hline
  $X$ & $Z$ & 1770 & 1901 & 1709 & 1861\\ 
 \hline
  $Y$ & $X$ & 1865 & 1911 & 1815 & 1652\\ 
 \hline
  $Y$ & $Y$ & 2973 & 819 & 761 & 2728 \\ 
 \hline
  $Y$ & $Z$ & 1725 & 1901 & 1655 & 1835\\ 
 \hline
  $Z$ & $X$ & 1766 & 1800 & 1690 & 1735\\ 
 \hline
  $Z$ & $Y$ & 1810 & 1762 & 1756 & 1686 \\ 
 \hline
 $Z$ & $Z$ & 526 & 3015 & 2824 & 673 \\ 
 \hline
\end{tabular}
\end{center}
The readout transformation matrix is measured to be:
\begin{equation*}
\begin{pmatrix}
p_{00}\\p_{01}\\p_{10}\\p_{11}
\end{pmatrix}=
\begin{pmatrix}
0.951 & 0.041 & 0.044 & 0.002 \\
0.025 & 0.935 & 0.001 & 0.043 \\
0.023 & 0.001 & 0.930 & 0.040 \\
0.001 & 0.023 & 0.025 & 0.915 \\
\end{pmatrix}
\begin{pmatrix}
n_{00}\\n_{01}\\n_{10}\\n_{11}
\end{pmatrix},
\end{equation*}
where $\{p_{ab}\}$ and $\{n_{ab}\}$ are the measured and actual populations, respectively, associated with Ion 1 in the $\ket{a}$ state and Ion 3 in the $\ket{b}$ state. After performing maximum likelihood tomography, the resulting density matrix for our system is:
\begin{equation*}
\hat{\rho}=
\begingroup
\setlength\arraycolsep{10pt}
\begin{pmatrix}
0.041+0.000i & 0.003+0.004i & 0.001+0.000i & -0.003+0.002i \\ 
0.003-0.004i & 0.455+0.000i & 0.323+0.015i & 0.003+0.001i \\ 
0.001-0.000i & 0.323-0.015i & 0.425+0.000i & 0.004+0.002i \\ 
-0.003-0.002i & 0.003-0.001i & 0.004-0.002i & 0.080+0.000i \\ 

\end{pmatrix}
\endgroup.
\end{equation*}
This matrix is ordered according to the state vector $\begin{pmatrix}\ket{00}&\ket{01}&\ket{10}&\ket{11}\end{pmatrix}^T$ where we adopt a convention of $\ket{\text{Ion~1},\text{Ion~3}}$ for labelling. The absolute values of this matrix are graphically depicted in Extended Data Fig.~10d.

\subsubsection{Probabilistic Quantum State Teleportation}
Data presented in this subsection is used to derive the results presented in Fig.~S4. The table below presents raw (uncorrected) Pauli-basis measurement results on Ion 1 after teleportation of different target states prepared on Ion 2 with a two-sided acceptance window size of $W=1~\mu$s. Here, $N_{a}$ corresponds to the number of measurement outcomes where Ion 1 was in the $\ket{a}$ state.

\begin{center}
\begin{tabular}{|c|c|c|c|}
 \hline
 Ion 2 target state & Ion 1 readout basis & $N_{0}$ & $N_{1}$\\
 \hline\hline
  $\ket{+X}$ & $X$ & 137 & 33\\ 
 \hline
  $\ket{+X}$ & $Y$ & 94 & 91\\ 
 \hline
  $\ket{+X}$ & $Z$ & 99 & 81\\ 
 \hline
  $\ket{-X}$ & $X$ & 39 & 139 \\ 
 \hline
  $\ket{-X}$ & $Y$ & 82 & 80\\ 
 \hline
  $\ket{-X}$ & $Z$ & 92 & 73 \\ 
 \hline
  $\ket{+Y}$ & $X$ & 95 & 82\\ 
 \hline
  $\ket{+Y}$ & $Y$ & 132 & 37 \\ 
 \hline
  $\ket{+Y}$ & $Z$ & 84 & 72\\ 
 \hline
  $\ket{-Y}$ & $X$ & 75 & 84 \\ 
 \hline
  $\ket{-Y}$ & $Y$ & 39 & 133  \\ 
 \hline
  $\ket{-Y}$ & $Z$ & 75 & 70 \\ 
 \hline
  $\ket{0}$ & $X$ & 82 & 76 \\ 
 \hline
  $\ket{0}$ & $Y$ & 68 & 92 \\ 
 \hline
 $\ket{0}$ & $Z$ & 140 & 12 \\ 
 \hline
  $\ket{1}$ & $X$ & 83 & 86\\ 
 \hline
  $\ket{1}$ & $Y$ & 93 & 72  \\ 
 \hline
 $\ket{1}$ & $Z$ & 25 & 154 \\ 
 \hline
\end{tabular}
\end{center}
The readout transformation matrix is measured to be:
\begin{equation*}
\begin{pmatrix}
p_{0}\\p_{1}
\end{pmatrix}=
\begin{pmatrix}
0.986 & 0.029\\
0.014 & 0.971 \\
\end{pmatrix}
\begin{pmatrix}
n_{0}\\n_{1}
\end{pmatrix},
\end{equation*}
where $\{p_{a}\}$ and $\{n_{a}\}$ are the measured and actual populations, respectively, associated with Ion 1 in the $\ket{a}$ state. For each initial state, Ion~1's density matrix after teleportation is derived through the following equation:

\begin{equation*}
\hat{\rho}=\frac{1}{2}\left[\mathbb{I}+\langle\hat{X}\rangle\hat{X}+\langle\hat{Y}\rangle\hat{Y}+\langle\hat{Z}\rangle\hat{Z}\right],
\end{equation*}

where $\hat{X}$, $\hat{Y}$, and $\hat{Z}$ are the Pauli matrices, and $\langle\hat{X}\rangle$, $\langle\hat{Y}\rangle$ and $\langle\hat{Z}\rangle$ are the measured Pauli operator expectation values, corrected for readout infidelity, obtained from the preceding measurement results. The resulting density matrices are:

\begin{align*}
&\hat{\rho}_{\ket{+X}}=
\begingroup
\setlength\arraycolsep{10pt}
\begin{pmatrix}
0.544+0.000i & 0.312-0.001i  \\ 
0.312+0.001i & 0.456+0.000i
\end{pmatrix}
\endgroup
,&\hat{\rho}_{\ket{-X}}=
\begingroup
\setlength\arraycolsep{10pt}
\begin{pmatrix}
0.552+0.000i & -0.301+0.001i  \\ 
-0.301-0.001i & 0.448+0.000i
\end{pmatrix}
\endgroup,\\ \\
&\hat{\rho}_{\ket{+Y}}=
\begingroup
\setlength\arraycolsep{10pt}
\begin{pmatrix}
0.532+0.000i & 0.031-0.286i  \\ 
0.031+0.286i & 0.467+0.000i
\end{pmatrix}
\endgroup
,&\hat{\rho}_{\ket{-Y}}=
\begingroup
\setlength\arraycolsep{10pt}
\begin{pmatrix}
0.510+0.000i & -0.037+0.293i  \\ 
-0.037-0.293i & 0.490+0.000i
\end{pmatrix}
\endgroup,\\ \\
&\hat{\rho}_{\ket{0}}=
\begingroup
\setlength\arraycolsep{10pt}
\begin{pmatrix}
0.932+0.000i & 0.012+0.086i  \\ 
0.012-0.086i & 0.068+0.000i
\end{pmatrix}
\endgroup
,&\hat{\rho}_{\ket{1}}=
\begingroup
\setlength\arraycolsep{10pt}
\begin{pmatrix}
0.116+0.000i & -0.017-0.059i  \\ 
-0.017+0.059i  & 0.884+0.000i
\end{pmatrix}
\endgroup.
\end{align*}

 These matrices are ordered according to the state vector $\begin{pmatrix}\ket{0}&\ket{1}\end{pmatrix}^T$. The Bloch vectors corresponding to these density matrices are graphically depicted in Fig.~S4c.

\subsection{Multiplexed Quantum Repeater}

This paper focuses on establishing a single entanglement link, however, for scalable long-range quantum networking a quantum repeater will be required to overcome the exponential reduction in entanglement rate with channel length due to photon absorption. The theory associated with such quantum repeater platforms in a non-multiplexed context is well established \cite{briegel1998}. For solid state platforms these repeater schemes typically rely on auxiliary nuclear-spin quantum memories combined with deterministic gates between electronic and nuclear qubits; we previously demonstrated this functionality for our platform using Vanadium nuclear spins \cite{Ruskuc2022}. In this section we propose two different multiplexed quantum repeater protocols.

\subsubsection{Scheme 1}
As with a typical quantum repeater, a single long-distance entanglement link (distance $L$) is subdivided into $2^n$ shorter links of length $L/2^n$. At each node we have $N$ ions that can be used for multiplexing. The sequence proceeds as follows:
\begin{enumerate}
\item First we perform multiplexed entanglement attempts on every second link (i.e. between nodes 1~and~2, 3~and~4, etc.). We assume that qubit memory times are sufficiently long to enable deterministic entanglement \cite{Humphreys2018a}, enabling parallelized preparation of $N$ Bell pairs for each link in a time $L/(c2^np_\text{link})$, where $p_\text{link}$ is the probability of entanglement success between two nodes, and $c$ is the speed of light.
\item At each intermediate node, we swap the spin state from each $^{171}$Yb ion to its respective local Vanadium nuclear spin memory.
\item Next, we establish entanglement on the remaining links (i.e. between nodes 2~and~3, 4~and~5, etc.). We again do this repeatedly in a multiplexed fashion until each of these links contains $N$ entangled Bell pairs.
\item Finally, we perform deterministic Bell state measurements (BSMs) between all electronic and nuclear spin memories.
\end{enumerate}
This leads to a total of $N$ end-to-end entanglement links between the two end nodes in the quantum repeater chain. The average time to establish a single entanglement link is then given by:
\begin{equation}
T_\text{entangle}=\frac{2L}{c2^nNp_\text{link}}.
\end{equation}
We have assumed that the time for an entanglement attempt is dominated by the photon travel time between nodes, and that the duration of the swap gates and BSMs is negligible. We can see that this protocol boosts the entanglement rate by a factor of $N$ compared to a non-multipexed repeater. However, one potential downside to this approach is that entanglement must be stored in quantum memories for a duration $L/(c2^np_\text{link})$, both in electronic memories while waiting for all $2^n$ links to be established, and also in nuclear memories during establishment of entanglement links in Step 3. Therefore, we propose an alternative scheme which utilizes multiplexing to also reduce the quantum memory duration requirements.

\subsubsection{Scheme 2}

In the second scheme a single long-distance entanglement link (distance $L$) is also subdivided into $2^n$ shorter links of length $L/2^n$, the sequence is depicted in Fig.~S8 and proceeds as follows:
\begin{enumerate}
\item First we perform multiplexed entanglement attempts on all links in parallel (inter-node entanglement). We repeat this multiple times until each intermediate node deterministically attains two entanglement links: one to the node on the left, one to the node on the right. Note that these links will utilize two different $^{171}$Yb ions at each intermediate node (Fig.~S8a).
\item In each intermediate node, we swap the spin state from each of the two $^{171}$Yb ions involved in an entanglement link to their respective local Vanadium nuclear spin memories (Fig.~S8b).
\item Next, we establish entanglement between the pair of $^{171}$Yb ions within each intermediate node using the protocol demonstrated in Extended Data Fig.~10. We repeat these attempts until entanglement is deterministically generated on all intra-node links (Fig.~S8c).
\item Finally, we perform deterministic Bell state measurements (BSMs) between all electronic and nuclear spin memories involved in the entanglement chain (Fig.~S8d), leading to a single end-to-end entanglement link (Fig.~S8e).
\end{enumerate}

Crucially, even though the intra-node entanglement in Step~3 is probabilistic, it can be performed repeatedly until success without destroying the previously generated inter-node entanglement links. Furthermore, since these $^{171}$Yb ions are co-located in the same cavity, there is no need to utilize multiplexing to boost the entanglement rate, hence Step~3 can deterministically connect ions within the same node. The total time to establish an end-to-end entanglement link via the quantum repeater chain is given by:
\begin{equation}
T_\text{entangle}=\frac{2L}{c2^nNp_\text{link}}+\frac{\tau_\text{int}}{p_\text{int}}.
\end{equation}
Where $c$ is the speed of light, $N$ is the number of ions in each node utilized for multiplexing, $p_\text{link}$ is the probability of entanglement success between two nodes, $\tau_\text{int}$ is the time per entanglement attempt between two ions in the same node (limited by the ion initialization time) and $p_\text{int}$ is the probability of entanglement success for ions in the same node. We have again assumed that the time for an inter-node entanglement attempt is dominated by the photon travel time, and that the duration of the swap gates and BSMs is negligible. Note here that the memory time requirements have been substantially reduced: entangled states only need to be stored for a duration of at most $2L/(c2^nNp_\text{link})$ during Step 1 on electronic spins and a duration $\tau_\text{int}/p_\text{int}$ during Step 3, on both nuclear and electronic spins. Therefore, when the intra-node entanglement rate is considerably faster than the non-multiplexed inter-node entanglement rate, this protocol will outperform Scheme~1 presented previously.

\clearpage
\newpage
\renewcommand{\figurename}{Fig.}
\renewcommand{\thefigure}{S\arabic{figure}}
\setcounter{figure}{0}

\begin{figure}[h!]
\includegraphics[width=180mm]{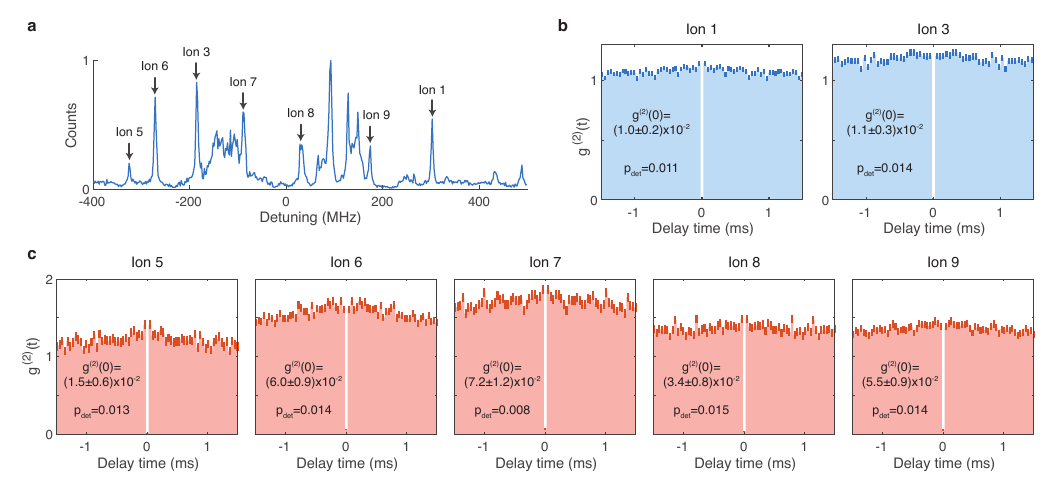}
\caption{{\bf Device~1 second order photon correlation measurements} {\bf a,} Normalized photoluminescence spectrum measured using Device~1, Ions~1 and 3 are used in this work. We additionally characterized Ions 5--9 in this device. Note that ion frequencies change slightly after each thermal-cycle of our cryostat (attributed to a changing strain environment in the crystal); therefore this spectrum looks slightly different compared to Fig.~1b. {\bf b,} Second order correlation function, $g^{(2)}(t)$, of photons emitted from Ions~1 and 3. We observe $g^{(2)}(0)=(1.0\pm0.2)\times 10^{-2}$ for Ion~1 and $g^{(2)}(0)=(1.1\pm0.3)\times 10^{-2}$ for Ion~3. {\bf c,} We characterized an additional 5 ions in this device to demonstrate the scalability of our multiplexing protocol. Each of these has a value of $g^{(2)}(0)<0.1$. In all cases, the correlation functions are normalized to have a value of 1 at very large delay times. We have also labelled the single photon detection efficiencies, $p_\text{det}$, for each of the ions.}
\end{figure}

\begin{figure}[h!]
\includegraphics[width=180mm]{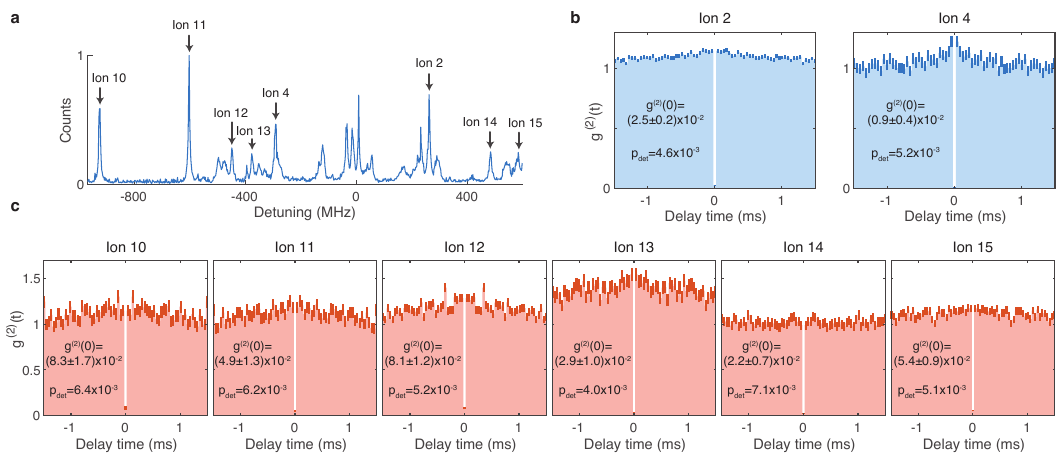}
\caption{{\bf Device~2 second order photon correlation measurements} {\bf a,} Normalized photoluminescence spectrum measured using Device~2, Ions~2 and 4 are used in this work. We additionally characterized Ions 10--15 in this device. Note that ion frequencies change slightly after each thermal-cycle of our cryostat (attributed to a changing strain environment in the crystal); therefore this spectrum looks slightly different compared to Fig.~1b. {\bf b,} Second order correlation function, $g^{(2)}(t)$, of photons emitted from Ions~2 and 4. We observe $g^{(2)}(0)=(2.5\pm0.2)\times 10^{-2}$ for Ion~2 and $g^{(2)}(0)=(0.9\pm0.4)\times 10^{-2}$ for Ion~4. {\bf c,} We characterized an additional 6 ions in this device to demonstrate the scalability of our multiplexing protocol. Each of these has a value of $g^{(2)}(0)<0.1$. In all cases, the correlation functions are normalized to have a value of 1 at very large delay times. We have also labelled the single photon detection efficiencies, $p_\text{det}$, for each of the ions.}
\end{figure}

\begin{figure}[h!]
\includegraphics[width=160mm]{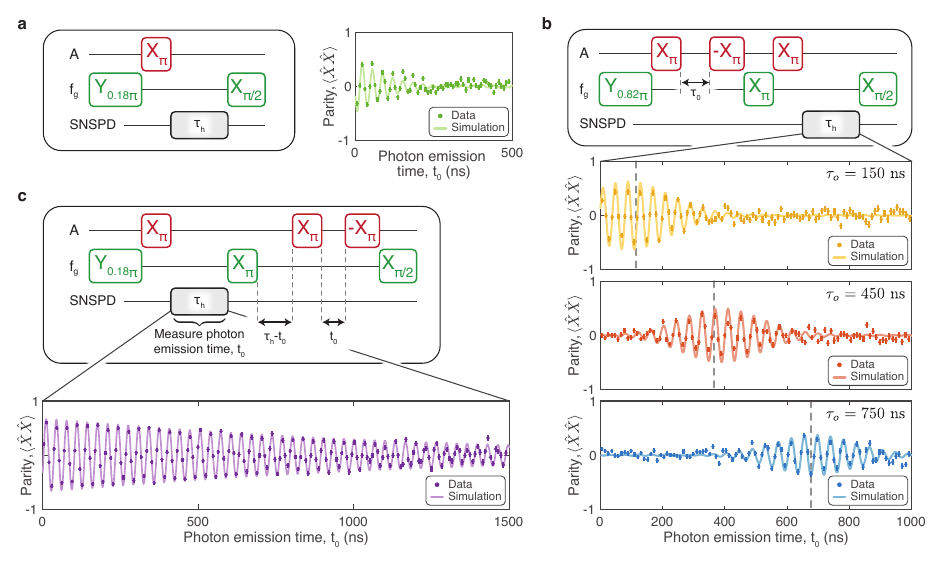}
\caption{{\bf Influence of photonic coherence in entanglement heralding protocols}. {\bf a,} Ramsey entanglement scheme, reproduced from Fig. 2b, Panel \textcircled{1}. The entangled state coherence is probed via a measurement of the $X$-basis parity expectation value, $\langle\hat{X}\hat{X}\rangle$, correlated with the photon emission time, $t_0$. The coherence oscillates at the ions' optical frequency difference, $\Delta\omega_{12}\approx2\pi\times31~$MHz and undergoes a Gaussian envelope decay with $185\pm15$~ns timescale, limited by the ions' optical Ramsey coherence times. {\bf b,} In this sequence, prior to entanglement heralding, optical phase is accumulated for a fixed duration, $\tau_0$, between two consecutive $A$ transition $\pi$ pulses. After a qubit $\pi$ pulse, the ions are excited and, during subsequent free evolution, the optical phase is compensated. Maximal coherence occurs when a heralding photon is detected at $t_0=\tau_0$. The three panels plot measurement results with $\tau_0=$150, 450 and 750~ns from top to bottom, respectively. We see that the coherence still exhibits Gaussian decay limited by the optical Ramsey coherence time. {\bf c,} Using the dynamic rephasing protocol described in the main text, we observe a significant increase in heralded coherence which exhibits exponential decay with a $1/e$ timescale of $970\pm30~$ns, now limited by the Purcell-enhanced optical lifetimes of the two ions (reproduced from Extended Data Fig.~6). In all cases, markers and solid lines correspond to experimental data and simulations, respectively, as detailed in Supplementary Information Section~G.}
\end{figure}

\begin{figure}[h!]
\includegraphics[width=0.48\textwidth]{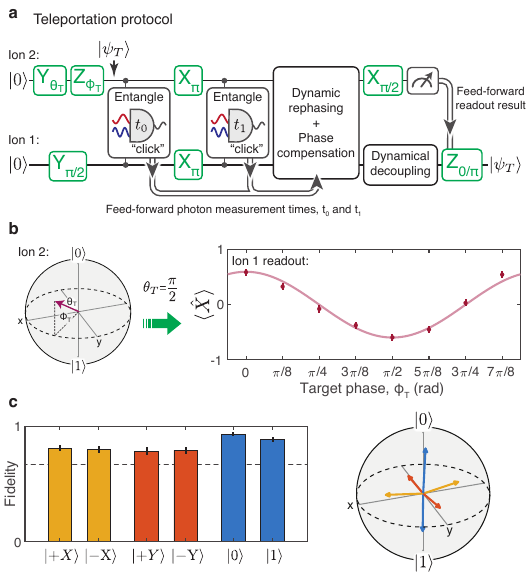}
\caption{{\bf Probabilistic quantum state teleportation between two remote $^{171}$Yb spin qubits}. {\bf a,} A target state, $\ket{\psi_T}=\cos(\theta_T/2)\ket{0}+e^{i\phi_T}\sin(\theta_T/2)\ket{1}$, prepared on Ion 2, is teleported to a remote qubit, Ion 1, which starts in $1/\sqrt{2}(\ket{0}+\ket{1})$ ($\theta_T$ and $\phi_T$ are angles that encode the target state). Both ions are optically excited and a photon is detected at time $t_0$. A $\pi$ pulse is applied to both qubits and a second optical excitation is followed by detection of a second photon at time $t_1$ leading to a quantum state $\ket{\psi(t_0,t_1)}=\cos(\theta_T/2)\ket{10}+e^{i(\phi_T+\phi(t_0,t_1))}\sin(\theta_T/2)\ket{01}$, where $\phi(t_0,t_1)=\Delta \omega_{12} (t_0-t_1)$. Dynamic rephasing and phase compensation are performed to cancel the stochastic phase, $\phi(t_0,t_1)$. Ion 2 is read out in the $X$ basis; meanwhile, dynamical decoupling preserves Ion 1's coherence. A qubit $Z$-rotation on Ion 1, by an angle of $0$ or $\pi$ conditioned on Ion 2's feed-forwarded measurement result, completes the quantum state transfer. Finally, Ion 1 is measured in the $X$, $Y$ or $Z$ bases to perform tomography of the single qubit state. {\bf b,} We benchmark this quantum state teleportation protocol by preparing Ion 2 in equally weighted superposition states with polar angle $\theta_T=\pi/2$ and variable azimuthal angle, $\phi_T$. After teleportation, the projection along the $X$-axis, $\langle\hat{X}\rangle$, representing the coherence of Ion 1's qubit, is plotted against $\phi_T$. {\bf c,} We estimate the fidelity of this protocol by independently preparing the six cardinal Bloch sphere states ($\ket{\pm X}, \ket{\pm Y}, \ket{0}, \ket{1}$) on Ion 2. Each state is teleported onto Ion 1 and measured. The fidelities of the six teleported states are plotted in the left panel with an average value of $0.834\pm0.011$ which exceeds the classical bound of $2/3$ (dashed line). The right panel plots the corresponding Bloch vectors of the six teleported states.}
\end{figure}

\begin{figure}[h!]
\includegraphics[width=90mm]{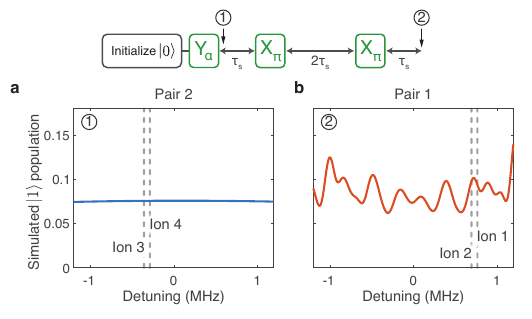}
\caption{{\bf Simulated effect of qubit drive detuning in multiplexed entanglement experiments}. We simulate the qubit control pulse sequence used for the multiplexed entanglements presented in Fig.~3 to investigate the effects of microwave detuning. {\bf a,} Simulated qubit population is plotted against pulse detuning after preparation of a weak superposition via a small rotation labelled $Y_{\alpha}$, this corresponds to the qubit populations when attempting to entangle Pair~2. Detunings corresponding to Ion~3 and 4 qubit frequencies are labelled. {\bf b,} Simulated qubit population is plotted against pulse detuning after preparation of a weak superposition followed by two dynamical decoupling periods (with inter-pulse separation $2\tau_s=5.8~\mu$s), this corresponds to the qubit populations when attempting to entangle Pair~1. Ion~1 and 2 qubit frequencies are labelled. Notice the considerable increase in $\ket{1}$ state population for small detuning when entangling Pair~1 compared to Pair~2.}
\end{figure}

\begin{figure}[h!]
\includegraphics[width=107mm]{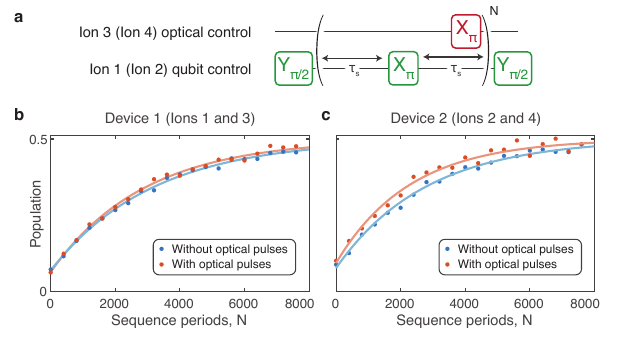}
\caption{{\bf Measuring the effect of cross-talk during multiplexed entanglement measurements}. {\bf a,} We investigate the effect of optical excitation of Pair 2 ions on the coherence of Pair 1. Specifically, for Device~1 (Device~2) we apply an XY-8 dynamical decoupling sequence to Ion~1 (Ion~2), with inter-pulse separation $2\tau_s=5.9~\mu$s, whilst concurrently applying optical $\pi$ pulses to Ion~3 (Ion~4) at the centre of each free evolution period. We measure the coherence of Ion~1's (Ion~2's) qubit as a function of the number of sequence periods, $N$. {\bf b,} For Device~1 we measure the coherence of Ion~1's qubit with, and without, optical pulses applied to Ion~3 leading to exponential decays with $1/e$ constants of $N_{1/e}=2890\pm60$ and $N_{1/e}=3170\pm50$ for these two cases, respectively. {\bf c,} For Device~2 we measure the coherence of Ion~2's qubit with, and without, optical pulses applied to Ion~4 leading to exponential decays with $1/e$ constants of $N_{1/e}=2290\pm70$ and $N_{1/e}=2800\pm110$ for these two cases, respectively.}
\end{figure}

\begin{figure}[h!]
\includegraphics[width=180mm]{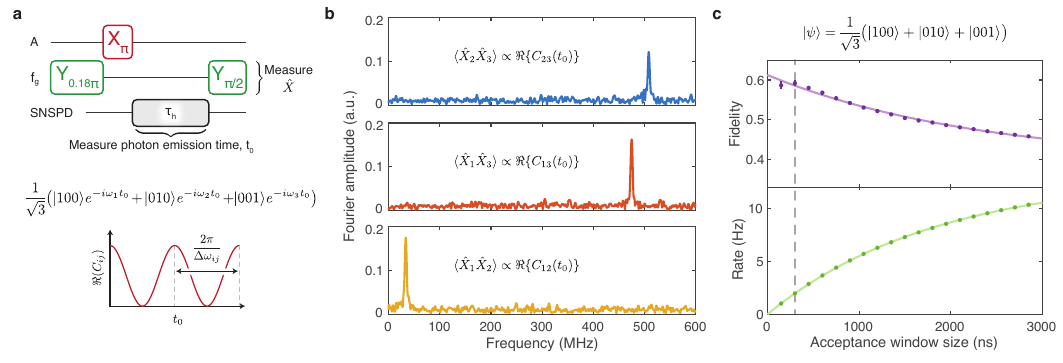}
\caption{{\bf Tripartite W state: additional results} {\bf a,} Tripartite W states are generated using a single photon heralding protocol, whereby  each Ion's qubit is prepared in a weak superposition state, all three ions are resonantly optically excited and entanglement is heralded by the detection of a single photon at stochastic time $t_0$. Subsequently, all three qubits are read out in the $X$ basis. The tripartite W state contains three quantum phases depending on the optical transition frequency of the excited qubit, $\omega_i$, multiplied by $t_0$. There are six quantum coherences, $C_{ij}=\Trace\{\hat{\rho}\ket{01}_{ij}\bra{10}_{ij}\otimes\ket{0}_k\bra{0}_k\}$, where $i,j,k$ correspond to permutations of qubits 1, 2 and 3. We note that the real part of the coherence term $\real\{C_{ij}\}$ will exhibit oscillations at the optical frequency difference between Ions $i$ and $j$ with photon measurement time, i.e. $\real\{C_{ij}(t_0)\}\propto\cos(\Delta\omega_{ij}t_0)$ where $\Delta\omega_{ij=}\omega_j-\omega_i$. {\bf b,} We probe the time dependence of these coherence terms by measuring two-qubit $X$-basis parity expectation values and noting that, for the ideal W state, $\langle\hat{X}_i\hat{X}_j\rangle\propto\real\{C_{ij}(t_0)\}$. We plot the Fourier transform of each parity measurement, the three panels correspond to $C_{23},~C_{13}$ and $C_{12}$ from top to bottom. In each case we observe an oscillation at the corresponding optical frequency difference, i.e. $\approx2\pi\times508$~MHz, $\approx2\pi\times475$~MHz and $\approx2\pi\times34$~MHz, respectively. {\bf c,} After performing dynamic rephasing and real-time phase compensation we obtain a deterministic phase-corrected W state. We characterize the rate and fidelity of these phase-corrected tripartite W states for varying photon acceptance window sizes and find values that range from $\mathcal{F}=0.59\pm0.01$ and $\mathcal{R}=1.0~$Hz for a 150~ns window size to $\mathcal{F}=0.458\pm0.003$ and $\mathcal{R}=10$~Hz for a 2850~ns window size.}
\end{figure}

\begin{figure}[h!]
\includegraphics[width=115mm]{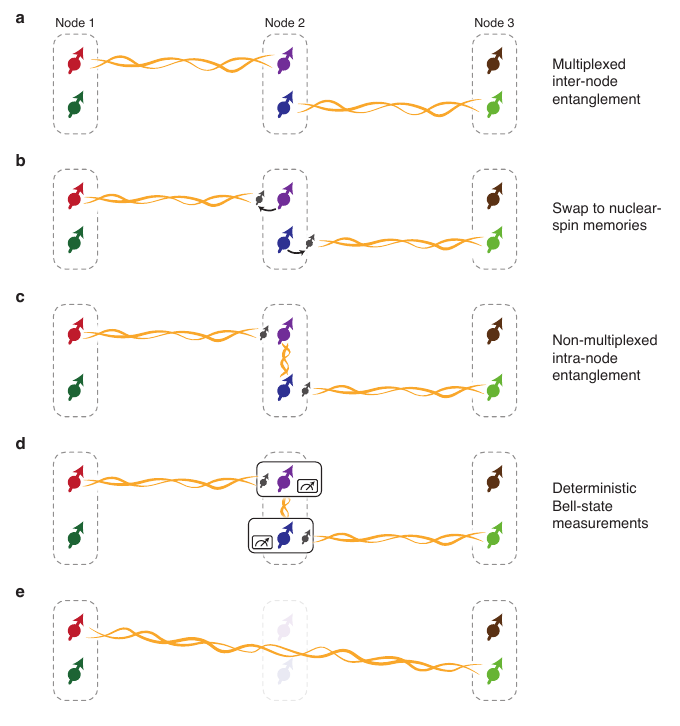}
\caption{{\bf Multiplexed quantum repeater protocol} {\bf a,} Entanglement is established between all nodes in parallel using the multiplexed protocol detailed in Fig.~3. Note that each intermediate node now contains two entanglement links: one to the node on the left, one to the node on the right, utilizing two different $^{171}$Yb ions. {\bf b,} The state of the $^{171}$Yb qubits at each intermediate node is swapped to their local Vanadium memories. {\bf c,} Entanglement is established between pairs of $^{171}$Yb qubits within each intermediate node. Note that heralding attempts can be repeated until success without destroying the previously generated inter-node entanglement. {\bf d,} Bell state measurements are performed on each pair of $^{171}$Yb and Vanadium qubits. {\bf e,} This leads to a single end-to-end entanglement link.}
\end{figure}

\clearpage

\section*{Acknowledgements}
This work was funded primarily by the Air Force Office of Scientific Research Grant No. FA9550-22-1-0178 and the Institute of Quantum Information and Matter, an NSF Physics Frontiers Center (PHY-1733907) with support from the Moore Foundation. We also acknowledge funding from NSF 2210570 and NSF 2137984. The device nanofabrication was performed in the Kavli Nanoscience Institute at the California Institute of Technology. A.R. acknowledges support from the Eddleman Graduate Fellowship. C.-J.W. acknowledges support from the J. Yang and Family Foundation and a Taiwanese government scholarship to study abroad. E.G. acknowledges support from the National Science Foundation Graduate Research Fellowship under Grant No. 2139433 and the National Gem Consortium. S.L.N.H. acknowledges support from the AWS Quantum Postdoctoral Fellowship. J.C. acknowledges support from the Terman Faculty Fellowship at Stanford. We thank E. Paul for help with the experimental setup, J. Thompson and M. T. Uysal for discussion related to the entanglement protocol, J. Borregaard, D. Lukin, T. Xie, M. Lei, R. Fukumori, E. Liu, and B. Grinkemeyer for useful discussions, and J. Rochman, T. Zheng, S. Gu and B. Baspinar for help related to nanofabrication.

\end{document}